\documentclass[onecolumn,authoryear]{els-mrw} 
\usepackage{amsmath,amssymb,amsfonts,amsthm,makeidx,graphicx}
\usepackage{txfonts}
\usepackage{helvet}
\usepackage{xcolor}
\usepackage{amsmath}
\numberwithin{equation}{section}

\newcommand{\la}{\langle}
\newcommand{\ra}{\rangle}
\newcommand{\lla}{\la\!\la}
\newcommand{\rra}{\ra\!\ra}
\newcommand{\beq}{\begin{eqnarray}}
\newcommand{\eeq}{\end{eqnarray}}
\renewcommand{\d}{\partial}
\newcommand{\eps}{\epsilon}
\newcommand{\half}{{1\over 2}}

\newcommand{\mn}{{\mu\nu}}

\newcommand{\Tr}{{\rm Tr}}

\newcommand{\bfL}{\mbox{{\boldmath $L$}}}
\newcommand{\bfA}{\mbox{{\boldmath $A$}}}

\newcommand{\bfx}{\mbox{{\boldmath $x$}}}

\newcommand{\bfV}{\mbox{{\boldmath $V$}}}

\newcommand{\bfp}{\Vec{p}}

\newcommand{\bfR}{\mbox{{\boldmath $R$}}}

\newcommand{\e}{{\rm e}}
\newcommand{\bfPhi}{\mbox{{\boldmath $\Phi$}}}

\newcommand{\btau}{\mbox{{\boldmath $\tau$}}}
\newcommand{\bpi}{\mbox{{\boldmath $\pi$}}}
\newcommand{\bbeta}{\mbox{{\boldmath $\beta$}}}

\newcommand{\balpha}{\mbox{{\boldmath $\alpha$}}}

\newcommand{\bfone}{\mbox{{\boldmath $1$}}}

\begin{document}
%\hfill{\rm YITP-26-49}

\chapter{Chiral Symmetry and Its Restoration in QCD}\label{chap1}
%\begin{Large}{\bf Chiral Symmetry and Its Restoration in QCD}\label{chap1}
%\end{Large}
%
%\vspace{1cm}

\author[1]{Teiji Kunihiro}%
%\author[2]{Second Author}%

%\author[1,2]{Third Author}%

\address[1]{\orgname{Yukawa Institute for Theoretical Physics}, 
\orgdiv{Kyoto University}, \orgaddress{606-0892 Kitashirakawa Sakyoku, Kyoto, Japan}}
%\address[2]{\orgname{Name of Institute}, \orgdiv{Division or Department}, 
%\orgaddress{Address of Institute}}

%\articletag{Chapter Article tagline: update of previous edition,, reprint..}

\maketitle

%\begin{glossary}[Glossary]
%\term{QCD} 
%\term{SSB}
%\term{EoS} 
%\end{glossary}

\begin{keywords}
chiral symmetry\sep  spontaneous symmetry breaking \sep chiral anomaly
\sep degeneracy in spectral functions  \sep restoration of chiral symmetry \sep
effective restoration of U(1)$_A$ symmetry \sep lepton-pair production in relativistic heavy-ion collisions \sep quark-hadron continuity \sep mesic atoms \sep
equation of state with parity doubling
\end{keywords}

\begin{glossary}[Nomenclature]
\begin{tabular}{@{}lp{34pc}@{}}
QCD &Quantum Chromodynamics\\
SSB & Spontaneous Symmetry Breaking\\
EoS & equation of state\\
%Partitioning Method Equilibrium\hfill\break Partitioning Method\\
%ERA &Ecological Risk Assessment\\
%HC &Hazardous Concentration\\
\end{tabular}
\end{glossary}

\begin{abstract}[Abstract]
%\begin{itemize}
Chiral symmetry is an approximate symmetry of QCD in the light-quark sector.
 Its spontaneous breaking in the QCD vacuum explains why the pion is anomalously
 light and why it plays a central role in nuclear physics, including the tensor 
component of the nuclear force and the saturation properties of nuclei. 
This article introduces chirality through the Dirac equation and shows 
how a fermion mass mixes left- and right-handed components, in close analogy
 with the Bogoliubov--Valatin theory of superconductivity. We then explain 
spontaneous symmetry breaking, the Nambu--Goldstone theorem, and the role of
 the chiral condensate as an order parameter. The axial $U(1)_A$ anomaly and
 its consequences, especially for the $\eta'$ meson, are discussed. Chiral
 effective models, including Nambu--Jona-Lasinio-type models, linear sigma 
models with anomaly terms, and parity-doublet models for baryons, are reviewed
 as tools for describing hadron properties and the equation of state of dense 
matter. Finally, we discuss how chiral symmetry may be partially restored 
at finite temperature and/or baryon density, and summarize experimental
 probes such as deeply bound pionic atoms, dilepton production 
in relativistic heavy-ion collisions, $\eta'$-mesic nuclei, and baryonic observables.
\end{abstract}

%%%%%Large%%%%%%
%\begin{Large}

\textbf{Objectives}
\begin{itemize}
\item Understand why the pseudoscalar nature of the pion is central
 to nuclear forces and why this property is rooted in chiral symmetry
 and its spontaneous breaking in QCD.
\item Learn the distinction between chirality and helicity through
 the Dirac equation, and understand how a fermion mass mixes 
left- and right-handed components.
\item Understand spontaneous symmetry breaking and the Nambu--Goldstone theorem, 
emphasizing that the symmetry is broken by the vacuum 
rather than by the Hamiltonian or Lagrangian.
\item Learn how the approximate chiral symmetry of light-quark QCD
 is spontaneously broken by the nonzero chiral condensate 
$\langle\bar q q\rangle$, and why the pion and other light 
pseudoscalar mesons are pseudo-Nambu--Goldstone bosons.
\item Understand the axial $U(1)_A$ anomaly and its consequences
 for the $\eta'$ mass and for the pattern of chiral symmetry restoration.
\item Learn how Nambu--Jona-Lasinio-type models, linear sigma models, and parity-doublet
 models implement chiral symmetry, explicit breaking, and anomaly effects.
\item Understand, using the Hellmann--Feynman theorem, why the presence of
thermally excited pions at finite temperature and nucleons in nuclear matter
reduces the magnitude of the chiral condensate.
\item Become familiar with major experimental probes of chiral restoration
 in hot and dense matter, including pionic atoms, dilepton spectra, mesic nuclei,
 and baryonic observables.
\end{itemize}

\section{Introduction}\label{chap1:sec1}
\subsection*{From Nuclear Physics to Quantum Chromodynamics}
A nucleus is a femtometer-scale quantum many-body system in which protons and neutrons are bound despite the Coulomb repulsion among protons. The existence of such systems immediately points to a strong attractive interaction among nucleons. Yukawa's meson theory provided the first field-theoretical explanation of this nuclear force and predicted a new particle, later identified with the pion. The pion is remarkable not only because it is the lightest hadron, but also because it is a pseudoscalar particle with negative intrinsic parity. 
This property generates a characteristic tensor component of the nuclear force,
 proportional to 
$S_{12}(\mathbf{\sigma}_1, \mathbf{\sigma}_2, \mathbf{r}):=\mathbf{\sigma}_1\cdot\mathbf{r} \mathbf{\sigma}_2\cdot\mathbf{r}
-\mathbf{\sigma}_1\cdot\mathbf{\sigma}_2/3$,
which causes the intricate structure of the deuteron and plays an essential role
in nuclear many-body dynamics. Indeed,
because of this property, the only two-nucleon bound system, the deuteron,
has the spin $J=1$ and the orbital angular momentum $L=0$ with a small mixing of $L=2$  (\cite{Preston:1975}).
This mixing of the angular momentum is caused by the tensor force,
because $S_{12}(\mathbf{\sigma}_1, \mathbf{\sigma}_2, \mathbf{r})$ is
% an operator composed of the spin operators $\mathbf{\sigma}_{1,2}$ and 
%the relative coordinate $\mathbf{r}$ of the two nucleons, 
composed of the spin and coordinate operators of rank-2 spherical tensors, respectively.
It is remarkable that this seemingly unfavorable parity property
is also of primary importance 
 in realizing the saturation of binding energy and 
density in heavy nuclei and nuclear matter,
 together with other ingredients such as
 short-range repulsion, intermediate-range attraction, many-body correlations, 
and three-nucleon forces (\cite{Bethe:1971xm,Preston:1975,Kohno:2015hha}).
%as well as the existence and the property of the deuteron
% mentioned above. 

From the modern viewpoint, the special role of the pion is a consequence of the approximate chiral symmetry of Quantum Chromodynamics (QCD) and its spontaneous breaking in the vacuum. The pion is light because it is a pseudo-Nambu--Goldstone boson. Thus, phenomena that first appear as empirical features of nuclear physics are deeply connected with the nonperturbative structure of the QCD vacuum. This article explains this connection and reviews how chiral symmetry and its restoration are described theoretically and explored experimentally.

\subsection*{Puzzling structure of the vacuum of QCD and hadron properties}

In quantum field theory, the vacuum is not empty space but 
the lowest-energy state of the system, and
the matter being observed is the excited states
on top of the ground state, the vacuum. 
Conversely, determining the structure of the vacuum is
equivalent to determining the elementary excitations observed as matter.
In QCD-based hadron physics,
hadrons are elementary excitations built on the QCD vacuum that is 
determined by QCD dynamics.
QCD-based hadron physics, however,  presents several conceptual puzzles:

 The (classical) QCD Lagrangian (\cite{cheng1994gauge}, \cite{georgi2009weak})
 has a form quite similar to that of QED (quantum electrodynamics) as
\beq
{\cal L}_{QCD}^{cl} =  \bar{q} (i \gamma_{\mu} D^{\mu} - m_q ) q
 - {1 \over 4} F_{\mu\nu}^a F^{\mu\nu}_a \ \ ,
\label{eq:classkcalQCD}
\eeq
where $q$ denotes the quark field with three colors and $N_f$
 flavors
$q$=($u$, $d$, $s$, $\cdot \cdot \cdot $), $m_q$ is a mass matrix for current
quarks  $m_q$=diag.($m_u, m_d, m_s, \cdot \cdot \cdot $),
 $D_{\mu} (\equiv \partial_{\mu} - i g \lambda_C^a A_{\mu}^a$)
 is a covariant derivative
with colored gauge field $A_{\mu}^a$ ($a$=1 $\sim $ 8), 
 $g$ the strong coupling constant
and $\lambda_C^a$ the SU(3)-color Gell-Mann matrix. 
%  ($[\lambda_C^a,\,\lambda_C^b]=i f_{abc}\lambda_C^c$,
%tr($\lambda_C^a  \lambda_C^b$)=$\delta^{ab}/2$).  
The field strength $F_{\mn}^a$ is defined as 
$F_{\mn}^a = \partial_{\mu}A_{\nu}^a -
\partial_{\nu}A_{\mu}^a + gf_{abc} A_{\mu}^b A_{\nu}^c$.
QCD Lagrangian is written solely in terms of 
colored quark and gluon fields,
 while only their colorless composite states, i.e., {hadrons} 
are physical states and observable 
(\cite{povh2008particles}, \cite{cheng1994gauge}, \cite{georgi2009weak}).
This implies that the QCD vacuum is a non-perturbative state determined by
QCD dynamics, and hadrons are its elementary excitations. 
In addition to that, 
although QCD Lagrangian is invariant under chiral transformation\footnote{A clear
definition of it is given in a later section.},
the QCD vacuum does not respect its symmetry, and 
there are no apparent traces of this symmetry in low-energy hadron spectra.

These seemingly puzzling features are consequences of the fact that 
the chiral symmetry is {\em spontaneously broken} in the QCD vacuum.
Moreover, just as superconductivity in metals tends to disappear {with increasing temperature $T$ and/or magnetic-field strength $B$} (\cite{schrieffer2018theory}, \cite{tinkham2004introduction}), 
one may expect that some interesting changes occur in 
the nature of the QCD vacuum and the properties of hadrons
 as  elementary excitations on top of the vacuum
 by the environmental changes characterized by $T$, the baryon density $\rho$, and so on.

\subsection*{The purpose of this article}
 This article provides an overview  of chiral symmetry
in QCD, the notion of spontaneous symmetry breaking (SSB),
and its consequences on the properties of some specific hadrons.
Then we shall discuss how chiral symmetry is (partially)
restored in the hot and/or dense QCD matter, and also {briefly review how
possible signatures of chiral restoration are searched for in experiments}.
In the baryon sector, restoration of chiral symmetry may be reflected in parity doubling
and can influence the equation of state of dense matter as realized in the interior of neutron stars.
%how  the possible restoration of chiral symmetry is 
%verified or not in various experiments.
Effects of a strong magnetic field on the hadron properties as well as the QCD vacuum
will not be treated in the present article.

%We refer to (\cite{Hayano:2008vn,Fukushima:2013rx}) 
%as related review articles with references  up to some years ago:
%Both of them  give a concise account of the chiral symmetry of QCD and its
%breaking and restoration to be seen in QCD medium.
%A related but unique review is (\cite{Rapp:2009yu}), which focuses on 
% the vector and axial vector mesons as probes of 
%the chiral restoration in the QCD vacuum\footnote{This subject is also discussed in 
%the above two reviews.}.
%In comparison with these reviews, 
%the present article is largely based on heuristic arguments
%focusing on the conceptual aspects of the 
%problem, and hence should be more pedagogical.

\section{A Chirality-Based View of the Dirac Equation}
%A nonstandard view on the Dirac Equation and the Emergence of Chirality}

In this section, 
we derive the Dirac equation in a form that naturally displays left- and 
right-handed components
%we give a non-standard view of 
%the Dirac equation in a manner that naturally exhibits chirality, 
as given in
(\cite{Nambu:1961tp}),
and thereby clarify the conceptual origin of chiral symmetry.
% and 
%its analogous structure to the Bogoliubov-Valatin approach  
%the superconductivity in metals (\cite{schrieffer2018theory,tinkham2004introduction}).

\subsubsection*{From Pauli Theory to Dirac equation}

We begin with 
%a two-component Pauli spinor $\phi(x)$
quantum mechanics of a nonrelativistic spin-$1/2$ particle,
the free Hamiltonian of which is given by (\cite{sakurai:1967})
\begin{equation}
H_0=
\begin{pmatrix}
\frac{\mathbf{p}^2}{2m} & 0 \\
0 & \frac{\mathbf{p}^2}{2m}
\end{pmatrix}=\frac{\mathbf{p}^2}{2m}\mathbf{1}_2
=\frac{1}{2m}(\boldsymbol{\sigma}\cdot\mathbf{p})^2,
\end{equation}
where $\mathbf{p}=-i\nabla$, $\mathbf{1}_n$ denote $n$-dimensional Identity matrix, and
$\boldsymbol{\sigma}$ the Pauli matrices.
We use natural units, $\hbar=c=1$.
%Using an example, we shall show how this representation 
%leads to a nontrivial thing. 
Upon minimal substitution
$\mathbf{p} \rightarrow \mathbf{p}-q\mathbf{A},
\,H \rightarrow H - qA_0$,
one obtains the Pauli Hamiltonian
\begin{equation}
H_{\rm Pauli}
=\left[
\frac{1}{2m}(\mathbf{p}-q\mathbf{A})^2+qA_0\right]\bfone_2
-\frac{q}{2m}\boldsymbol{\sigma}\cdot\mathbf{B},
\end{equation}
with $\mathbf{B}=\nabla\times\mathbf{A}$.
The magnetic moment is therefore
\(
\boldsymbol{\mu} = \frac{q}{m}\frac{\boldsymbol{\sigma}}{2}
\),
corresponding to the gyromagnetic ratio $g=2$;
it is noteworthy that a nonrelativistic treatment with the Pauli spin matrix
has led to  $g=2$, which anticipates a deeper relativistic origin of it.

With the use of the above trick, the relativistic energy-momentum relation
$E^2-\mathbf{p}^2=m^2$
suggests a factorization:
%\begin{equation}
$(E-\boldsymbol{\sigma}\cdot\mathbf{p})
(E+\boldsymbol{\sigma}\cdot\mathbf{p})=m^2$ .
%\end{equation}
Promoting $E$ and $\mathbf{p}$ to operators,
$E \rightarrow i\partial_0,\,
\mathbf{p} \rightarrow -i\nabla$,
we obtain a relativistic wave equation for a particle
with a spin $1/2$ (\cite{sakurai:1967})
\begin{equation}
(i\partial_0-\boldsymbol{\sigma}\cdot(-i\nabla))
(i\partial_0+\boldsymbol{\sigma}\cdot(-i\nabla))
\phi(x)
=
m^2\phi(x).
\end{equation}
Defining two independent two-component spinors
%\begin{equation}
$\psi^{(L)}(x):= \phi(x)$ and
$\psi^{(R)}(x):=\frac{1}{m}
(i\partial_0-i\boldsymbol{\sigma}\cdot\nabla)\phi(x)$,
we obtain the coupled equation
\begin{align}
i(\partial_0+\boldsymbol{\sigma}\cdot\nabla)\psi^{(R)}
&= m\psi^{(L)}, \\
i(\partial_0-\boldsymbol{\sigma}\cdot\nabla)\psi^{(L)}
&= m\psi^{(R)} .
\end{align}
If we define the four-component  spinor by
\begin{equation}
\psi_C(x):=
\begin{pmatrix}
\psi^{(R)}(x) \\
\psi^{(L)}(x)
\end{pmatrix},
\end{equation}
the coupled equation is rewritten in a matrix form as
\beq
i\begin{pmatrix}
\d_0\bfone_2 + \boldsymbol{\sigma}\cdot\nabla &0 \\
0 &\partial_0\bfone_2 -\boldsymbol{\sigma}\cdot\nabla
\end{pmatrix}\psi_C(x)=
\begin{pmatrix}
0 & m\bfone_2 \\
m\bfone_2 &0 
\end{pmatrix}\psi_{C}(x).
\label{eq:coupl-Weyl}
\eeq
Now using the sum $\psi_+:=(\psi_R+\psi_L)/\sqrt{2}$
 and difference $\psi_-:=(\psi_R-\psi_L)/\sqrt{2}$,
we define new four-component spinor by
\[
\psi_D(x):=
\begin{pmatrix}
\psi_+(x) \\
\psi_-(x)
\end{pmatrix}=P\begin{pmatrix}
\psi^{(R)}(x) \\
\psi^{(L)}(x)
\end{pmatrix}=P\psi_C(x), \quad \quad
P:=\frac{1}{\sqrt{2}}\begin{pmatrix}
\bfone_2 & \bfone_2 \\
\bfone_2 & -\bfone_2
\end{pmatrix}=P^{-1},
\]
which diagonalizes the mass matrix as
\[
P\begin{pmatrix}
0 & \bfone_2 \\
\bfone_2 & 0
\end{pmatrix}P^{-1}=\begin{pmatrix}
\bfone_2 & 0 \\
0 & -\bfone_2
\end{pmatrix}=:\gamma^0.
\]
Similarly, we introduce the other gamma matrices as
\beq
\gamma^k=\begin{pmatrix}
 \Vec{0} & \sigma^k \\
-\sigma^k & \Vec{0}
 \end{pmatrix},\,\, \quad (k=1,\,2,3).
\eeq
The gamma matrices $\gamma^{\mu}$\, ($\mu=0,1,2,3$) in these forms are called
{\em the Dirac representation}.
The four gamma matrices  $\gamma^\mu$ satisfy the 
following anti-commutation relations irrespective of the representation;
\(
\{\gamma^\mu,\gamma^\nu\}=2g^{\mu\nu}
\)
with metric signature $(+,-,-,-)$.
Applying $P$ on Eq.~(\ref{eq:coupl-Weyl}) from the left, we
have the equation for $\psi_D(x)$ as follows
\beq
(i\gamma^\mu \partial_\mu - m)\psi_D(x)=0,
\eeq
which is nothing but the Dirac equation for a Dirac particle, which we call
a quark\footnote{One-flavor quark.}.

\subsubsection*{Notion of chirality for massless quarks}

Substituting $\psi_C= e^{-iE_pt+i\mathbf{p}\cdot \mathbf{r}}\tilde{\psi}_C(p)$
into Eq.~(\ref{eq:coupl-Weyl}), we have the eigenvalue equation 
\beq
\begin{pmatrix}
\boldsymbol{\sigma}\cdot \mathbf{p} & m\\
m & -\boldsymbol{\sigma}\cdot \mathbf{p}
\end{pmatrix}\tilde{\psi}_C(p)=E_p\tilde{\psi}_C(p),
\label{eq:weyl-mom-1}
\eeq
giving the well-known  energy eigenvalues
$E=\pm\sqrt{p^2+m^2}$.
In the massless case ($m=0$),  Eq.~(\ref{eq:weyl-mom-1}) is reduced to
\begin{eqnarray}
\boldsymbol{\sigma}\cdot \mathbf{p}\gamma_{C5}\tilde{\psi}_C
=E_P\,\tilde{\psi}_C, \quad \quad 
\gamma_{C5}:=\begin{pmatrix}
\bfone_2 & 0\\
0 & -\bfone_2
\end{pmatrix}.
\label{eq:weyl-mom}
\end{eqnarray}
 The eigenvalues $\pm 1$ of $\gamma_{C5}$
are called {\em chirality}, and the respective eigenstates read
$
\begin{pmatrix}
1 \\
0
\end{pmatrix}=:\psi_R$ and
$\begin{pmatrix}
0 \\
1
\end{pmatrix}
=:\psi_L$.
Of course, these are also
the energy eigenstates with the eigenvalues  
$E^{\pm}_{p}=\pm p$.
In terms of the {\em helicity} operator
$\boldsymbol{\sigma}\cdot \hat{\mathbf{p}}$\, ($\hat{\mathbf{p}}:=\mathbf{p}/p$), 
we have
\[
\boldsymbol{\sigma}\cdot \hat{\mathbf{p}}\,\psi_R = \psi_R,
\qquad
\boldsymbol{\sigma}\cdot \hat{\mathbf{p}}\,\psi_L = - \psi_L.
\]
Thus,  massless spin-$1/2$ states are simultaneous eigenstates of 
 the chirality and helicity with equal eigenvalues; the particle 
with helicity $+$ ($-$) has a spin parallel (anti-parallel) to the momentum,
implying that the direction of spin rotation is right-handed (left-handed),
 and hence the subscript $R$ ($L$).  
When the mass $m$ is nonzero, Eq.~(\ref{eq:weyl-mom-1}) shows that the eigenstates 
with chiralities $\pm 1$ mix with each other, and 
 the chirality is no longer a good quantum number, 
in contrast to the helicity\footnote{See p.240 of (\cite{Hatsuda:1994pi})}.

The structure of Eq.~(2.11), and the role played by the quark mass $m$, 
closely correspond to the basic equation and the gap $\Delta$
in the Bogoliubov--Valatin theory of superconductivity (\cite{schrieffer2018theory,tinkham2004introduction}),
as noticed by Nambu and Jona-Lasinio (\cite{Nambu:1961tp}). There is, however, apparently 
an important difference between the gap $\Delta$ 
in the superconductors and the mass $m$ because the former is generated dynamically, while 
the latter was thought to be given in the Hamiltonian from the outset.
The notion of the spontaneous breaking of chiral symmetry provides a clue
to the dynamical origin of the quark mass.

Here, we remark that the matrix $\gamma_{C5}$ takes the following form
in the Dirac representation where $\gamma^0$ is diagonal, 
\[
P\gamma_{C5}P^{-1}=\begin{pmatrix} 0 & \bfone_2\\
                  \bfone_2 & 0\end{pmatrix}=:\gamma_5.
\]
The representation of the gamma matrices in which $\gamma_{C5}$ is diagonal is
called {\em the Weyl or Chiral representation}\footnote{Needless to say, 
the eigenvalues of $\gamma_5$
coincide with those of $\gamma_{C5}$, and are called chirality.}.

\section{Some symmetry properties of QCD}
 
We focus on the three lightest quarks, i.e., the u, d, and s quarks, 
whose masses \footnote{The $\overline{\textrm{MS}}$ masses at the scale
$\mu=2$\,GeV.} 
%well below the $\Lambda_{\textrm{QCD}}\sim 200$-$300$ MeV
are, according to the Particle Data Group (\cite{ParticleDataGroup:2024cfk}), 
$m_u  =  (2.16 \pm 0.07)\, {\rm MeV},\,
m_d=(4.70 \pm 0.07)\, {\rm MeV}, \,
m_s  =  (93.5 \pm 0.8)\, {\rm MeV}$.
It is known that the dynamics of these light quarks is, in many low-energy contexts,
 well approximated by that given in the massless limit\footnote{It may
be related to the fact that the energy scale at which 
chiral symmetry breaking sets in is around 1.4 GeV and is much larger
than the light-quark masses (\cite{Manohar:1983md}).},
 which is called the {\em chiral limit}. A remark is that the chiral limit
 is quantitatively less accurate for phenomena involving the strange quark, because
the strange-quark mass is much larger than
the u- and d-quark masses.

For completeness, the c, b, and t quarks are much heavier than the light quarks and
their masses in the same definition are about
1.3, 4.2, and 170\, GeV, respectively.
The chiral symmetry is relevant 
to the light three quarks (\cite{cheng1994gauge}, \cite{georgi2009weak}).
We define quark-mass-related quantities that are used in later discussions:
\[
\hat{m}:=(m_u+m_d)/2=3.49 \pm 0.07\, {\rm MeV}, \qquad
m_s/\hat{m}=27.33\substack{+0.18 \\ -0.14}.
\]

\subsection{Chiral symmetry}
%In what follows, we write the quark field as $q_f^a$ 
%with the indices $a$ and $f$ referring color and flavor, respectively.
%We adopt the following representation of the gamma matrices:
%\[
%\gamma^0 =
%\begin{pmatrix}
%\mathbf{1}_2 & \vec{0} \\
%\vec{0} & -\mathbf{1}_2
%\end{pmatrix},
%\qqud
%\gamma^k =
%\begin{pmatrix}
%\vec{0} & \sigma^k \\
%-\sigma^k & \vec{0}
%\end{pmatrix},
%\qquad (k=1,2,3),
%\]
%where $\mathbf{1}_n$ denotes the $n\times n$ identity matrix, and $\sigma^k$ are 
%the Pauli matrices.

For introducing the chiral symmetry (\cite{cheng1994gauge}, \cite{georgi2009weak}), 
we define the left- and right-handed quark fields by
\[
q_L=P_L\,q,
%\frac{1-\gamma_5}{2}\, q
%   =\begin{pmatrix} u_L \\ d_L \\ s_L \end{pmatrix},
\quad
q_R=P_R\,q;\,\,
\gamma_5\, q_L = -\,q_L, \,\gamma_5\, q_R = +\,q_R,
\qquad P_{R/L} = \tfrac{1}{2}(1\pm \gamma_5),
%\frac{1+\gamma_5}{2}\, q
 %  =\begin{pmatrix} u_R \\ d_R \\ s_R \end{pmatrix}.
\]
where $P_L$ and $P_R$ are the projection operators onto chirality $-1$ and $+1$, 
respectively:$P_L+P_R=1;\,P_L^2 = P_L,\,P_R^2 = P_R,\,P_L P_R = P_R P_L = 0$.
%The fields $q_L$ and $q_R$ are eigenstates of $\gamma_5$ with eigenvalues $-1$ and $+1$,
% respectively:
Note that
% $(q_L)^\dagger = q^\dagger \frac{1-\gamma_5}{2}$, 
$\overline{q_L} = (q_L)^\dagger \gamma^0 = q^\dagger \frac{1-\gamma_5}{2}\gamma^0
 = \overline{q}\frac{1+\gamma_5}{2} = \overline{q}P_R$ and 
$\overline{q_R} = \overline{q}\,P_L.$
%Notices that the subscripts $L$ and $R$ are interchanged.

\subsubsection*{Two-flavor case}

 For pedagogical reasons, 
we first consider the two-flavor case ($N_f = 2$) consisting of the
$u$ and $d$ quarks;
$q_L=\,^t(u_L,\, d_L),\,q_R=\, ^t(u_R,\, d_R)$.
 The color indices play no role here, and will not be explicitly written.

First we define
$\mathbf{t} = (\tau^0, \boldsymbol{\tau})$ where
$\tau^0:=\bfone_2$ and  $\boldsymbol{\tau}=(\tau^1, \tau^2, \tau^3)$ being
 the Pauli matrices for isospin with the normalization 
 ${\rm tr}\, t^a t^b = 2\, \delta^{ab}.$
Then, the chiral transformation is defined with the four real parameters
$\boldsymbol{\theta}_{L,R}$ by
\begin{align}
q_L \rightarrow q'_L = U(\boldsymbol{\theta}_L)\, q_L=:U_L\, q_L ;\quad
q_R \rightarrow q'_R = U(\boldsymbol{\theta}_R)\, q_R=:U_R\, q_R,
\label{eq:u2-u2-chiral}
\end{align}
with $[U(\boldsymbol{\theta}) = \e^{\, i\, \boldsymbol{\theta}\cdot\mathbf{t}/2}$\,
($\boldsymbol{\theta}\cdot\mathbf{t} \equiv \sum_{a=0}^{3} \theta_a\, \tau^a$).
It is to be noted that  
the left- and right-handed fields $q_L$ and $q_R$ 
transform in mutually independent representation spaces, and  
the matrices $U_L$ and $U_R$ represent two $2\times 2$
unitary transformations with the independent parameters $\boldsymbol{\theta}_L$ 
and $\boldsymbol{\theta}_R$.   Then the chiral transformations form
the group U(2)$\times$U(2). When the representation vector space is specified, 
one also writes as U(2)$_L\,\times$U(2)$_R$. 

There are two important subgroups in U(2)$_L\,\times$U(2)$_R$: 
If we restrict the parameters to 
$\boldsymbol{\theta}_L=\boldsymbol{\theta}_R=:\boldsymbol{\theta}$,
the quark field is transformed as
\[
q=q_L+q_R\,\rightarrow\,q'= U(\boldsymbol{\theta})\, q_L+U(\boldsymbol{\theta})\, q_R=
U(\boldsymbol{\theta})\, q=\e^{i\theta_0/2}\e^{i\boldsymbol{\theta}\cdot\boldsymbol{\tau}/2}\,q.
\]
which is a phase and isospin transformation.

Now we are in a position to consider the relevance of the chiral transformation to 
QCD (\cite{cheng1994gauge}, \cite{georgi2009weak}).
%First, we note that the antiquark fields are transformed under this transformation as
%$\overline{q_L} \rightarrow \overline{q'_L}
%  = \overline{q_L}\, U^{\dagger}(\boldsymbol{\theta}_L)=\overline{q_L}\,U_L^{\dag},\,
%\overline{q_R} \rightarrow \overline{q'_R}
%  = \overline{q_R}\, U^{\dagger}(\boldsymbol{\theta}_R)=\overline{q_R}\,U_R^{\dag}$.
%P_R\gamma^{\mu}P_L=P_RP_R\gamma^{\mu}=P_R\gamma^{\mu}=P_R\gamma^{\mu}P_L, \quad
%P_L\gamma^{\mu}P_L
We first note that the vector current 
$V^{\mu}:= \overline{q}\gamma^{\mu}q= \overline{q}(P_R+P_L)\gamma^{\mu}(P_R+P_L)q
= \overline{q_R}\gamma^{\mu}q_R
 + \overline{q_L}\gamma^{\mu}q_L$ 
 is invariant under the chiral transformation;
\[
V^{\mu}\, \rightarrow
{V'}^{\mu}
%= \overline{q}\gamma^{\mu}q
= \overline{q_R}U_R^{\dag}\gamma^{\mu}U_Rq_R
 + \overline{q_L}U_L^{\dag}\gamma^{\mu}U_Lq_L=\overline{q_R}\gamma^{\mu}q_R
 + \overline{q_L}\gamma^{\mu}q_L=V^{\mu}.
\]
Quite similarly, one sees that the kinetic term in (\ref{eq:classkcalQCD}) is
 invariant under the chiral transformation by writing 
$\bar{q}i\gamma^{\mu}D_{\mu}q=\overline{q_L}i\gamma^{\mu}D_{\mu}q_L+
\overline{q_R}i\gamma^{\mu}D_{\mu}q_R$.
On the other hand,
the quark mass term in (\ref{eq:classkcalQCD}) is transformed as
\[
\bar{q}m_qq
 = \bar{q}m_q\,(P_R+P_L)q
% = \bar{q}P_R\,m_q q + \bar{q}m_qP_Lq
 = \overline{q_L}m_qq_R + \overline{q_R}m_qq_L\, \rightarrow\, 
\overline{q_L}U_L^{\dag}m_qU_Rq_R + \overline{q_R}U_R^{\dag}m_qU_Lq_L,
\]
which is invariant only for the special case with 
$\boldsymbol{\theta}_L=\boldsymbol{\theta}_R$ and $m_u=m_d$, but 
{\em not} invariant under the generic chiral transformation
with $\boldsymbol{\theta}_L\not=\boldsymbol{\theta}_R$.
Thus, the full classical QCD Lagrangian (\ref{eq:classkcalQCD}) is 
chirally invariant and has a U(2)$_L\times$U(2)$_R$ symmetry 
in the chiral limit.
%\footnote{QCD Lagrangian is always invariant 
%even off the chiral limit.}.

\subsubsection*{Three-flavor case}
Next, let us include the strange quark and consider the case with three flavors.  
In this case, the chiral transformation is defined by 
%given as follows ($i,\,j = u,\,d,\,s$):
%\beq
%P_L q_i \equiv q_{iL}
%\rightarrow (U_L)_{ij} q_{jL},\quad
%P_R q_i \equiv q_{iR}
%\rightarrow (U_R)_{ij} q_{jR}.
%\label{eq:su3-chiral-tr}
%\eeq
changing  $U_L$ and $U_R$ to
$U_L :=  \exp\left( i\sum_{k=0}^{8} \theta_{L,R}^{k} \lambda_{k}/2\right)
%\, \boldsymbol{\theta}_L \cdot \boldsymbol{\lambda} / 2 \right)
      \equiv U(\boldsymbol{\theta}_L), \quad
U_R := \exp\left( i\sum_{k=0}^{8} \theta_{L,R}^{k} \lambda_{k}/2\right)
%\, \boldsymbol{\theta}_R \cdot \boldsymbol{\lambda} / 2 \right)
      \equiv U(\boldsymbol{\theta}_R)$.
%\label{eq:su3-chiral-tr}
%\[\boldsymbol{\theta}_{L,R} \cdot \boldsymbol{\lambda}
%   = \sum_{k=0}^{8} \theta_{L,R}^{k} \lambda_{k}.\]
The flavor Gell-Mann matrices $\lambda_{\alpha}$ $(\alpha = 0, 1, 2, \dots, 8)$
are the generators of the unitary group of degree three:
$\left[ \lambda_{\alpha}/2,\ \lambda_{\beta}/2 \right]=i\, f_{\alpha\beta\gamma}\, (\lambda_{\gamma}/2)$.
These generators are normalized as
$\mathrm{tr}\,\lambda_{\alpha} \lambda_{\beta}  = 2 \delta_{\alpha\beta}$\,
($\alpha, \beta = 0, 1, 2, \dots, 8$).
As is easily verified, the classical QCD Lagrangian with three flavors (u,\, d,\,s ) has 
U(3)$_L\times$U(3)$_R$ symmetry in the chiral limit.
We note that U(3)$_L\times$U(3)$_R$ has subgroups
U(3)$_V$ given by the restricted parameters
${\theta}^k_L = {\theta}^k_R$.
Also, the transformation with another restriction of the parameter
$-{\theta}^k_L = {\theta}^k_R$ 
%\equiv \beta^k$
is called the axial transformation and denoted by
 U(3)$_A$,
although this notation is slightly abusive, 
since purely axial transformations do not by themselves form a subgroup.

The Noether currents $\bar{q} \gamma_{\mu} \lambda^a/2 q=:V_{\mu}^a$ for the U(3)$_V$ 
transformation in QCD read
\beq
\d^{\mu}V_{\mu}^a=
%\partial_{\mu}(\bar{q} \gamma_{\mu} \lambda^a/2 q)}   = 
i \sum_{i,j=u,d, s} \bar{q}_i (m_i - m_j)  \lambda^a/2 q_j, \quad
(a=0 \sim 8),
\eeq
which tells us that if the current quark masses are the same, 
nine U(3)$_V$ charges are conserved and form a nonet.
Similarly, 
the octet Noether current $\bar{q} \gamma_{\mu} \gamma_5 \lambda^a/2 q=: A_{\mu}^a$ 
for the axial SU(3)$_A$ transformation 
reads
\beq 
\d^{\mu}A_{\mu}^a=
%:={\partial_{\mu}(}  =  
i \sum_{i,j=u, d, s} \bar{q}_i (m_i + m_j) \gamma_5 \lambda^a/2 q_j,\quad 
(a=1 \sim 8),
\eeq
which shows that the presence of the current quark masses violates the 
invariance under the axial SU(3)$_A$ transformation.
It is said that the current quark mass breaks the 
chiral symmetry {\bf explicitly}.
We remark that the forms of the above equations are not altered in 
the quantum theory. 
However, the Noether current 
$\bar{q} \gamma_{\mu} \gamma_5 q=:A_{\mu}^0$ for the singlet axial U(1)$_A$ 
transformation does have an anomalous additional term 
generated by quantum effects due to gluons,
 known as  the {\em axial anomaly} (or  U(1)$_A$  anomaly)
(\cite{Bell:1969ts,Adler:1969gk,Fujikawa:1979ay,Fujikawa:1980eg}), as follows:
\beq
\partial^{\mu}A_{\mu}^0   = 
i \sum_{i=u, d, s} \bar{q}_i 2m_i \gamma_5 q_i + 
\frac{3 g^2}{16\pi^2} F_{\mn}^a \tilde{F}^{\mn}_a, \quad
\tilde{F}_a^{\lambda \rho}:={1 \over 2} \epsilon^{\mn \lambda \rho} F_{\mn}^a,
\label{eq:QCD-anomaly-eq}
\eeq
%{\partial_{\mu}D^{\mu} = \Theta_{\mu}^{\mu}} & = &
%(1+ \gamma_m) \sum_i^{N_f} \bar{q}_i m_i q_i + {\beta \over 2 g}
% F_{\mn}^a F^{\mn}_a  \ \ ,\eeq
% and $\Theta_{\mn}$ is the energy momentum tensor of QCD.
implying that the axial U(1)$_A$ symmetry is {\em explicitly} broken 
even in the chiral limit. Thus 
 QCD has a U(1)$_V\times$SU(3)$_L\times$SU(3)$_R$ symmetry, where U(1)$_V$
 corresponds to the quark number conservation\footnote{In the following, 
 we may not write this trivial symmetry in an explicit way.}. 

\subsection{Non-perturbative nature of the QCD vacuum}
 
 Because of the non-perturbative interactions among quarks and gluons,
the ground state of QCD has a non-trivial structure.
For instance, the current-algebra relation called Gell-Mann-Oakes-Renner (GOR)
relation (\cite{cheng1994gauge}, \cite{georgi2009weak}) reads
\beq
f_{\pi}^2 m_{\pi^{\pm}}^2 \simeq
 - \hat{m} \la :(\bar{u}u + \bar{d}d): \ra
\label{eq:GOR}
\eeq
where
%$\hat{m}$=$(m_u+m_d)/2$ is an averaged mass of $u$ and $d$ quarks, and 
$\la :O:\ra$ denotes the vacuum expectation value of a normal-ordered operator $O$,
$m_{\pi}$ is the pion mass and $f_{\pi}$ denotes 
the pion decay constant, which is defined through the transition amplitude 
given by\footnote{We shall say more on the important implications of this equation, later.}
\beq
\la 0\vert A^a_{\mu}(x)\vert \pi^b(p)\ra=if_{\pi} \delta^{ab}p_{\mu}\e^{-ip\cdot x}.
\label{eq:def-pion-decay-const}
\eeq
By taking the known values $f_{\pi}\simeq 93$\, MeV
 and $\hat{m}(2$GeV)$\simeq 3.5$\, MeV, we have
$\la: \bar{u}u: \ra \simeq \la: \bar{d}d: \ra \sim
[- (260) {\rm MeV}]^3$ at $\mu^2= 2$ GeV.
That is, the QCD vacuum has 
a non-vanishing expectation value (condensation) of the quark and anti-quark pairs.
 Note that the operators are normal-ordered, and 
hence the expectation values of them would vanish 
if the vacuum were the unperturbed one.
 Thus, {the presence of these nonzero vacuum expectation values indicates
the nonperturbative nature of the QCD vacuum}. 

\subsection{Basic commutation relations of quark operators
and SU(3)$_L\times$SU(3)$_R$ chiral symmetry}

The set of the generators of the U(3)$_L\times$U(3)$_R$ are given by
($a = 0,\,1,\dots,8$);
\beq
Q^a(x^0) :=\int d^3\bfx\, q^{\dagger}(x)\,(\lambda^a/2)\, q(x)
 =\int d^3\bfx\, V^a_0 (x_0, \bfx), \quad 
Q_5^a(x^0):= \int d^3\bfx\, q^{\dagger}(x)\,\gamma_5\,(\lambda^a/2)\, q(x) 
= \int d^3\bfx\, A^a_0(x_0, \bfx) ,
\eeq
with $\lambda^0=\sqrt{2/3}I_3$.
For $a = 1,\,2, \dots,8$, they satisfy the following commutation relations,
\beq
\left[Q^a(x_0),\, V^b_{\mu}(x)\right] = i f_{abc}\, V^c_{\mu}(x), \quad
\left[Q^a(x_0),\, A^b_{\mu}(x)\right] = i f_{abc}\, A^c_{\mu}(x), \quad
\left[Q_5^{a}(x_0),\, V^b_{\mu}(x)\right] = i f_{abc}\, A^c_{\mu}(x), \quad 
\left[Q_5^{a}(x_0),\, A^b_{\mu}(x)\right] = i f_{abc}\, V^c_{\mu}(x), 
\label{eq:su-3-current-algebra}
\eeq
and the integration over $\bfx$ of these relations leads to
\beq
\left[Q^a,\, Q^b\right] = i f_{abc}\, Q^c, \qquad
\left[Q^a,\, Q_5^b\right] = i f_{abc}\, Q_5^c, \qquad
\left[Q_5^{a},\, Q_5^b\right] = i f_{abc}\, Q^c, \label{eq:su-3-algebra}
\eeq
where the argument $x_0$ is suppressed.
Then the left- and right-handed 
generators defined by
$Q_L^a=(1/2)\cdot(Q^a-Q_5^a)$ and $Q_R^a=(1/2)\cdot(Q^a+Q_5^a)$
satisfy the following commutation relations
\[
\left[Q_L^a,\, Q_L^b\right] = i f_{abc}\, Q_L^c, \qquad
\left[Q_R^a,\, Q_R^b\right] = i f_{abc}\, Q_R^c, \qquad
\left[Q_L^{a},\, Q_R^b\right] =0,
\]
%We note that the last two equations of (\ref{eq:su-3-current-algebra}) imply
%that the vector current is transformed to the axial vector current 
%by the axial transformation and vice versa.
which show that $Q_L^a$ and $Q_R^a$ ($a = 1,\,2,\dots,8$) 
form the generators of the 
SU(3)$_L \times$SU(3)$_R$ group, hence the name of 
SU(3)$_L\times$SU(3)$_R$.

%Of course,  $Q^0$ and $Q_5^a$ satisfy similar commutaion relations; for instance,

For pedagogical reasons and later convenience, let us take the two-flavor case, where
the commutation relations 
(\ref{eq:su-3-algebra}) is reduced to the
 SU(2)$_L \times$SU(2)$_R$ group 
with the replacement of the structure constants $f_{abc}\,\to\, \eps_{abc}$ 
and the matrices $\lambda^a\, \to \, \tau^a$,\,( $\tau^0=I_2$). 
Then the following transformation properties of the  scalar 
$\bar{q}(x)q(x)=:\hat{\sigma}(x)$ and 
pseudoscalar densities $\bar{q}(x)i\gamma_5\tau^a/2q(x)=:\hat{\pi}^a(x)$
hold for $a, b=1\,\sim\, 3$ on account of the above commutation relations:
\beq
\left[Q_5^a(x^0),\, \hat{\sigma}(x)\right]=i\hat{\pi}^a(x)
%\bar{q} q(x)\right]
%=i \bar{q}i\gamma_5\tau^a/2q, 
\quad
\left[Q_5^a(x^0),\, \hat{\pi}^b(x)\right]
%\bar{q}\, i\gamma_5\,(\tau^b/2)\, q(x)\right]
%= \, \bar{q}\, \left\{\lambda^1/2,\, \lambda^1/2\right\} q(x)
%=-i\delta^{ab} \half (\bar{q}q)
=-i\delta^{ab} \half \hat{\sigma}(x)= -i\delta^{ab} \half(\bar{u}u+\bar{d}d).
\label{eq:CCR-for-SSB} 
\eeq
For later convenience, we also write down some commutation relations
involving $Q_5^0$:
%\beq
$[Q_5^0(x^0),\, 
%\bar{q}(x)i\gamma_5\tau^a/2q(x)]
\hat{\pi}^a(x)]=-i\bar{q}(x)\tau^aq(x)$,\, ($a=0,\,1,\,2,\, 3$),
%[Q_5^0(x_^0),\, \bar{q}(x)i\gamma_5\tau^0/2q(x)]=-i\bar{q}(x)q(x)],\, (a=1\sim 3)
%\eeq
which implies that
\beq
[Q_5^0(x^0),\, \bar{q}(x)i\gamma_5\tau^3/2q(x)]=-i(\bar{u}(x)u(x)-\bar{d}(x)d(x)), \quad
[Q_5^0(x^0),\, \bar{q}(x)i\gamma_5\tau^0/2q(x)]=-i(\bar{u}(x)u(x)+\bar{d}(x)d(x)),
\label{eq:UA1-comm-rel-1}
\eeq
or
%\beq
$[Q_5^0(x^0),\, \bar{u}(x)i\gamma_5u(x)]=-i\bar{u}(x)u(x)$, and
$[Q_5^0(x^0),\, \bar{d}(x)i\gamma_5d(x)]=-i\bar{d}(x)d(x)$.
%\label{eq:UA1-comm-rel-2}
%\eeq

%which means that the pseudoscalar density is transfomed into the 
%scalar density by the axial transformation.

%We consider the situation in which the current quark masses of the $u$, $d$, and $s$ quarks can be neglected (the \textbf{chiral limit}\index{chiral limit@chiral limit}). 

\subsection{The notion of spontaneous breaking of chiral symmetry: Wigner 
versus {Nambu--Goldstone} phases}
\label{subsection:Wigner-NG phases}

For the moment, we consider the extreme case of the  {\em chiral limit}, 
where QCD is invariant under the SU(3)$_L \times$SU(3)$_R$ 
chiral transformation.
We collectively denote the generators of this chiral symmetry, 
$Q^a$ and $Q_5^a$ ($a = 1, 2, \dots, 8$), by $\tilde{Q}^{\alpha}$, 
and the corresponding Noether currents by $\tilde{j}_{\mu}^{\alpha}(x)$.
Let the QCD vacuum be denoted by $\lvert 0 \rangle$.  
Then the chiral symmetry is realized in either of the following two ways, the Wigner or {Nambu--Goldstone} realization
(\cite{cheng1994gauge}).
\subsubsection*{Wigner realization}
It is said that 
the vacuum state $\lvert 0 \rangle$ is in the Wigner phase, if
the following equality holds, 
\beq
\tilde{Q}^{\alpha}\, \lvert 0 \rangle = 0, \qquad \forall a.
\label{eq:ssb-wigner}
\eeq
Because of (\ref{eq:ssb-wigner}), some interesting equalities between 
the vacuum expectation values of 
some specific correlators with opposite-parity operators 
hold, which ensures 
the degeneracy of the scalar/pseudoscalar and vector/axial vector modes in the Wigner phase:
Recalling the Jacobi identity 
$[[A, B], C]+[[B, C], A]+[[C, A], B]=0$, 
we have 
\beq
0= \la 0\rvert[[\hat{\pi}^b(x), \hat{\sigma}(0)], Q_5^a(x_0)]\rvert 0\ra 
=-\la 0\rvert[[ \hat{\sigma}(0), Q_5^a], \hat{\pi}^b(x)]\rvert 0\ra-
\la 0\rvert[[Q_5^a, \hat{\pi}^b(x)], \hat{\sigma}(0)]\rvert 0\ra
= i\la 0\rvert[\hat{\pi}^a(0), \hat{\pi}^b(x)]\rvert 0\ra+
i\delta_{ab}\la 0\rvert[\hat{\sigma}(x), \hat{\sigma}(0)\rvert 0\ra,\nonumber
\eeq
where (\ref{eq:CCR-for-SSB}) has been used together with the fact that
 $Q_5^a\lvert 0\ra=0=\la 0\rvert Q_5^a$. Putting $a=b$,
we have the equality of the vacuum expectation values of the scalar and 
pseudoscalar density operators as
\beq
\la 0\rvert[\hat{\sigma}(x), \hat{\sigma}(0)\rvert 0\ra=
\la 0\rvert[\hat{\pi}^a(x), \hat{\pi}^a(0)]\rvert 0\ra.
\label{eq:vac-sigma-pi-ccr}
\eeq
Similarly, using (\ref{eq:su-3-current-algebra}), 
we have the equality of the vector and axial vector currents 
\beq
\la 0\rvert[V_{\mu}^a(x), V_{\nu}^a(0)\rvert 0\ra
=\la 0\rvert[A_{\mu}^a(x), A_{\nu}^a(0)]\rvert 0\ra.
\label{eq:vac-vector-avector-ccr}
\eeq
As will be shown in \S\ref{subsection:spectral function}, 
these correlator equalities 
 {imply the exact equality of the corresponding spectral functions} for the $0^{++}$ scalar modes ($\sigma$-meson channel) 
and the isovector pseudoscalar
mode (pion channel) as well as the isovector  vector and axial vector modes like
the $\rho$ and $a_1$ mesons.

\subsubsection*{Nambu--Goldstone (NG) realization}
%This corresponds to the case where Eq.~(\ref{eq:ssb-wigner}) does not hold.  
%Formally, 
The Nambu--Goldstone realization corresponds to the case in which at least one generator does not annihilate the vacuum, i.e., 
\beq
\tilde{Q}^{\alpha}\, \lvert 0 \rangle \neq 0,\qquad \textrm{for some}\,\, \alpha.
\label{eq:ssb-ng}
\eeq
In this case, we say that the symmetry is 
{\bf spontaneously broken, and the vacuum is in the NG phase}. 
We emphasize that this notion of `symmetry breaking' does not refer to a breaking of the symmetry of the Hamiltonian operator (or Lagrangian) under a transformation, but rather to a property of the vacuum state as the lowest-energy state of the Hamiltonian. 
%\end{enumerate}

\begin{figure}[bht]
\centering
\includegraphics[width=.65\textwidth]{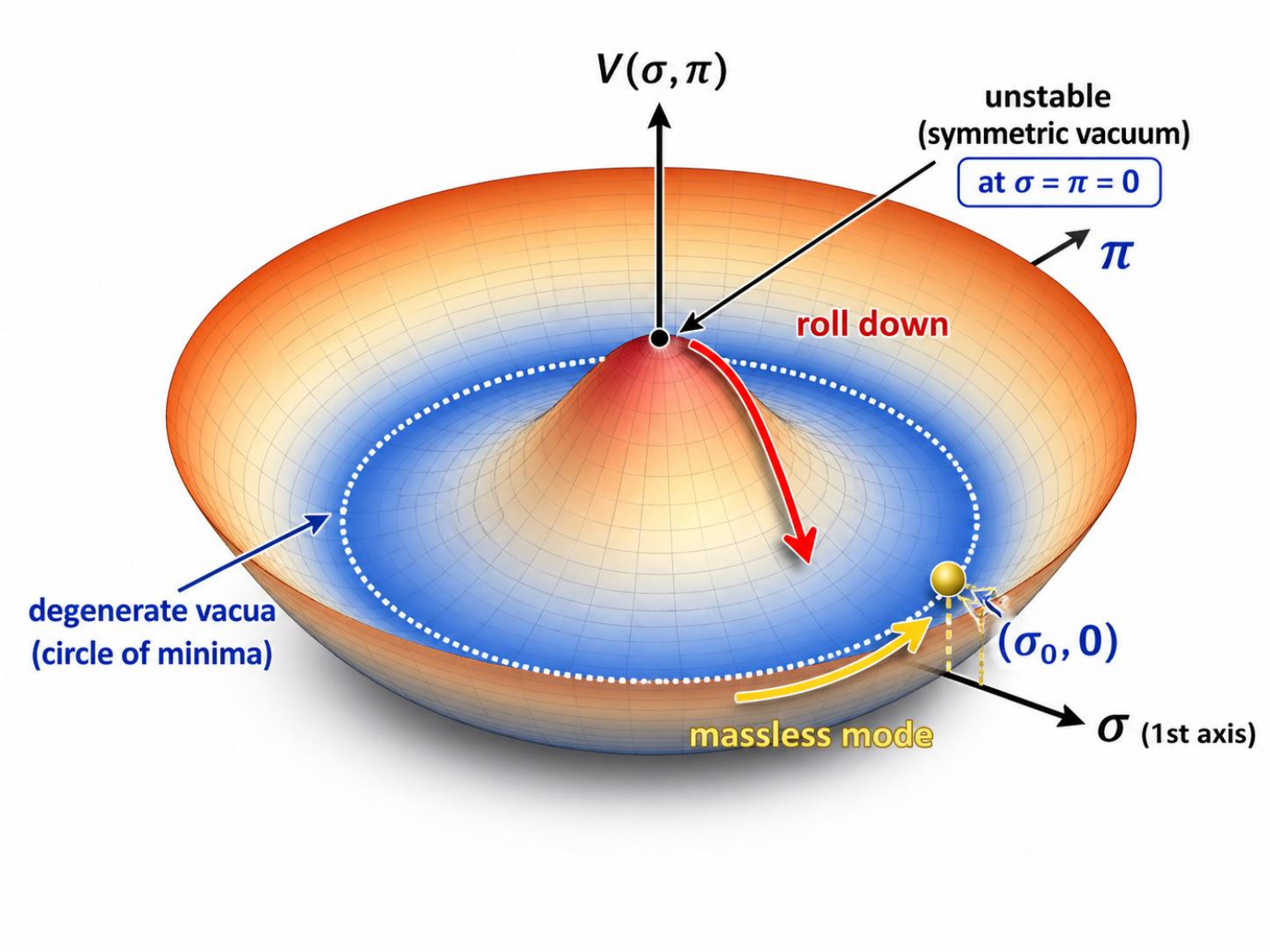}
\caption{A schematic description of the spontaneous symmetry breaking in
(\ref{eq:U1-model}). 
When $\mu^2<0$, the origin $(\sigma=0,\pi=0)$ is the local maximum of the 
energy, and hence an unstable state.
The vacuum energy $V(\sigma, \pi)$ 
 is a function of $\sqrt{\sigma^2+\pi^2}$, symmetric for the rotation around 
the vertical axis, 
and so is the lowest energy state, as is denoted by a dotted circle along  the bottom.
%The genuine vacuum state can be chosen
 One may choose a representative vacuum at
 $(\sigma=\sigma_0, \pi=0)$, and any rotation of the field
from the vacuum state leads to an excited state, 
which  is a situation where the U(1) symmetry is
spontaneously broken. 
On top of the chosen vacuum there can be two types of excitation modes; one 
is a rotational mode, which would be massless, and the other is a radial oscillation or 
$\sigma$ mode.}
\label{fig:3-d-eff-pot}
\end{figure}

To see the situation schematically, let us take the following Lagrangian consisting of
two scalar fields ($\sigma+i\pi=:\phi$),
\beq
\mathcal{L}
=
\frac{1}{2}\,(\partial_{\mu}\sigma\,\partial^{\mu}\sigma\,+\,
\partial_{\mu}\pi\,\partial^{\mu}\pi)
 -\frac{1}{2}\,\mu^{2}\,(\sigma^2\,+\,\pi^2)
 -g (\sigma^2+\pi^2)^2,  \quad \quad (g\,>\,0)
\label{eq:U1-model}
\eeq
which is invariant under U(1) transformation
$\phi\,\to\, \e^{i\theta}\phi$, since 
$\sigma^2+\pi^2=\vert\phi\vert^2$.
It may be instructive to note that the U(1) transformation is 
{equivalent to a two-dimensional orthogonal transformation} as
\[
\begin{pmatrix}
\sigma\\
\pi
\end{pmatrix}\,\to \,
\begin{pmatrix}
\sigma'\\
\pi'
\end{pmatrix}=
\begin{pmatrix}
\cos\theta & -\sin\theta\\
\sin\theta& \cos\theta
\end{pmatrix}\,
\begin{pmatrix}
\sigma\\
\pi
\end{pmatrix}.
\]
When $\mu^2<0$, the vacuum energy $V(\sigma, \pi)$ as a function of ($\sigma, \pi$) 
may have the form as shown in Fig.~\ref{fig:3-d-eff-pot}, which is invariant under rotations along 
the vertical axis: The lowest energy state is located at $(\sigma\not=0, \,\pi\not=0)$, 
and infinitely degenerate along the bottom circle denoted  by a dotted
line. The vacuum state can be chosen at $(\sigma_0\not= 0,\, \pi_0=0)$. Then any infinitesimal 
rotation along the flat bottom $\delta\sigma=0,\,\delta\pi=\delta\theta \sigma_0$
would change the state, which is equivalent to Eq.~(\ref{eq:ssb-ng}), and 
 it is said that {\em the U(1) symmetry is spontaneously broken}. 
A quite similar figure of the vacuum energy will be shown for a realistic chiral 
effective theory of QCD in \S~\ref{section:NJL-2f}. 

%Conversely, if $\tilde{Q}^{\alpha}$ does exist, the translational invariance
%of the vacuum and the fact that the vacuum is an eigenstate of the four-momentum 
%with $P_{\mu}=0$ imply the Wigner realization (\ref{eq:ssb-wigner}) {\cal A}{kugo2}.

In the case of QCD, the chiral symmetry is spontaneously broken. 
Let us see how this observation is obtained, and its physical consequences, briefly.
%$\langle 0 \lvert :\!\bar{q}q\!: \rvert 0 \rangle \neq 0$.  
%In this case, for any $a$, the relation
%\beq
%Q_5^a \lvert 0 \rangle = 0
%\label{eq:false-SSB}\eeq
%is \textbf{impossible}.  
%Indeed, 
Taking the vacuum expectation value of both sides of the normal-ordered version
of Eq.~(\ref{eq:CCR-for-SSB}), we have
\[
\langle 0 \lvert \left[ Q_5^a(x^0),\, :\!\bar{q} \gamma_5 \tau^b q(x)\!: \right] 
\rvert 0 \rangle
=
i\, \delta^{ab}\, 
\langle 0 \lvert :\!\bigl(\bar{u}u(x)+\bar{d}d(x)\bigr)\!: \rvert 0 \rangle 
\neq 0,
\]
where the nonzero quark condensate, inferred for example from the GOR relation (\ref{eq:GOR}), provides evidence that the QCD vacuum is in the Nambu--Goldstone realization.
%However, if Eq.~(\ref{eq:false-SSB}) were valid, the left-hand side would become
%\[\langle 0 \lvert Q_5^a(x^0)\, :\!\bar{q} i\gamma_5 \tau^b q(x)\!: 
%\rvert 0 \rangle -\langle 0 \lvert :\!\bar{q} i\gamma_5 \tau^b q(x)\!: \, Q_5^a(x^0) 
%\rvert 0 \rangle = 0,\]which contradicts the above nonzero result.  
%Thus, even in the chiral limit, it is natural to conclude that the chiral symmetry generated by $Q_5^a$ is spontaneously broken in the QCD vacuum.  
Thus, we have 
\[
Q_5^a \lvert 0 \rangle \neq 0, \qquad
(a = 1, 2, 3),
\]
which can be extended to $a=1,\ldots,8$ \footnote{In fact,  the generator $\tilde{Q}^{\alpha}$ 
cannot be properly defined in the NG realization.
Indeed, using the translational invariance, 
%of both $\lvert 0 \rangle$ and $\tilde{Q}^{\alpha}(0)\lvert 0 \rangle$, 
we obtain
$\langle 0 \lvert \tilde{Q}^{\alpha}(0)\,\tilde{Q}^{\alpha}(0) \rvert 0 \rangle
% = \int d^3 \mathbf{x}\, \langle 0 \lvert \tilde{j}^{\alpha}_{0}(\mathbf{x}, 0)\,\tilde{Q}%^{\alpha}(0) \rvert 0 \rangle 
%= \int d^3 \mathbf{x}\, \langle 0 \lvert \tilde{j}^{\alpha}_{0}(\mathbf{0}, 0)\,\tilde{Q}%^{\alpha}(0) \rvert 0 \rangle 
= \langle 0 \lvert \tilde{j}^{\alpha}_{0}(\mathbf{0}, 0)\,\tilde{Q}^{\alpha}(0) \rvert 0 \rangle
\int d^3 \mathbf{x}\,\to\, \infty.$ 
Thus the spatial integral diverges, and $\tilde{Q}^{\alpha}$ cannot be 
defined as an operator.}.
Thus, we arrive at an important conclusion:
{\bf The chiral symmetry of QCD is spontaneously broken in the vacuum,
and the scalar condensate $\la \bar{q}q\ra$ 
plays the role of the order parameter with which the vacuum distinguishes 
 the NG and Wigner phases.}
We note that this conclusion is also obtained from (\ref{eq:def-pion-decay-const}):
If we integrate the current in the l.h.s. of (\ref{eq:def-pion-decay-const})
over $\bfx$, then we have an axial charge $Q_5^a$ that acts on the vacuum state
$\la 0\vert$, which should not vanish because $f_{\pi}\not=0$ in the r.h.s.,
implying that the chiral symmetry is spontaneously broken. This argument also
tells us that {\bf the value of the pion decay constant $f_{\pi}$ also plays the role
of the order parameter of the chiral transition.}
A warning is in order here: In reality, owing to the non-vanishing current quark masses,
the chiral transition is not a genuine {first- or second-order} phase transition with some observables
having a singularity, but inevitably becomes a crossover transition with a gradual change of 
the chiral condensate or $f_{\pi}$. This is actually seen in lattice QCD simulations at finite temperature
 (\cite{Aoki:2006we,Ding:2015ona,Aoki:2020noz}).
%%%pseudo-critical %%%

Next we make a couple of remarks on the breaking of 
U(1)$_A$ symmetry.
\begin{itemize}
\item The equality (\ref{eq:UA1-comm-rel-1}) tells us  that the presence of the 
vacuum chiral condensate
implies that $U(1)_A$ is also broken. 
%in addition to its explicit breaking by the anomaly
%also U(1)$_A$ is also broken:
{Although this observation may appear redundant because the anomaly already breaks
U(1)$_A$ explicitly,} one of the current issues  of the debates on the
nature of the chiral transition {\em at {nonzero} temperature}  
is the role of {\em effective restoration} of
U(1)$_A$ symmetry at finite temperature; 
see for instance (\cite{Tomiya:2016jwr,Aoki:2021qws}), and
the item 3 below.
\item
In QCD, the vector-type isospin symmetry is not spontaneously broken.  
This is known as the Vafa--Witten theorem (\cite{Vafa:1983tf});
$Q^a \lvert 0 \rangle = 0$ \,
($a = 1, 2, \dots, 8$).
Let ${\cal H}$ denote the set of generators that satisfy Eq.~(\ref{eq:ssb-wigner}).  
If we take any two elements in ${\cal H}$ and denote them by  
$\tilde{Q}^a = Y^a$ and $\tilde{Q}^b = Y^b$, then
$[Y^a,\, Y^b]\lvert 0 \rangle 
= Y^a Y^b \lvert 0 \rangle - Y^b Y^a \lvert 0 \rangle = 0$.
Thus the set ${\cal H}$ forms a closed Lie algebra.  
For the case $Y^a = Q^a$, we have
\[
[Y^a,\, Y^b] 
= [\tilde{Q}^a,\, \tilde{Q}^b]
= i f_{abc} Q^c,
\]
which generates SU(3)$_V$.
In this case, it is said that 
{\bf the SU(3)$_L\times$SU(3)$_R$ symmetry is spontaneously broken to 
SU(3)$_V$.}
\end{itemize}

Then, 
\begin{enumerate}
\item 
according to the {Nambu--Goldstone} theorem 
(\cite{Nambu:1961tp,Goldstone:1961eq,Goldstone:1962es}), 
massless pseudoscalar bosons, which 
are called the {Nambu--Goldstone} (NG) bosons,  emerge,
the number of which coincides with that of the generators of the broken symmetry,
i.e., 8 in our case. 
\item In reality, the three quarks have small current quark masses that
explicitly break the chiral symmetry, which makes the NG bosons slightly massive.
Thus they are identified as the low-lying pseudoscalar mesons, i.e.,
three pions ($\pi^{\pm}(140), \pi^0(135)$), four Kaons (K$^{\pm}(494)$, K$_0(498)$, 
$\bar{\rm K}_0(498)$) and
$\eta(548)$. 
\item A remark is in order concerning flavor symmetry:
In the two-flavor case, for instance, 
the isospin symmetry is not spontaneously 
broken even in the NG phase of SU(2)$_L\times$SU(2)$_R$ chiral symmetry,
 which implies that the identities (\ref{eq:vac-sigma-pi-ccr}) and
 (\ref{eq:vac-vector-avector-ccr}) also hold 
for different isospin indices ($a\not=b$) as
%$\la 0\rvert[\hat{\sigma}(x), \hat{\sigma}(0)\rvert 0\ra=
\[
\la 0\rvert[\hat{\pi}^a(x), \hat{\pi}^a(0)]\rvert 0\ra=
\la 0\rvert[\hat{\pi}^b(x), \hat{\pi}^b(0)]\rvert 0\ra, \quad
\la 0\rvert[V_{\mu}^a(x), V_{\nu}^a(0)\rvert 0\ra=
\la 0\rvert[V_{\mu}^b(x), V_{\nu}^b(0)\rvert 0\ra, \quad
\la 0\rvert[A_{\mu}^a(x), A_{\nu}^a(0)]\rvert 0\ra
=\la 0\rvert[A_{\mu}^b(x), A_{\nu}^b(0)]\rvert 0\ra,
\]
irrespective of the phase of the QCD vacuum.
\end{enumerate}

An important remark is in order concerning the ninth pseudoscalar boson. 
 The anomaly equation (\ref{eq:QCD-anomaly-eq}) shows that U(1)$_A$ is not an
exact symmetry of QCD even in the chiral limit, which would have implied
that there is no massless  NG boson in the singlet channel.
 The anomaly equation (\ref{eq:QCD-anomaly-eq}) is, however, merely an operator equality
and holds irrespective of the properties of the vacuum state.
 The way the physical states are affected by the axial anomaly
depends on the properties of the vacuum. 
Indeed, tunneling between topologically distinct gauge-field configurations, described
semiclassically by instantons (\cite{tHooft:1976snw}), leads to the $\theta$-vacuum
structure of QCD, reminiscent of a Bloch state in a periodic potential. Consequently,
the anomaly term entering the Ward--Takahashi identity can have a nonvanishing vacuum
matrix element, so that the flavor-singlet pseudoscalar need not be a massless
Nambu--Goldstone boson (\cite{Fujikawa:2004cx}).

Now 
the explicit breaking of the SU(3)$_V$ symmetry due to the mass difference among 
u, d, and s leads to some patterns of flavor mixing,
and the flavor-singlet state $\eta_0=\bar{q}i\gamma_5\lambda_0q/\sqrt{2}$ 
is slightly mixed
with the flavor octet state $\eta_8=\bar{q}i\gamma_5\lambda_8q/\sqrt{2}$ as
\[
\eta'=\sin\theta_{\eta'}\,\eta_8+\cos\theta_{\eta'}\,\eta_0
\]
 with $\theta_{\eta'}\simeq -10^{\circ}-20^{\circ}$.
Thus the $\eta'(958)$ is found to be the physical manifestation 
of the would-be ninth NG boson,  while 
$\eta(548)$ is almost an octet state with a small mixing of the singlet state;
 \[
\eta(548) =\cos\theta_{\eta}\,\eta_8-\sin\theta_{\eta}\,\eta_0
\] where
$\theta_{\eta}$ can be {different from} $\theta_{\eta'}$ but still
$\theta_{\eta} \simeq -10^{\circ}-20^{\circ}$.

{But why is the $\eta'(958)$ so massive---indeed, slightly heavier than the nucleon?}
%The local operator identity (\ref{eq:QCD-anomaly-eq})
%alone  does not determine the large singlet mass $m_{\eta_0}$.  
 {Witten and Veneziano
(\cite{Witten:1979vv,Veneziano:1979ec}) derived, in the chiral and large-$N_c$ limits,
a relation between the singlet mass and the pure-Yang--Mills topological susceptibility,}
%\beq
$ m_{\eta_0}^2 \simeq \frac{2N_f}{f_0^2}\,\chi_{\rm YM}$ with
$\chi_{\rm YM}
 = \int d^4x\,\la 0|T\,Q(x)Q(0)|0\ra_{\rm YM,conn}$
%       \label{eq:WV-chiral-limit}
where
$ Q(x)=\frac{g^2}{32\pi^2}
 F_{\mu\nu}^a(x)\widetilde F_a^{\mu\nu}(x)$ is the topological charge
%\label{eq:chi-YM-def}
and 
 $F_0$ denotes the decay constant in the
chiral limit.  Subsequently,
{a useful leading-order finite-mass extension of the mass formula for three flavors}
was written down (\cite{Christos:1984tu,Smit:1987un}) as
\beq
 m_{\eta'}^2+m_\eta^2-2m_K^2
 \simeq \frac{6\,\chi_{\rm YM}}{f_\pi^2},
\label{eq:Witten-Veneziano}
\eeq
with an assumption $f_0\simeq f_\pi$.  This formula is 
essentially based on the mass matrix of the {pseudoscalar} mesons
 developed by
Kawarabayashi--Ohta and Di Vecchia and collaborators
(\cite{Kawarabayashi:1980dp,DiVecchia:1981a}). The relation (\ref{eq:Witten-Veneziano})
{is often also referred to as the Witten--Veneziano relation.}
  For a fuller discussion of QCD topology, $\theta$ dependence,
the Witten--Veneziano mechanism, and recent lattice results for the
topological susceptibility $\chi_{\rm YM}$, see the review by Bonanno,
Bonati, and D'Elia (\cite{Bonanno:2026topology}).

\section{Chiral effective Lagrangians}

In the pre-QCD era, various effective Lagrangians with (approximate) 
chiral symmetry written in terms of hadron fields and/or quark fields were
 proposed and
provided important insights into the phenomenology.
% which   often lead to some physical ideas on the underlying dynamics.
Even apart from the historical interest, 
such chiral effective theories are instructive and still  
provide  convenient tools to develop physical ideas which often lead to
good phenomenology.

\subsection{Chiral quark models}
\label{subsection:chiral-quarl-models}

To construct a chirally invariant\index{chiral invariance} Lagrangian in terms
 of the quark fields, 
it is convenient to introduce the following bilinear form $\Phi_{ij}$ 
($i, j = u,\, d,\, s$) of the quark fields,
%represented as a $3\times 3$ matrix :
\beq
(\bfPhi)_{ij}=\Phi_{ij}:= \bar{q}_j(1-\gamma_5) q_i 
 = 2\,\overline{q_{jR}}\, q_{iL}
 = (\bar{q}_j q_i + i\,\bar{q}_j \gamma_5 q_i)/2.
\label{eq:Phi-su3-def}
\eeq
Note the orders of the indices.
Under the SU(3)$_L \times$SU(3)$_R$ transformation,  
$\Phi_{ij}$ and $(\Phi^\dagger)_{ij}:= \bar{q}_j(1+\gamma_5) q_i = 2\,\overline{q_{jL}} q_{iR}$  
(with $\overline{q_L} q_R = (\overline{q_R} q_L)^\dagger$) transform
as the $(3,\bar{3})$ and $(\bar{3},3)$ representations, respectively;
$\Phi_{ij} \rightarrow (U_L)_{ik}\, \Phi_{kl}\, (U_R^\dagger)_{lj}$,\,
$(\Phi^\dagger)_{ij} \rightarrow (U_R)_{ik}\, (\Phi^\dagger)_{kl}\, (U_L^\dagger)_{lj}$.
Similarly, under U(1)$_L \times$U(1)$_R \simeq$ U(1)$_V \times$U(1)$_A$
transformation;
$\Phi_{ij} \rightarrow e^{i(\theta_{0L}-\theta_{0R})/2}\, \Phi_{ij}$,\,
$(\Phi^\dagger)_{ij} \rightarrow 
e^{-i(\theta_{0L}-\theta_{0R})/2}\, (\Phi^\dagger)_{ij}$.
%\label{eq:Phi-tr-law}
Then, {\rm det}\,$\bfPhi$  transforms under a U(3)$_L \times$U(3)$_R$ 
transformation as 
{\rm det}\,$\bfPhi\, \to \, e^{-i(\theta_{0L}-\theta_{0R})/2}{\rm det}\,\bfPhi$,
which tells us {\rm det}\,$\bfPhi$ is not invariant unless
$\theta_{0L}=\theta_{0R}$, and accordingly invariant under 
U$_V$(1)$\times$SU(3)$_L \times$SU(3)$_R$ but 
breaks U(1)$_A$ symmetry.

Now it is readily verified that the series of quark operators defined by
\beq
I_n = {\rm tr}(\bfPhi \bfPhi^{\dagger})^n,\qquad (n = 1, 2, 3, \ldots), 
\eeq
are invariant under the U(3)$_L \times$U(3)$_R$ transformations.
$I_n$ can be cast into a more familiar form by expanding 
$\bfPhi$ in terms the Gell-Mann matrices $\lambda^a$\, ($a=0,\,1,\,\dots\,\, 8$) 
that make a complete set for  $3\times 3$ matrices:
\[
\bfPhi =\sum_{a=0}^8\, \phi_a\lambda^a; \quad
\Phi_a=\frac{1}{2}{\rm Tr}[\bfPhi\lambda^a]=\overline{q_{iR}}\, q_{jL}(\lambda^a)_{ij}
=\overline{q_{R}}\,\lambda^a q_{L}=\half(\sigma^a+i\pi^a),
\] 
with
$\sigma^a:=\bar{q}\lambda^aq$ and $\pi^a:=\bar{q}i\gamma_5\lambda^a q$.
Thus, $I_1$, the lowest-dimensional four-fermion interaction, for instance, 
can be rewritten in a familiar form as
\beq
I_1=\sum_{a, b=0}^8\Phi_a\Phi_b^{\dag}{\rm Tr}[ \lambda^a\lambda^b] 
=2\sum_{a=0}^8\Phi_a\Phi_a^{\dag}= \frac{1}{2}
\sum_{a=0}^{8} \left[
(\bar{q}\lambda^a q)^2 + (\bar{q} i\gamma_5 \lambda^a q)^2
\right]=:I_S.
\label{eq:3-fl-NJL-U3-symm}
\eeq
%which is the four-fermion interaction
%first introduced by 
%by  Nambu-Jona-Lasino (NJL)'s four-fermion interaction.

%Since 
%the U(1)$_A$ symmetry in explicitly
%broken in QCD as the quantum theory, as shown by Eq.~(\ref{eq:QCD-anomaly-eq}),
%it is desirable 
%a low-energy effective theory of QCD takes into account this fact.

So far, we have restricted our discussions to scalar and pseudoscalar modes.
We next turn to the vector modes.
Corresponding to the case of the scalar fields (\ref{eq:Phi-su3-def}), 
we define the left- and right-handed vector fields by
\beq
(\bfL_{\mu})_{ji}=L_{\mu\,ij}:=2\bar{q}_i\gamma_{\mu}P_Lq_j=V_{\mu\,ji}-A_{\mu\,ji},\quad
(\bfR_{\mu})_{ji}=R_{\mu\,ij}:=2\bar{q}_i\gamma_{\mu}P_Rq_j=V_{\mu\,ji}+A_{\mu\,ji},
\quad
(i,\,j=u, d, s)
\eeq
where 
$V_{\mu\,ji}:=\bar{q}_i\gamma_{\mu}q_j=(\bfV_{\mu})_{ji}$ and 
$A_{\mu\,ji}:=\bar{q}_i\gamma_{\mu}\gamma_5 q_j=(\bfA_{\mu})_{ji}$. 
They are transformed under the chiral transformation as
$\bfL\,\to\, U_L\bfL U_L^{\dag}$ and
$\bfR\,\to\, U_R\bfR U_R^{\dag}$, respectively. 
Then it is readily verified that the following
quantities are invariant under 
chiral U(3)$_L\times$U(3)$_R$ transformations and under 
 Lorentz transformations:
${\cal I}_L:={\rm Tr}(\bfL_{\mu}\bfL^{\mu})$ and
${\cal I}_R:={\rm Tr}(\bfR_{\mu}\bfR^{\mu})$.
In describing the physical world, however, 
the use of the vector $\bfV_{\mu}$ 
and axial vector field $\bfA_{\mu}$ is more convenient, which can be
made by taking the sum of ${\cal I}_L$ and ${\cal I}_R$ as
${\cal I}_L+{\cal I}_R=2{\rm Tr}\, (\bfV_{\mu}\bfV^{\mu}+\bfA_{\mu}\bfA^{\mu})=:2{\cal I}_V$.
As in the scalar case, ${\cal I}_V$ can be rewritten in terms of the Gell-Mann matrices;
\beq
{\cal I}_V=\sum_{a=0}^8\,\left[(\bar{q}\gamma_{\mu}\lambda^aq)^2\,+\,
(\bar{q}\gamma_{\mu}\gamma_5\lambda^aq)^2\right].
\label{eq:3-fl-NJL-U3-symm-V}
\eeq
Note that the summation is taken from $a=0$.

%%%%%%July 6%%%%%%
\subsection{Generalized Nambu-Jona-Lasinio Model}

Summing the scalar (\ref{eq:3-fl-NJL-U3-symm}) and the
vector (\ref{eq:3-fl-NJL-U3-symm-V}) interaction together with the kinetic term, we
have 
\beq
{\cal L}_{GNJL} = \bar{q}\, (i\gamma \cdot \partial- {\bf m}) \, q
+G_S I_S-G_VI_V,
\label{eq:GNJL}
\eeq
which is sometimes called the generalized Nambu-Jona-Lasinio 
(GNJL) model
(\cite{Vogl:1991qt}), and its intimate relevance to QCD was extensively 
analyzed (\cite{RuizArriola:1991gc,Ebert:1994mf,Bijnens:1995ww}), where the following points 
were clarified: (i)~ The GNJL
model is a kind of realization of the chiral quark picture 
(\cite{Manohar:1983md}).\, (ii)~The numerical values of 
the low-energy constants $L_i$'s in the nonlinear chiral 
Lagrangian (\cite{Gasser:1983yg,Gasser:1984gg})
can be well reproduced by the GNJL model (\ref{eq:GNJL}).

A remark is in order: Gasser and Leutwyler (\cite{Gasser:1983yg})
showed that {\em the renormalizable} linear $\sigma$ model cannot
reproduce the low-energy coupling constants $L_i$'s because the very
renormalizability causes a constraint on some $L_i$'s. The GNJL model 
is certainly
a linear realization of the chiral symmetry, but non-renormalizable, and 
the Gasser-Leutwyler theorem
does not apply to the NJL-type Lagrangian in four dimensions; 
see (\cite{Shizuya:1979bv}).

\subsubsection{Nambu-Jona-Lasinio model with axial anomaly: three-flavor case}
\label{subsection:NJL-anomaly-3f}

It is worth mentioning that the Lagrangian
(\ref{eq:3-fl-NJL-U3-symm}) constructed solely from polynomials of 
$\bfPhi\bfPhi^{\dag}$ is 
% Even if one restricts oneself 
%to bilinear forms in $\Phi$ (i.e., four-fermion interactions 
%in terms of the quark fields), then 
%has a structure that if any interaction 
%terms are constructed in terms of the quark fields so as to be 
not only U(1)$_V\times$SU(3)$_L \times$SU(3)$_R$ invariant but 
 automatically U(3)$_L \times$U(3)$_R$ invariant.
%This fact plays an important role when representing the 
%axial anomaly in the low-energy chiral Lagrangian.
To implement the axial anomaly of QCD in the chiral effective Lagrangian, 
it is necessary to introduce operators 
that are invariant under U(1)$_V \times$SU(3)$_L \times$SU(3)$_R$ 
but not under U(1)$_A$.  
As was shown in \S\ref{subsection:chiral-quarl-models},
the lowest-dimensional operators that satisfy this 
requirement are ${\rm det}\Phi$ 
and its Hermitian conjugate ${\rm det}\Phi^{\dagger}$.  
Their Hermitian combination,
\beq
I_D = {\rm det}\Phi + {\rm det}\Phi^{\dagger},
\label{eq:det-int-1}
\eeq
and/or their analytic functions thus represent effective interactions
 among quarks that encode 
the axial anomaly (\cite{Kobayashi:1970ji,Mirelli:1976ww}). Indeed 
(\ref{eq:det-int-1}) can be derived as a one-instanton-mediated interaction
(\cite{tHooft:1976snw}); see also (\cite{Pisarski:2019upw}). 
%%%%Discussion on ln det 

Even if the mass term is flavor diagonal, it is not invariant under a general chiral transformation 
with $\boldsymbol{\theta}_L \neq \boldsymbol{\theta}_R$,
even if it forms a flavor-SU(3)$_f$ singlet 
(i.e., $\sum_i \bar{q}_i q_i = \sum_i (\Phi_{ii} + (\Phi^{\dagger})_{ii})/2$), 
Thus, a popular chiral effective Lagrangian 
incorporating the axial anomaly as well as the explicit
breaking of chiral symmetry due to the small quark mass
%incorporating these ingredients 
takes the form  
(\cite{Kobayashi:1971qz,Mirelli:1976ww,Kunihiro:1987bb,Hatsuda:1994pi,Takizawa:1989sv,Vogl:1991qt,Klevansky:1992qe})\footnote{See 
also (\cite{Bernard:1987sg}).}
\begin{equation}
\begin{aligned}
{\cal L} = \bar{q}\, i\gamma \cdot \partial \, q
+ \sum_{a=0}^{8} 
\frac{g_{S}}{2}\left[(\bar{q}\lambda_a q)^2 
+ (\bar{q} i\lambda_a \gamma_5 q)^2\right]
- \bar{q}\, {\bf m}\, q  
 + g_{D}\left[ \det \bar{q}_i (1-\gamma_5) q_j 
+ \det \bar{q}_i (1+\gamma_5) q_j \right]  
\equiv{}& {\cal L}_0 + {\cal L}_{S} + {\cal L}_{SB} + {\cal L}_{\rm KMT} ,
\label{eq:KMNJL}
\end{aligned}
\end{equation}
which incorporates the following three essential aspects of 
non-perturbative QCD dynamics, i.e.,  
(i) the approximate SU(3)$_L\times$SU(3)$_R$  chiral symmetry,\,
(ii) the axial anomaly induced through ${\cal L}_{\rm KMT}$, and  
(iii) the explicit breaking of chiral symmetry due to the current quark mass term 
${\cal L}_{SB}$.

Using the model Lagrangian Eq.~(\ref{eq:KMNJL}), one can quantitatively and clearly understand many low-energy 
hadron phenomena as consequences of the interplay among these three effects.  
This complete NJL  Lagrangian with ${\cal L}_{\rm KMT}$ 
was first considered by Kobayashi and Maskawa in 1970, and
the form of a determinant was derived as an instanton-induced 
interaction by 't Hooft in 1976.  It is therefore referred to as
the Kobayashi-Maskawa-'t Hooft term or KMT term in short.

%Accordingly, the Lagrangian~(\ref{eq:KMNJL}) may suitably be called the 
%Kobayashi-Maskawa-Nambu-Jona-Lasinio (KM-NJL) model.

\subsubsection{Nambu-Jona-Lasinio model with axial anomaly: two-flavor case}
\label{section:NJL-2f}
In the following, we shall describe,  using a two-flavor NJL model, 
how the quark mass is dynamically generated through the 
spontaneous breaking of chiral symmetry, which leads to an emergence of the pion as an 
NG boson to be envisaged in a geometrical way in terms of the symmetry property of the vacuum energy.

For the two-flavor case, the SU(2)$_L\times$SU(2)$_R$ NJL model with the quark mass term is given by
\beq
{\cal L} = \bar q i\gamma\cdot \partial q 
 \,+\, g\big[(\bar q q )^2 + (\bar q\, i\gamma_5 \boldsymbol{\tau}\, q )^2\big]- \bar{q}{\bf m}\, q,
%= {\cal L}_0 + {\cal L}_1 .
\label{eq:su2-NJL}
\eeq
where, ${\bf m} = {\rm diag}(m_u, m_d)$ is the current-quark mass matrix;
we  assume the isospin symmetry $m_u=m_d=:{m}$ for simplicity. 
The first two terms have the SU(2)$_L\times$SU(2)$_R$ symmetry, which is broken 
by the current quark mass term. 
The coupling constant $g$
with a positive value has
a mass dimension $-2$, which implies that the theory is unrenormalizable and has
a cut-off $\Lambda$ as an important ingredient of the model (\cite{Hatsuda:1994pi}). 
It is  convenient 
 to parameterize $g$  with a dimensionless and dimensionful parameters $G$ and $m_0$ as(\cite{Hatsuda:1994pi})
\[
2g=(G/m_0)^2.
\]

The fact that the Lagrangian (\ref{eq:su2-NJL}) 
 takes care of the axial anomaly, and is only 
U$_V$(1)$\times$SU(2)$_L\times$SU(2)$_R$ invariant is readily seen as follows:
The instanton-induced two-flavor interaction 
takes the form (\cite{Hatsuda:1985ey}),
$I^{(2)}_D:={\rm det}\Phi + {\rm det}\Phi^{\dagger}[=(1/2)\cdot [
(\bar q q )^2 + (\bar q\, i\gamma_5 \boldsymbol{\tau}\, q )^2-
(\bar q\,  \boldsymbol{\tau}\, q )^2-(\bar q i\gamma_5 q )^2]$, 
while the U(2)$_L\times$U(2)$_R$-invariant  interaction reads
$I_1 = {\rm tr}(\bfPhi \bfPhi^{\dagger})=(1/2)\cdot[(\bar q q )^2 + 
(\bar q\, i\gamma_5 \boldsymbol{\tau}\, q )^2+
(\bar q\,  \boldsymbol{\tau}\, q )^2+(\bar q i\gamma_5 q )^2]$. Then,
their squared sum $2(I_1+I_D)$ leads to the interaction term of (\ref{eq:su2-NJL}).
%the familiar four-quark interaction up to the coupling constant,
%which is found to be SU(2)$_L\times$SU(2)$_R$-invariant
%but breaks U(1)$_A$ symmetry.

%Let us first take the simple case with $m=0$, i.e., the case in the chiral limit.
Now let us find the vacuum state of (\ref{eq:su2-NJL}).
The true vacuum is the lowest energy state of the system,
and the vacuum energy with the possible vacuum expectation values or condensates
$ \la\bar{q}q\ra=:-(m_0^2/G) \sigma$ and $\la\bar{q}i\gamma_5\tau q\ra=:-(m_0^2/G) \pi$,
is given by the effective potential 
%given in the path-integral formalism
(\cite{cheng1994gauge,Peskin:1995ev}).
Let ${\cal V}(\sigma, \pi)$
be the effective potential in the one-loop approximation, which 
is equivalent to the vacuum energy  
in the (self-consistent) mean-field approximation
% with the constituent quark mass 
 familiar in the 
quantum many-body theory (\cite{Kunihiro:1983ej, Hatsuda:1994pi}).

\begin{figure}[bht]
\centering
\includegraphics[width=.8\textwidth]{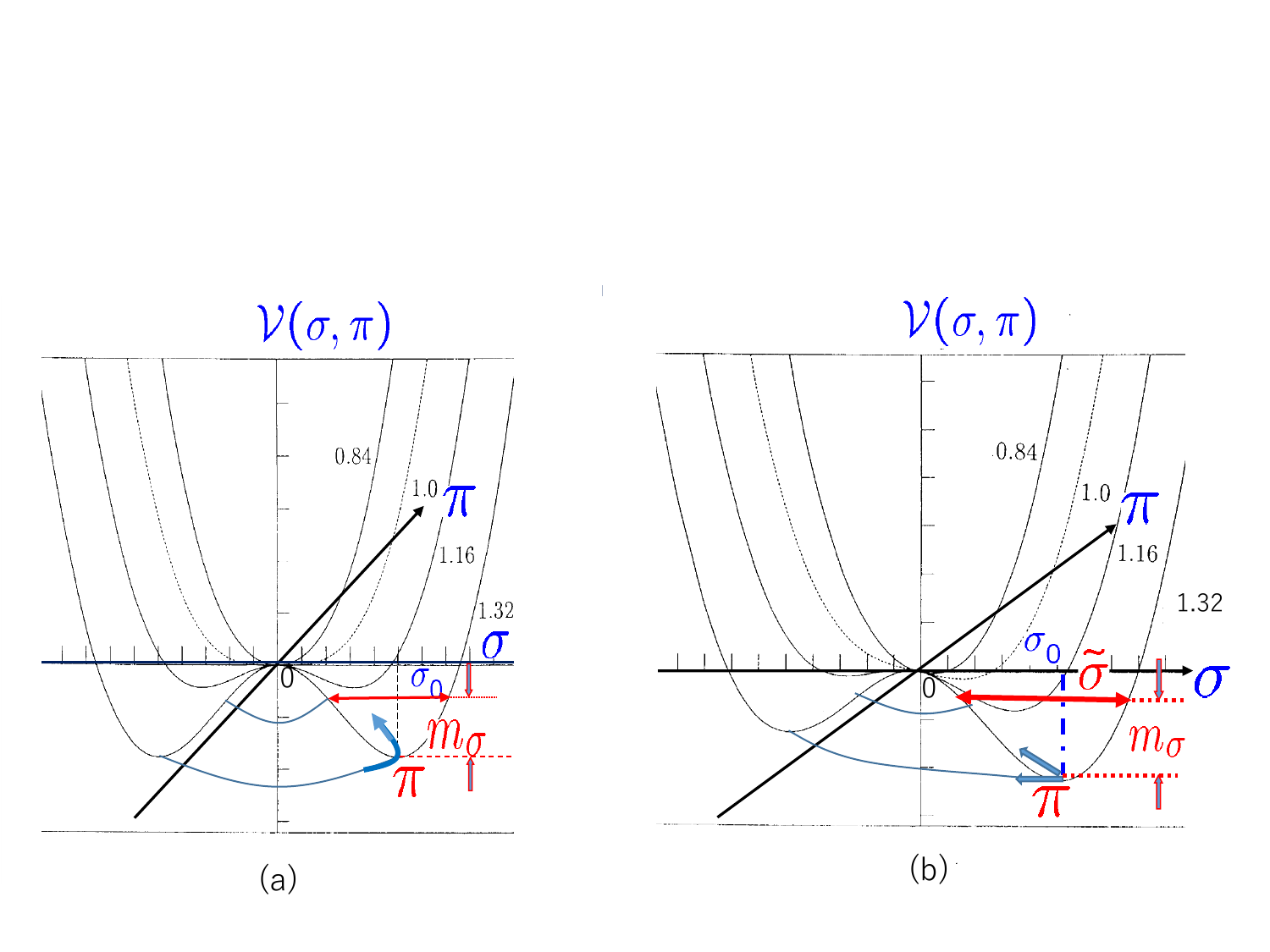}
\caption{(a)\, The effective potential ${\cal V}(\sigma, \pi)$ in the chiral limit,
which is a function of $\sqrt{\sigma^2+\pi^2}$, where $\pi$ is an abbreviation of 
$\Vec{\pi}=(\pi_1,\pi_2,\pi_3)$.
The attached numbers are $g/g_c=:\alpha$. The $\sigma_0$ 
corresponds to the chiral condensate in the vacuum. 
The effective potential in the vacuum is just flat 
in the $\pi$-field direction because of the chiral symmetry, as shown in the Figure, 
which makes the pion massless. The 
$\sigma$ meson is the amplitude fluctuation of the order parameter
$\la \bar{q}q\ra \sim \sigma$, and the mass $m_{\sigma}$ of it reflects the amount of the 
curvature of the effective potential at $\sigma=\sigma_0$. Conversely,
when the system is approaching the critical point by, say, raising temperature, 
the curvature
tends to be smaller, and one can expect that $m_{\sigma}$  becomes smaller.\, \, (b)\, The same as (a)
but off the chiral limit with $m=5.5$ MeV. 
The effective potential is slightly tilted toward the $\sigma$ direction, and 
the small slope in the $\pi$-field direction 
reflects in the small but {nonzero} mass of the pion.}
\label{fig:Eff-pot-sigma-pi}
\end{figure}

The right panel of Fig.~\ref{fig:Eff-pot-sigma-pi}
shows ${\cal V}(\sigma, \pi)$
for some values of the coupling constant;
the  numbers attached to the respective curves 
are those of the `normalized coupling constant' $\alpha:=gN_CN_f\Lambda^2/\pi^2=:g/g_c$
where $g_c:\pi^2/N_cN_f\Lambda^2$, $N_C=3$, and $N_f=2$
 with $c$ standing for `critical', at which value the true vacuum with $\sigma=\pi=0$ becomes
unstable in the chiral limit; see the left panel.
One sees that the lowest energy state is given for $\pi=0$ but $\sigma\not=0$ when
$m\not=0$. In this case, ${\cal V}(\sigma, \pi=0)$ consists of 
the (negative) energy of the Dirac sea
for the quarks with the constituent quark mass 
\beq
M_q:=-2g\la \bar{q}q \ra+m
\label{eq:def-Const-Mass-off}
\eeq
 and an additional positive term 
with which the double counting of the interaction is avoided 
(\cite{Kunihiro:1983ej, Hatsuda:1994pi}).
%${\cal V}(\sigma, \pi)$
%the effective potential 
{The value $\sigma_0$ that minimizes the effective potential is determined by the stationary condition}
$
\d{\cal V}(\sigma, \pi=0)/\d\sigma\vert_{\sigma=\sigma_0}=0$,
which is reduced to
Eq.~(\ref{eq:def-Const-Mass-off}) with
\beq
\la \bar{q}q \ra=2 N_c N_f\int_{p<\Lambda}
\frac{d^3 p}{(2\pi)^3}\frac{-M_q}{E_p},\quad \quad E_p:=\sqrt{M_q^2+p^2}.
\label{eq:q-condensate-2fl}
\eeq
The resultant equation is called
the 'gap equation' determining the constituent quark mass self-consistently.
%and reduced to
%Off the chiral limit,
%the 'gap equation' for the constituent quark mass
%$M_q:=-2g\la \bar{q}q\ra+m$ reads 
%\beq
% {m}/{M_q}=1 - (2gN_cN_f/\pi^2) \int _0^{\Lambda} \, {p^2dp}/{E_p}.
%\label{eq:NJL2-gap}
%\eeq

%$M_q:=-2g\la \bar{q}q\ra$\, 
The physics of the SSB in the NJL model is better seen in the chiral limit. 
 The left panel of Fig.~\ref{fig:Eff-pot-sigma-pi}
shows ${\cal V}(\sigma, \pi)$;
%for some values of the coupling constant;
the  numbers attached to the respective curves 
are the same as in the right panel.
%those of $\alpha:=gN_CN_f\Lambda^2/\pi^2=:g/g_c$
%where $g_c:\pi^2/N_cN_f\lambda^2$, $N_C=3$, and $N_f=2$
% with $c$ standing for `critical'. 
One sees that 
${\cal V}(\sigma, \pi)$ in this case
%e chiral limit
%the effective potential 
%($m=0$) 
is  a function of $\sqrt{\sigma^2+\pi^2}$ only, and hence
symmetric for
rotations in the $\pi$-field direction in the figure.
As the coupling constant is increased, ${\cal V}(\sigma, \pi)$
%the effective potential 
tends to bend down and the 
origin at $\sigma=\pi=0$ ceases to be the lowest-energy state,  and 
the phase transition occurs at $g=g_c$ or $\alpha=1$:
For  
%$g\,>\,g_c$
$\alpha>1$, ${\cal V}(\sigma, \pi)$
%the effective potential 
{has its minimum at a nonzero value of $\sqrt{\sigma^2+\pi^2}$,
determined by the stationary condition.  This condition leads to} 
the 'gap equation' 
% determining the constituent quark mass self-consistently,
%\beq
$M_q:=-2g\la \bar{q}q\ra$\,
with  Eq.~(\ref{eq:q-condensate-2fl}).
%=:G_{\pi q}\sigma_0, 
%\label{eq:NJL2-gap-chiral-limit-0}
%\eeq
This equation for $M_q$ has a vanishing solution for {\em any} $g$, 
and for $\alpha>1$\,($g>g_c$) the nonzero solution $M_q=-2g\la\bar{q}q\ra$ 
emerges, which {\em is} the solution giving the lower energy
of the effective potential; 
the {nonzero} chiral condensate $\la\bar{q}q\ra$
implies that the chiral symmetry is spontaneously broken,
 and accordingly the system is in the {Nambu--Goldstone} phase.  

% For later convenience, we note that
% the gap equation in the chiral limit
% is reduced to 
%\beq
%1= (2gN_cN_f/\pi^2) \int _0^{\Lambda}\,{p^2dp }/E_p.
%\eeq
%which has a finite solution $M_q$ for $g\,>\,g_c$ (as well as a vanishing one for 
%all $g$)
%for $M_q\not=0$.
% and hence a finite chiral quark condensate
%$\la \bar{q}q\ra$. 
%Notice that the finiteness of the  chiral condensate 
%implies that the chiral symmetry is spontaneously broken,
% and the system is in the Nambu-Goldstone phase.  

Let us denote the massless and massive states by $\vert 0\ra$ and 
$\vert \sigma_0\ra$, respectively. Then the two states
 are related through the following equation (\cite{Hatsuda:1994pi}),
\beq
\vert \sigma _0\rangle =\prod _{{\bf p};r=R, L; f=u,d} 
\exp[-\frac {\theta_{\bf p}}{2}(a_0^{\dag}({\bf p},r,f)b_0^{\dag}(-{\bf p},r,f)\,-\,{\rm h.c.})
]\,\vert 0\ra ,
\quad \quad (\tan\theta_{\bf p}:=M/{\bf p}),
\eeq 
where the operators $a_0^{\dag}({\bf p},R,f)$ ($b_0(-{\bf p},R,f)$)
 %in $a_{_M}^{\dag}({\bf p},+1)$ 
creates (destroys) one chirality,
 while {\em both} 
%$a_0^{\dag}({\bf p},R)$ and $b_0(-{\bf p},R)$
 create one helicity.  
%Hence the  state  created by $a_{_M}^{\dag}
%({\bf p},+1)$ is an eigenstate of the helicity but {\em not} of the chirality.  
%Similarly,  $b_{_M}^{\dag}(-{\bf p},+1)$ does  not have a 
%definite chirality but  {\em  destroys} one helicity. 
Thus, 
the massive state $\vert \sigma_0\ra$ created by
$\prod _{{\bf p},r} U({\bf p},r)$ operating on $\vert 0\ra$
%has  a definite helicity but 
is a \textbf{superposition of states with different chiralities}
 implying  {\em not} an eigenstate of chirality. 
Then it is said that \textbf{the chiral symmetry is spontaneously broken in 
  $\vert \sigma _0\rangle$}.

On account of the gap equation determining the constituent quark mass $M_q$,
% (\ref{eq:NJL2-gap}), 
%the dispersion relations determining 
it turns out that the masses squared of
 the pion and $\sigma$ meson are determined by 
the following equations, respectively:
\beq
\frac{\hat m}{M_q} = m_{\pi}^{\,2}\, 2gC \cdot 
F\!\left(\frac{M_q}{\Lambda},\, \frac{m_{\pi}}{2\Lambda}\right),\quad
\frac{\hat m}{M_q} = (m_{\sigma}^{\,2}-4M_q^{2})\, 2gC \cdot 
F\!\left(\frac{M_q}{\Lambda},\, \frac{m_{\sigma}}{2\Lambda}\right),
%\quad 
%F(a,b) &=& \int_{0}^{1} 
%\frac{x^{2}\, dx}{\sqrt{x^{2}+a^{2}}}\,
%\frac{1}{x^{2}+a^{2}-b^{2}},
\label{eq:2-flav-pi-sigma-mass}
\eeq
where $F(a, b)$ is a positive-definite integral.
The first of (\ref{eq:2-flav-pi-sigma-mass}) tells us that the pion is massless, while the 
second  $m_{\sigma}=2M$ in the chiral limit:
The finite current quark mass gives the pion a tiny mass and
$m_{\sigma}$ becomes slightly larger than $2M$ and enters the continuum.

\paragraph{The Nambu relation and the riddle 
with the $\sigma/f_0(500)$}
%: the quantum fluctuation of the order parameter versus. the $\pi$-$\pi$ resonance}}

The simple mass-ratio relation obtained in the chiral limit
\beq
m_{\pi}\,:\, M_q\,:\, m_{\sigma}\simeq 0\,:\, 1\,:\, 2
\label{eq:Nambu-rel}
\eeq
is called {\bf the Nambu relation}.
Note that this proportionality  holds as long as the system is in the NG phase,
and  has an important
implication for the behavior of the mesons when the chiral symmetry 
is restored as given by the decrease in the absolute value of the chiral condensate 
$\sigma_0$: Since the dynamical mass $M_q$
is essentially proportional to the chiral condensate,
 we can naively expect that the sigma meson mass $m_{\sigma}\sim 2M_q$ 
would decrease as the chiral symmetry is being restored.
 In fact, this behavior reflects the fact that the $\sigma$ meson describes the 
quantum fluctuations $\la(\bar{q}q-\la \bar{q}q\ra)^2\ra$ of the 
order parameter $\la \bar{q}q\ra$ 
around the minimum point of the effective potential, quite similar to
 {\bf the Higgs particle} in the standard model,
 and the curvature there
decrease or flatten along with the 
chiral restoration.

%\begin{figure}[bht]
%\centering
%\includegraphics[width=.4\textwidth]{blankfig}
%\includegraphics[width=.53\textwidth]{para-pi-para-sigma.pdf}
%\caption{}
%\label{figx}
%\end{figure}\vspace{-.35cm}

Here,  some words are in order 
on the relation  between the $\sigma$ meson as the quantum fluctuations 
of  the order parameter of the chiral phase transition
 and the intricate low-lying broad scalar resonance now listed as $f_0(500)$ in 
the PDG (\cite{ParticleDataGroup:2024cfk}).
The  $f_0(500)$ has been identified experimentally 
in the $\pi$-$\pi$ scattering through the sophisticated theoretical analysis 
respecting the chiral symmetry, the analyticity, the crossing symmetry, 
and so on (\cite{Pelaez:2015qba}): Indeed, the hadronic $0^+$ state may mix with 
 tetra quark states (\cite{Jaffe:1976ig,Maiani:2004uc}), a glue ball,
and so on(\cite{Pelaez:2015qba}).  
Moreover, the celebrated success of the chiral perturbation
theory (\cite{Gasser:1983yg,Gasser:1984gg}) in the non-linear realization of the 
chiral symmetry may imply that the curvature of the effective potential 
at the symmetry-breaking point $\sigma=\sigma_0$ is so big that the mass
of the quantized field of the fluctuations of the order parameter 
can be large,
say, larger than 1 GeV. In fact, it is shown (\cite{Oller:1998hw}) that,
with the use of the lowest and the second-order chiral
Lagrangian by Gasser and Leutwyler (\cite{Gasser:1983yg,Gasser:1984gg},
the energy dependence of the phase shift 
of the $\pi$-$\pi$ scattering in the $\sigma$ channel
(as well as the $\rho$-meson one) is reproduced in the 
inverse amplitude method (\cite{Truong:1988zp,Truong:1991gv}) in an excellent manner
 with a complex pole at $m_{\sigma}=442 -i454$ MeV: In other words,
the $\sigma$ meson or $f_0(500)$ is dynamically generated without 
the seed of an elementary $\sigma$; see also (\cite{Oller:1998zr}).  
Anyway,  elucidating the nature of the $\sigma$ meson or $f_0(500)$ 
should reveal the essential ingredients of QCD dynamics 
(\cite{Kunihiro:2003yj}) including the axial anomaly (\cite{Tomiya:2016jwr,Aoki:2021qws}).

\subsection{Three-flavor linear $\sigma$ model with anomaly term}
\label{subsection:3-flavor-sigma}

At sufficiently low energies/long wavelengths so that the quark 
substructure of hadrons is invisible, it 
would be legitimate  to 
approximate the composite quark field
$\Phi_{ij}(x)$ by local bosonic fields $\phi_{ij}(x)$ with
the same transformation property for the
 $\mathrm{U(3)}_L \times \mathrm{U(3)}_R$ $\phi(x)$ group;
$\phi \rightarrow L\,\phi\,R^{\dagger}$.
Then an $\mathrm{SU(3)}_L \times \mathrm{SU(3)}_R$-invariant Lagrangian 
up to the fourth order in $\phi(x)$ is given by 
(\cite{Schechter:1971qa,Carruthers:1971ldw})
\[
\mathcal{L}^{(0)}_{\sigma}
=
\frac{1}{2}\,\mathrm{tr}(\partial_{\mu}\phi\,\partial^{\mu}\phi)
 -\frac{1}{2}\,\mu^{2}\,\mathrm{Tr}(\phi\,\phi^{\dagger})
 -\lambda\,\left[\mathrm{Tr}(\phi\,\phi^{\dagger})\right]^{2}
 -\gamma\,\mathrm{Tr}\!\left[(\phi\,\phi^{\dagger})^{2}\right]
 +\tau\,(\det\phi + \det\phi^{\dagger}).
\]
The coefficient $\tau$ represents the strength of the axial anomaly.  
The vacuum stability requires that $\lambda > 0$ and $\gamma > 0$.  
Furthermore, it should be that $\tau > 0$ for the role of the
present anomaly to be consistent with QCD.
The sign of $\mu^{2}$ determines whether chiral symmetry is 
spontaneously broken.
%We now consider the following term, which 
We add an explicit symmetry-breaking term 
due to nonvanishing current quark masses 
in a minimal way as 
%\label{eq:LSB-def}
$
\mathcal{L}_{\sigma}^{\mathrm{SB}}=-\eps_u (\phi_u+\phi_d) -\eps_s \phi_s$\, 
$(\phi_u:=\phi_{11},\, \phi_d:=\phi_{22}, \phi_s:=\phi_{33})$,
where the isospin symmetry is assumed. Thus the total Lagrangian 
reads $\mathcal{L}_{\sigma}=\mathcal{L}^{(0)}_{\sigma}+\mathcal{L}_{\sigma}^{\mathrm{SB}}.$

We write the vacuum expectation value of the field $\phi$ as
$\langle \phi \rangle_{0} = \bar{\phi}
= \mathrm{diag}(\bar{\phi}_u=:\sigma_{u},\, \bar{\phi}_d=\sigma_{u},\, \bar{\phi}_s=:\sigma_{s})$,
where $\sigma_{u}$ and $\sigma_{s}$ are determined, in the classical
approximation, from the stationary conditions of the vacuum energy;
%\begin{equation}
%\label{eq:vac-energy-def}
$\langle -\mathcal{L}_{\sigma} \rangle \equiv V(\sigma_{u},\,\sigma_{s})$.
%\end{equation}
%Here we assume that $\sigma_{u}$ and $\sigma_{s}$ are real constants.
%Noting that$\langle \mathcal{L}_{\mathrm{SB}} \rangle 
%  = -\,2\eps_u\,\sigma_{u} - \eps_s\,\sigma_{s}$,a straightforward calculation yields
%\begin{equation}
%\label{eq:V-sigma-u-s}
%\begin{aligned}
%V(\sigma_{u},\,\sigma_{s})
%= 2(2\lambda + \gamma)\,\sigma_{u}^{4}
%   + (\lambda + \gamma)\,\sigma_{s}^{4} + 4\lambda\,\sigma_{u}^{2}\sigma_{s}^{2} 
% - 2\tau\,\sigma_{u}^{2}\sigma_{s}
%   + \frac{\mu^{2}}{2}\,(2\sigma_{u}^{2} + \sigma_{s}^{2})
%  + 2\eps_u\,\sigma_{u} + \eps_s\,\sigma_{s}.
%\end{aligned}
%\end{equation}
We note that $V(0,\,0)=0$.

As a simple case, let us consider the chiral limit $(c=d=0)$.
In this case, since $\sigma_u=\sigma_s\equiv\sigma_0$, 
the stationary condition above reduces to a single equation:
$\sigma_0\Bigl[-\frac{\mu^2}{2}-6\lambda\sigma_0^2-2\gamma \sigma_0^2+\tau \sigma_0\Bigr]=0$.
This cubic equation has the solutions $\sigma_0=0$ and $\sigma_0\neq 0$.
The latter nonzero solution is obtained from
\begin{equation}
\mu^2+4(3\lambda+\gamma)\sigma_0^{\,2}-2\tau\sigma_0=0,
\label{eq:sigma-scc-chi-limit}
\end{equation}
which has real-value solutions when $\mu^2<0$ which we assume here;
%\[\sigma_0=\frac{\tau\pm\sqrt{-4\mu^2(3\lambda +\gamma)+\tau^2}}{4(3\lambda+\gamma)}.
%\]
we denote the positive one by $\sigma_M$.
One can confirm that the vacuum energy 
at $(\sigma_u,\sigma_s)=(\sigma_M,\sigma_M)$ is negative
$V(\sigma_M,\sigma_M)\,<\,0=V(0, 0)$. 
So we take $\sigma_0=\sigma_M$.
% instead of $\sigma_0=0$.
%In the following, we assume $\mu^2<0$ and, even away from the chiral limit $(c\neq0,\ d\neq0)$,
%we take $\sigma_{u,s}>0$.

Now that the vacuum has been determined, let us study the excitation spectrum on top of this vacuum.
To this end, we decompose the field $\phi$ 
into the vacuum expectation value $\bar{\phi}$ and a fluctuation $\phi'$:
$\phi = \bar{\phi} + \phi'$.
The fluctuation part is further represented 
by Hermitian scalar fields $S$ and pseudoscalar fields $P$ as
$\phi' = \frac{S+iP}{\sqrt{2}} = \sum_{\alpha=0}^{8} \frac{S_\alpha + i P_\alpha}{\sqrt{2}} \lambda_\alpha$
with
$S^\dagger = S = \sum_{\alpha=0}^{8} S_\alpha \lambda_\alpha$ and 
$P^\dagger = P = \sum_{\alpha=0}^{8} P_\alpha \lambda_\alpha$,
where $\lambda_\alpha$ are the Gell-Mann matrices ($\lambda_0 \propto \mathbf{1}$).
Substituting $\phi$ expressed in this way into the Lagrangian ${\cal L}_{\sigma}$, 
and extracting the quadratic terms in $\phi'$, one can determine the masses of the excitation modes.
The masses of the scalar mesons and pseudoscalar mesons, $m_{S\,\alpha}^2$ and $m_{P\,\alpha}^2$, 
are given by the coefficients of $-\frac12 S_\alpha^2$ and $-\frac12 P_\alpha^2$, respectively.

First, let us consider the chiral limit. In this case, the mesons can be classified
according to the flavor SU(3) symmetry, and the masses simply split into those of
an octet and a singlet. Here, we focus only on the pseudoscalar mesons. The pions
($\pi^{+},\, \pi^{-},\, \pi^0$), kaons ($K^{+},\, K^{-},\, K^0,\, \bar{K}^0$), and the
eta meson ($\eta_8$) form the octet.
Their squared masses are given by
\beq
m^2_{P8} = \mu^2 + 4(3\lambda + \gamma)\sigma_0^2 - 2\tau\sigma_0 = 0.
\eeq
In the second equality, we used the condition for determining the vacuum expectation
value in the chiral limit, Eq.~(\ref{eq:sigma-scc-chi-limit}). 
Thus, the pseudoscalar octet becomes massless and 
constitutes the Nambu--Goldstone bosons.
On the other hand, 
the squared mass of
the singlet pseudoscalar $\eta_0$
becomes, on account of (\ref{eq:sigma-scc-chi-limit}) again,
\beq
m^2_{\eta_0}
  = \mu^2 + 4(3\lambda + \gamma)\sigma_0^2 + 4\tau\sigma_0
  = 6\tau\sigma_0 \, > \, 0,
\label{eq:eta-0-mass-chiral-limit}
\eeq
which is definitely nonvanishing. 
Recall that the parameter $\tau$ is the coefficient of the axial
anomaly term in the Lagrangian. Thus we find that 
the singlet pseudoscalar meson  $\eta_0$ remains massive even
in the chiral limit due to the axial anomaly.

Here, an important model-specific remark is in order.  In this linear
sigma model, Eq.~(\ref{eq:eta-0-mass-chiral-limit}) shows that the anomalous
singlet mass generated by the determinant interaction is proportional to
$\tau\sigma_0$.  Consequently, even with a fixed anomaly coupling $\tau$, the
$\eta_0$ mass decreases as the chiral order parameter $\sigma_0$ is reduced.
This illustrates why an in-medium decrease of the physical $\eta'$ mass, by
itself, should not be identified with a suppression of the U(1)$_A$ anomaly;
this point will be discussed again below.

\subsection{Chiral effective models with vector and axial vector interactions}
%As in the previous section, 
%at so sufficiently low energies/long wavelengths that the quark 
%substructure of hadrons is invisible, it is reasonable to 
%approximate 
The composite vector and axial-vector  fields
$V_{mu}^a(x)$ and $A_{mu}^a(x)$ would couple to the parity minus
vector mesons and parity plus axial vector mesons, respectively;
they are parity partners.
 For instance, the vector  modes for 
$a=1\sim 3$ may  be identified as the iso-vector 
$\rho (770)$\, ($ J^{PC}=1^{--}$)  
and the $a_1 (1260)$\,($J^{PC}=1^{++}$) mesons,
respectively. 
If it were to exist well defined 
vector and axial vector modes in 
the chiral symmetric phase with $f_{\pi}=0$,
they should form a chiral multiplet with a mass degeneracy.
In reality, they have clearly different masses. Is it possible
to account for the mass difference in terms of the finiteness of $f_{\pi}$,
an order parameter of chiral transition?
In a celebrated paper (\cite{Weinberg:1967kj}), S. Weinberg derived
two sum rules for the difference between the spectral 
functions in the vector and axial-vector channels, $\rho_V(s)$ and $\rho_A(s)$, 
on the basis of the chiral symmetry in the high-energy asymptotic 
state\footnote{For more comprehensive derivation of the sum rules, 
see also (\cite{Das:1967zze,sakurai1969currents,Kapusta:1993hq}).}, which read 
\beq
(\textrm{I})\, \, 
\int^{\infty}_0\,\frac{ds}{s}[\rho_V(s)-\rho_A(s)]=f_{\pi}^2, \quad
\quad (\textrm{II})\,\, \int^{\infty}_0\,ds[\rho_V(s)-\rho_A(s)]=0, 
\label{eq:sum-rules-I-II}
\eeq
%As a side remark, we refer to (\cite{Das:1967zze,sakurai1969currents}) for a 
%comprehensive derivation of the Weinberg sum rules 
% and 
%Retuning back to the Weinberg's argument, 
A mass formula 
between $m_{\rho}$ and $m_{A_1}$ was derived 
by assuming one-pole dominance of the spectral functions
 and the KSRF relation (\cite{Kawarabayashi:1966kd, riazuddin1966algebra}) 
\footnote{See (\cite{sakurai1969currents}) for a comprehensive 
account of the KSRF relation, and the derivation of the Weinberg's mass formula.} ;
the results read 
$m_{A_1}=\sqrt{2} m_{\rho}\simeq 1090$\, MeV, which deviates from the
empirical mass only by 10 \%.

We remark that Das, Mathur, and Okubo (\cite{Das:1967ek}) derived
another sum rule, known as the DMO sum rule, given by
\beq
(\textrm{III})
\int^{\infty}_0\,\frac{ds}{s^2}[\rho_V(s)-\rho_A(s)]=f_{\pi}^2\frac{\la r^2\ra_{\pi}}{3}-F_A, 
\label{eq:sum-rules-III}
\eeq
where $\la r^2\ra_{\pi}$ is the mean squared radius of the charged pions 
and $F_A$ denotes the coupling constant for the radiative pion decay.
Experimental tests have been made of the 
sum rules (I), (II), and (III) from the spectral data of
the decays of the $\tau(1776.84\pm 0.17)$ 
lepton (\cite{ALEPH:1998rgl,OPAL:1998rrm}), which show
that (III) is well satisfied within the kinematic range accessible in 
the $\tau$ decay, while
 the other two only approach saturation with large error bars.

\subsection{Parity-doublet model of baryons}
\label{section:parity-doublet baryons}

So far, we have exclusively dealt with chiral properties of mesons.
How about baryons? First of all, how are baryons
 transformed under the chiral transformation?
Here, we introduce  a linear $\sigma$  model which incorporates
 baryons (\cite{Detar:1988kn}), and discuss its properties relevant to 
chiral restoration.
This model allows a chirally invariant mass term $m_0$, so that a part of the nucleon mass need not vanish even when the chiral condensate is reduced.
 
\subsubsection*{Model}

Motivated by the results of lattice QCD simulations at finite temperature,
which shows parity doubling of the {\em screening masses}, i.e., 
the damping rates of the space correlation function,  of  
positive- and negative-parity nucleons 
(\cite{Detar:1987kae, Detar:1987hib, Gottlieb:1987gz}),
 the following  parity-doublet model for baryons 
{\em at finite temperature and density}was proposed 
(\cite{Detar:1988kn})\footnote{See (\cite{Jido:1998av,Jido:2001nt}) for 
a comprehensive and detailed account of the model and its possible 
phenomenological consequences in hadron physics in vacuum.}:
\beq
\label{lagdoubl}
{\cal L}
& =& \bar{\Psi} i\gamma\cdot\partial \Psi
     - g_1 \bar{\Psi}(\sigma + i\btau\cdot\bpi\, \rho_3 \gamma_5)\Psi
      + g_2 \bar{\Psi}(\rho_3\sigma + i\btau\cdot\bpi\,\gamma_5)\Psi
     - iM_0 \bar{\Psi}\rho_2\gamma_5\Psi \nonumber \\
&& \quad       + {\cal L}_M(\sigma, \bpi),
\eeq
where $\Psi ={}^t(\psi_1,\, \psi_2)$, and both $\psi_1$ and $\psi_2$ are
Dirac spinors. The matrices $\rho_i$ ($i=1,2,3$) denote the Pauli matrices
acting on the two-component spinor $(\psi_1,\,\psi_2)$. The fields $\psi_i$
($i=1,2$) represent the ``bare'' nucleon states with positive and negative
parity, respectively. The term proportional to $\rho_2$ in
Eq.~(\ref{lagdoubl}) mixes these two states and generates the physical
positive- and negative-parity eigenstates, $\Psi_{+}$ and $\Psi_{-}$.
This Lagrangian is invariant under the following extended chiral
transformation, in which the two baryon fields
$\Psi={}^t(\psi_1,\,\psi_2)$ transform oppositely under the axial part:
\beq
\delta \Psi
 = \frac{1}{2} i\, \balpha \cdot \btau \, \Psi
  + \frac{1}{2} i\, \bbeta \cdot \btau\, \rho_3 \gamma_5 \Psi.
\eeq
The factor $\rho_3$ implies opposite axial transformations of $\psi_1$ and
$\psi_2$, while their vector transformations are the same. This opposite
assignment of the chiral transformation is called the ``mirror assignment''
in the literature (\cite{Jido:2001nt}).

An important remark on this model concerns 't Hooft anomaly matching
(\cite{tHooft:1979rat,Frishman:1980dq,Coleman:1982yg}).
In the ordinary Nambu--Goldstone phase, the QCD chiral anomaly is carried by the
Nambu--Goldstone sector; for $N_f\geq 3$ this matching is encoded explicitly in
the Wess--Zumino--Witten term
(\cite{Wess:1971yu,Witten:1983tw}). Thus, the presence of massive
parity-doublet baryons in the chirally broken phase is not in conflict with
anomaly matching. Moreover, owing to the opposite chiral transformations of
$\psi_1$ and $\psi_2$ in the mirror assignment, the relevant triangle-anomaly
contributions of a complete mirror pair cancel, so that the mirror baryons
themselves carry no net anomaly. At nonzero temperature and/or chemical potential, however, the infrared
realization can be more involved because Lorentz invariance is lost
(\cite{Itoyama:1982up,Pisarski:1997bq,Hsu:2000by}). Gapless collective
modes, such as hydrodynamic modes, or, at finite density, particle--hole
excitations around a Fermi surface may then participate in anomaly matching.
Thus, a hypothetical confined phase with exact chiral symmetry cannot match
the QCD anomaly with massive mirror baryons alone, but additional gapless
infrared degrees of freedom may provide the required anomaly matching. Hence,
anomaly matching constrains the complete infrared realization of a chirally
restored confined phase rather than the parity-doublet model itself.

To obtain the physical positive- and negative-parity states, we diagonalize
the bilinear part of Eq.~(\ref{lagdoubl}). In the lowest-order
approximation, the eigenstates are found to be
$\Psi_{+}(p)
 = N \,{}^{t}\!\left( \psi_{+}(p),\, e^{-\theta}\gamma_{5}\psi_{+}(p) \right)$,\,
$\Psi_{-}(p)
 = N \,{}^{t}\!\left( -e^{-\theta}\gamma_{5}\psi_{-}(p),\, \psi_{-}(p) \right)$,
where the Dirac spinors $\psi_{\pm}$ satisfy
$\gamma\cdot p \, \psi_{\pm}(p) = M_{\pm}\, \psi_{\pm}(p)$ where 
\beq
M_{\pm}
 = \mp g_2 \sigma_0
   + \sqrt{(g_1\sigma_0)^2 + M_0^2 },
\label{eq:DeTKuni-masses}
\eeq
$e^{-\theta}  = {M_0}/[g_1\sigma_0 + \sqrt{(g_1\sigma_0)^2 + M_0^2}]$
and $N = [{1 + \exp(-2\theta)}]^{-1/2}$
assuring the normalization condition $\Psi_{\pm}\Psi_{\pm} = 1$.

\begin{figure}[htb]
\centering
\includegraphics[width=0.45\textwidth]{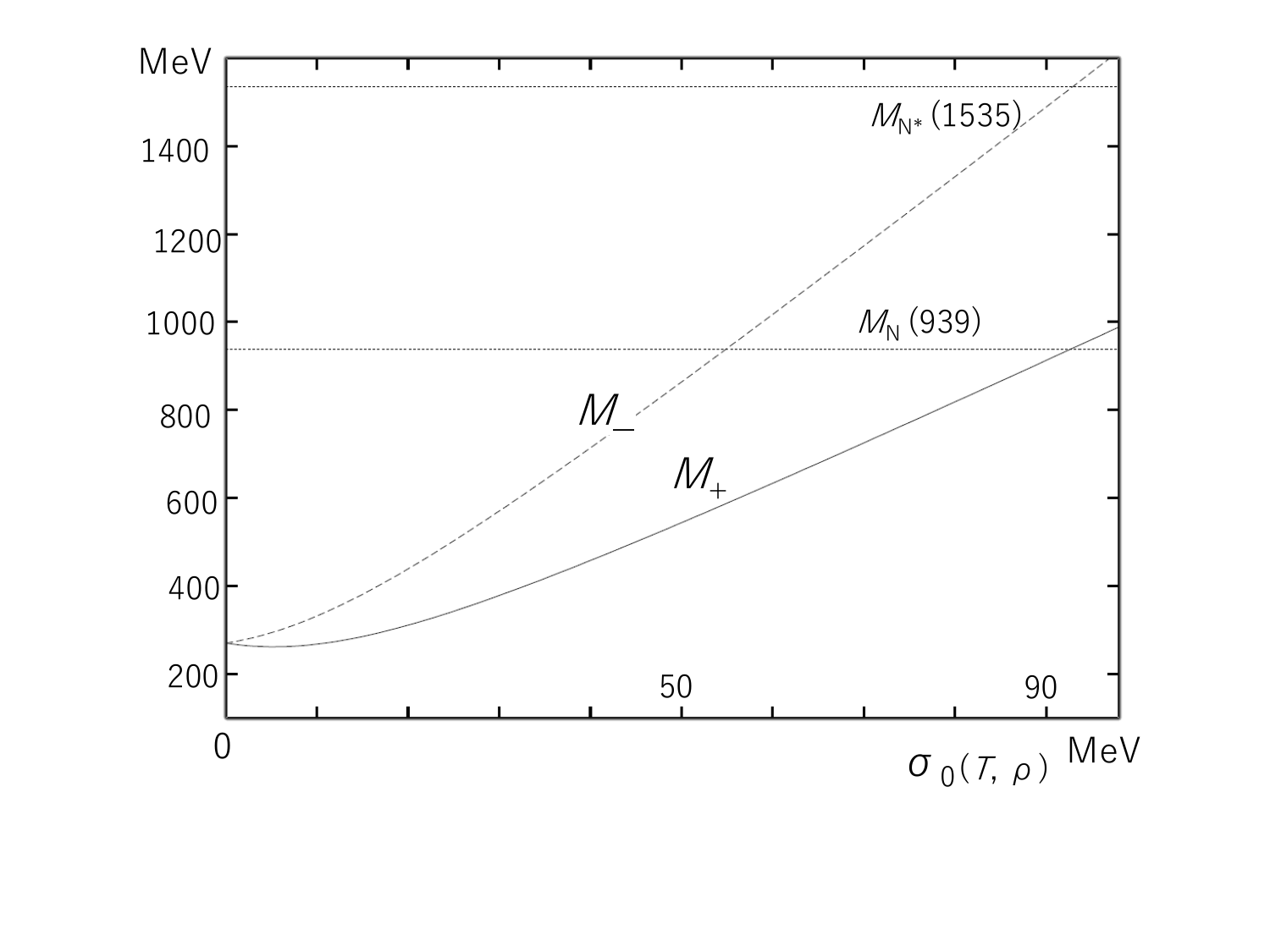}
\vspace{-.75cm}\caption{The $\sigma_0$ dependence of the mass $M_{+}$ ($M_-$) of the 
positive-(negative-)parity nucleon when $M_0=270$ MeV. The two horizontal lines
indicate the physical masses $M_N=939$ MeV  and $M_{N^*}=1535$ MeV in the vacuum. 
The temperature and density dependence of the chiral condensate
$\sigma_0$ in the equilibrium is made explicit because the model is 
supposed to be applied in the hot and/or dense matter where the Lorentz invariance is
lost.}
\label{DeTKuni-baryons}
\end{figure}

%\vspace[-.75cm]
Figure~\ref{DeTKuni-baryons} shows the $\sigma_0$ dependence 
of the masses $M_{+}$ and $M_-$
% of the positive-(negative-)parity nucleon 
given by Eq.~(\ref{eq:DeTKuni-masses}) with the parameters 
discussed below. However,  the value $M_0 = 270~{\rm MeV}$ and other parameter 
values used in the figure should not be taken as definite ones associated to 
the parity-doublet model. In fact,  applications of  extended parity-doubling 
models of baryons in three-flavors to the equation of state of neutron stars suggests
that an optimum value of $M_0$ should be much larger than the above value and 
might be as large as 500 to 900 MeV; see for instance (\cite{Gao:2026scv})
and related references cited in \S~\ref{section:baryon-neutron-star} below.
From the figure, one can  see the following:
(1)~The negative-parity state is heavier than the positive one ($M_{-} > M_{+}$) and 
as chiral symmetry is restored ($\sigma_0\,\to\,0$), both $M_{\pm}$ decrease
and the two baryons become degenerate since
the difference is given by $\Delta M_{\pm} :=M_{-} - M_{+} = 2 g_2 \sigma_0$.
However, the $\sigma_0$ dependence of $M_{-}$ is 
larger than $M_{+}$.
%\item 
(2)~Even when the chiral symmetry is restored,
      $\sigma_0 \rightarrow 0$, the masses do not vanish:
      $M_{\pm} \;\longrightarrow\; |M_0| \neq 0$,
      demonstrating the characteristic parity doubling of the model in accordance 
with the result of the lattice simulations (\cite{Detar:1987kae, Detar:1987hib, Gottlieb:1987gz}).
%\end{enumerate}

%A natural question is which physical baryon can be identified as
%the chiral partner $N_{-}$ of the nucleon.
Although it might be the case that $N_{-}$ corresponds to no known baryon
because of experimental difficulties,
let us simply identify $N_{-}$ with $N^{\ast}(1535)$.
The decay width of $N^{\ast}(1535)\, \to\, N + \pi$ is given by
$\Gamma_{\pi N} = \frac{3\, q_N (E_N + M_N)}{4\pi M_{N'}}g_{\pi NN^*}^{\,2}
  \simeq 70~{\rm MeV}$,
and using the experimental values
$M_{+}=939~{\rm MeV}$,\,$M_{-}=1535~{\rm MeV}$, and
$\sigma_0 = 93~{\rm MeV}$, we obtain $g_{\pi NN^*} = 0.70$, and thus
the model parameters are nicely determined as
$\sinh\theta = 5.5$,\,$g_1 = 13.0$,\,$g_2 = 3.2$, and $M_0 = 270~{\rm MeV}$.

The axial charges of the nucleon N and N$^{\ast}$ are calculated to be
% has a matrix structure and is given by
%\beq
%\hat{g}_A =
%\begin{pmatrix}
$(g_A)_{N}=\tanh \theta=-(g_A)_{N^{\ast}}$.
% & -1/\!\cosh\theta \\-1/\!\cosh\theta & -\tanh\theta
%\end{pmatrix}.\eeq
The fact that these values have opposite signs 
is a characteristic feature of this model. 
Indeed, a lattice QCD simulation with $u$ and $d$ quarks (\cite{Takahashi:2008fy}) 
shows that $(g_A)_{N^*N^*}$ is slightly negative or consistent with zero
within the numerical errors. Moreover, the fact that the coupling to 
$N$-$\eta$ is large suggests that lattice QCD simulations incorporating the strange
quark are indispensable for properly describing  $N^{*}(1535)$.

\subsubsection*{Model extension}
There have been  quite a few attempts to extend the original model
in various ways, such as incorporating 
the higher baryon states (\cite{PhysRevLett.84.3252}), 
strangeness baryons with/without 
 the anomaly terms, various meson fields, and so on 
(\cite{Nemoto:1998um,Steinheimer:2011ea,PhysRevD.92.054022,Olbrich:2015gln,Olbrich:2017fsd,Gao:2022klm,Kong:2023nue,Gao:2025eax}); see also references cited in these articles.
The axial charges provide another useful constraint on such extensions.
In an extended parity-doublet model containing
$N(939)$, $N(1440)$, $N(1535)$, and $N(1650)$, with mixing among several
chiral representations and derivative interactions, Yamazaki and Harada
(\cite{Yamazaki:2018stk}) found the sum rule
$ \sum_{i=1}^{4} g_A(N_i)=0$.
The individual physical axial charges need not, however, take their naive
mirror-assignment values. This sum rule is a consequence of the mirror
structure of the axial-charge matrix and should not be identified with the
triangle-anomaly cancellation discussed above, although both are rooted in the mirror structure of the baryon sector. More recently,
Kummer, Leupold and von Smekal (\cite{Kummer:2025kch}) introduced kinetic
mixing, generated by higher-derivative meson--baryon couplings, to overcome
the well-known difficulty of the standard two-state parity-doublet model in
reproducing the empirical nucleon axial charge $g_A>1$.

\subsubsection*{Test of the model}
An important problem is to test this model (\ref{lagdoubl}) 
hopefully by experiment or by QCD lattice simulations.
Some intensive and instructive discussions of the $N^{\ast}(1535)$ phenomenology 
as described by the model are given in (\cite{Nemoto:1998um,Jido:2001nt}).
Recently, a lattice QCD simulation (\cite{PhysRevD.92.014503})
has shown that 
as $T$ is increased in the hadron phase ($T<T_c$){, where $T_c$ is the pseudocritical 
temperature for the confinement--deconfinement} 
transition as described by the renormalized Polyakov loop,
the `mass' \footnote{The mass extracted by fitting the time correlation functions of
the nucleon operators with two exponential functions.}
 of the positive-parity nucleon is largely unaffected
by temperature, whereas the mass of the negative-parity nucleon shows a stronger 
temperature dependence and decreases rapidly as $T$ is raised{.  At sufficiently high temperature $T>T_c$,}
 the positive- and negative-parity contributions become approximately degenerate, signaling parity doubling.
All these features are not inconsistent  with  the prediction given by the 
present model (\ref{lagdoubl}): see Fig.~\ref{DeTKuni-baryons}
%2 (b) in (\cite{Jido:2001nt}) . 

\section{QCD matter at finite temperature and baryon density}

Next, we discuss what would be expected to occur in
hot and/or dense QCD matter, i.e., the matter composed of hadrons and/or 
quarks and gluons.
In the vacuum, chiral symmetry is spontaneously broken and the condensate $\langle\bar q q\rangle$ is nonzero.
If the environment is changed by heating the system or by adding baryons,
the vacuum structure is modified.  At low temperature and baryon density,
the presence of thermally excited pions and nucleons, respectively, reduces
the magnitude of the chiral condensate.  In the following subsections, we
show how this reduction follows directly from the Hellmann--Feynman theorem
(\cite{Hellmann1937,Feynman1939}).  Indeed,
the Hellmann--Feynman theorem was intensively used to extract scalar and
strange-quark contents of hadrons from their current-quark-mass dependence
%notably in early chiral-quark and NJL-model analyses
(\cite{Kunihiro:1988sqc,Bernard:1987sg,Nelson:1989sgc,Kunihiro:1990ts,Hatsuda:1991fml,Hatsuda:1990uw}).

\subsection{Preliminaries: the Hellmann--Feynman theorem and state normalization}
\label{section:H-F-theorem}

For a Hamiltonian $\hat H(\lambda)$ with a unit-normalized eigenstate
$|E,\gamma\rangle$, differentiation of
$(E-\hat H)|E,\gamma\rangle=0$ immediately gives the Hellmann--Feynman
relation
\beq
 \frac{dE}{d\lambda}
 =\langle E,\gamma|\frac{d\hat H}{d\lambda}|E,\gamma\rangle .
\label{eq:H-F-3}
\eeq
For one quark flavor, the current-mass dependent part of the QCD Hamiltonian
is $H_m=m_q\int d^3x\,\bar q q$; hence, for a unit-normalized hadron state
$|\widetilde h\rangle$ ($\la \widetilde h|\widetilde h\rangle=1$), one has
$\partial E_h/\partial m_q=\langle\widetilde h|\int d^3x\,\bar q q|\widetilde h\rangle$.
For a stable hadron at rest, $E_h=M_h$.

In relativistic field theory it is more usual to adopt the covariant
normalization $\langle h(\bfp')|h(\bfp)\rangle_{\rm c}=2E_{\bfp}(2\pi)^3
\delta^{(3)}(\bfp'-\bfp)$.  In a box of volume $V$, the corresponding
unit-normalized state is
$|\widetilde h\bfp)\rangle=|h(\bfp)\rangle/\sqrt{2E_{\bfp}V}$.
The same theorem can then be written compactly as
\beq
 \frac{\partial E_{\bfp}}{\partial m_q}
 =\frac{1}{2E_{\bfp}}\langle h(\bfp)|\bar q q|h(\bfp)\rangle_{\rm c},
 \qquad
 \frac{\partial M_h}{\partial m_q}
 =\frac{1}{2M_h}\langle h|\bar q q|h\rangle_{\rm c} .
\label{eq:HF-covariant}
\eeq
Note that the factor $1/(2M_h)$ is entirely a consequence of the state-normalization
convention, and if we had used the {non-covariant} normalization
$\langle h(\bfp')|h(\bfp)\rangle:=(2\pi)^3
\delta^{(3)}(\bfp'-\bfp)$, it does not appear (\cite{Kunihiro:1988sqc,Bernard:1987sg,Nelson:1989sgc,Kunihiro:1990ts,Hatsuda:1991fml,Hatsuda:1990uw}).
In the following we shall use the {\em {non-covariant} normalization} since it gives simpler
and intuitively understandable expressions.

  Strictly, $M_h=E_h-E_0$, so the scalar matrix element in
Eq.~(\ref{eq:HF-covariant}) is understood to be vacuum-subtracted.
% thissubtraction can be implemented by using the normal product $:\bar{q}q:$
% in hadronic matrix elements. 

\subsection{The case of low temperature with vanishing baryon density}

The Hellmann--Feynman theorem also applies to thermal expectation values.
If $F(T)$ denotes the Helmholtz free-energy density with the vacuum
contribution subtracted, the change of the condensate is
%\beq

$ \delta\langle\bar q q\rangle :=\langle\bar q q\rangle_T-\langle\bar q q\rangle_0
 ={\partial F(T)}/{\partial m_q}$; accordingly,
\beq
\langle\bar q q\rangle_T=\langle :\bar q q:\rangle_0
 +\frac{\partial F(T)}{\partial m_q}.
\label{cond-T}
\eeq

\begin{figure}[thb]
\centering
\includegraphics[width=.65\textwidth]{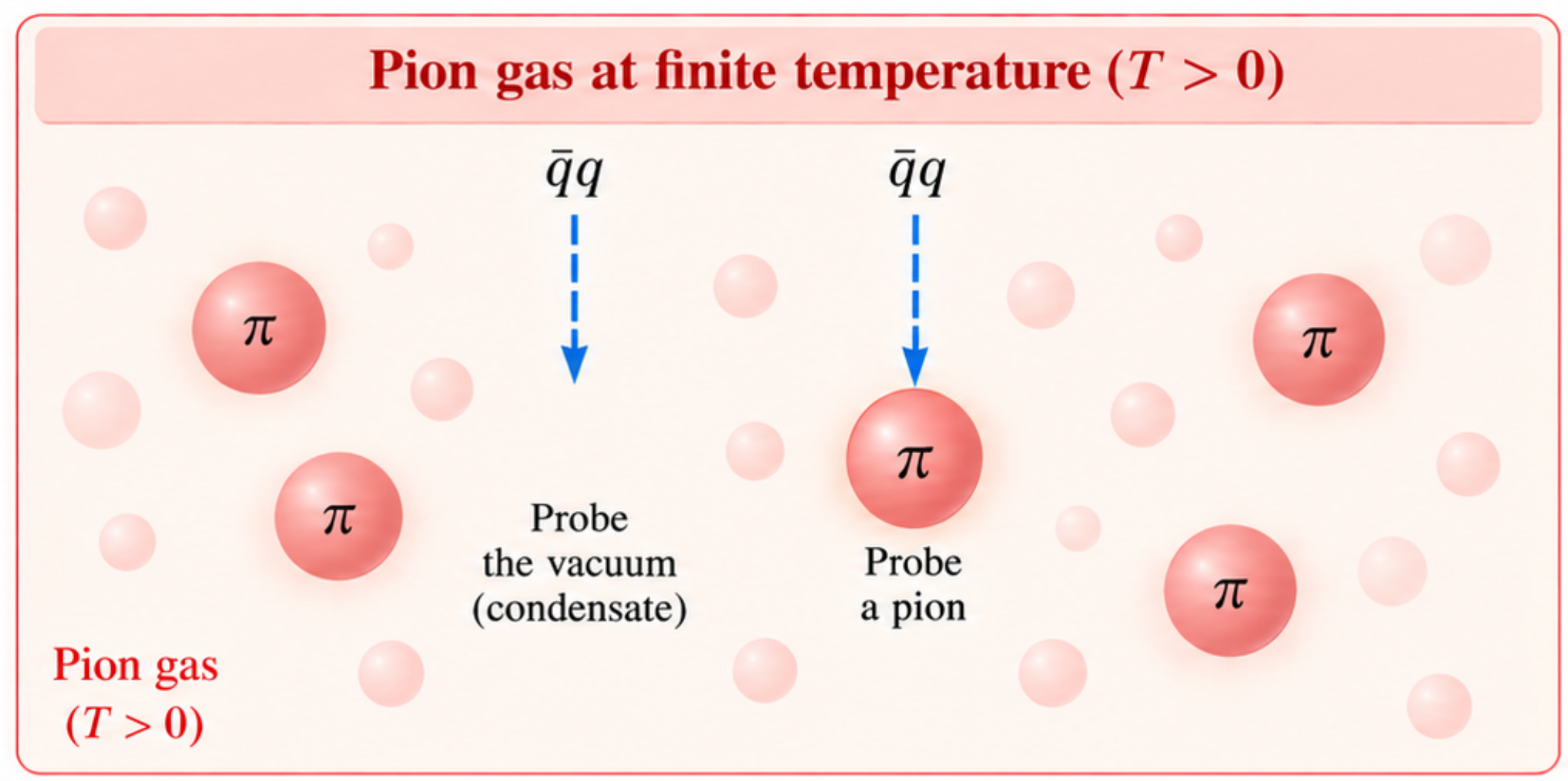}
\caption{Pictorial interpretation of Eq.~(\ref{pi-qcon2}).
%using covariantlynormalized pion states. 
{Thermally excited pions contribute positively to 
the chiral condensate} as given by 
$\langle\pi(k)|\bar q q|\pi(k)\rangle$ weighted by the thermal distribution.
It therefore reduces the magnitude of the negative vacuum chiral condensate.}
\label{Hellmann-Feynman-T:fig}
\end{figure}

At sufficiently low temperature the system is well approximated by a dilute,
noninteracting pion gas, for which
$F_{\pi\text{-gas}}=3T\sum_{\bf k}\ln(1-e^{-E_\pi(k)/T})$ with
$E_\pi(k)=\sqrt{m_\pi^2+k^2}$.  Therefore
\beq
 \delta\langle\bar q q\rangle
 \simeq 3\sum_{\bf k}\frac{\partial E_\pi(k)}{\partial m_q}\,n_\pi(k)
 =3\sum_{\bf k}\,\langle\pi(k)|:\bar q q:|\pi(k)\rangle \,n_\pi(k)
 =3\sum_{\bf k}\, \frac{E_{\pi}(k)}{m_{\pi}}\langle\pi|:\bar q q:|\pi\rangle \,n_\pi(k),
\label{pi-qcon2}
\eeq
where $n_\pi(k)=1/(e^{E_\pi(k)/T}-1)$ and the last matrix element is
{non-covariantly normalized}.
%  Equivalently, with unit-normalized pion states the
%factor $1/(2E_\pi)$ is absent and the spatially integrated scalar densityappears.
  Thus the thermal change of the condensate is simply the scalar
quark content of the thermally excited pions weighted by their Bose
distribution.  This also gives the intuitive interpretation shown in
Fig.~\ref{Hellmann-Feynman-T:fig}.
The scalar quark content of the pion may be estimated
from the GOR relation (\ref{eq:GOR}) as (\cite{Hatsuda:1990uw}); see also (\cite{Kunihiro:1990ts})
\[
\la \pi|\bar{q}q|\pi\ra
\simeq m_\pi/(m_u+m_d){\sim 7--10>0}.
\]
Thus, the presence of thermally
excited pions reduces the magnitude of the negative vacuum chiral condensate
as the temperature rises, signaling partial restoration of chiral symmetry.
As the temperature increases and the number of thermally excited pions grows,
the interactions among them can no longer be neglected.
Calculations that incorporate such interaction effects have been carried out
  (\cite{Gerber:1988tt}) for the case of the chiral limit with $N_f = 2$.

\subsection{The case of 
low baryon density at vanishing temperature}
\label{subsection4.2}

Let $|{\rm nm}\rangle$ be the nuclear-matter state at zero temperature.  In the
linear-density approximation,
\begin{eqnarray}
 \la {\rm nm} | {\cal H}_{\rm QCD} | {\rm nm} \ra
 = E_{\rm vac}+n_B\left[M_N+E_{\rm b}\right],
\label{en-density}
\end{eqnarray}
where $E_{\rm vac}$ is the vacuum energy density, $M_N$ the nucleon mass, and
$E_{\rm b}$ the binding energy per particle.  Neglecting the small
quark-mass dependence of $E_{\rm b}$, the Hellmann--Feynman theorem gives
the leading-density change of the light-quark condensate.

For later use, we first write the pion--nucleon sigma term with the non-covariant
normalization of the nucleon state,
$\la N(\bfp')|N(\bfp)\ra=(2\pi)^3\delta^{(3)}(\bfp'-\bfp)$:
\beq
 \sigma_{\pi N}
 =\hat m\,\frac{\partial M_N}{\partial\hat m}
 ={\hat m}\la N|\,:\bar uu+\bar dd:\,|N\ra,
 \qquad \hat m=\frac{m_u+m_d}{2}.
\label{eq:sigma-piN-cov}
\eeq
 Thus the corresponding dilute-medium
contribution is
$
{n_B}\, \la N|\,:\bar uu+\bar dd:\,|N\ra.
$
Using Eq.~(\ref{eq:sigma-piN-cov}) and the GOR relation
(\ref{eq:GOR}), one obtains
\beq
 \frac{\la\bar uu+\bar dd\ra_{n_B}}
      {\la\bar uu+\bar dd\ra_0}
 \simeq
 1-\frac{n_B\,\sigma_{\pi N}}{f_\pi^2m_\pi^2},
\label{cond-rho}
\eeq
which is independent of the normalization convention when expressed in terms
of the physical sigma term.  Thus, 
the presence of nucleons reduces the magnitude of the chiral condensate as
the density increases.  Since the empirical value of $\sigma_{\pi N}$ is
positive and of order several tens of MeV, the reduction is roughly
$30$--$35\%$ at normal nuclear density
$\rho_0\simeq0.17~{\rm fm}^{-3}$ for typical values of the sigma term.
%\footnote{In Refs.~(\cite{Kunihiro:1990ts,Hatsuda:1990uw}), quark and gluon
%contents of various baryons are evaluated using a chiral quark model.}

\begin{figure}[thb]
\centering
\includegraphics[width=.65\textwidth]{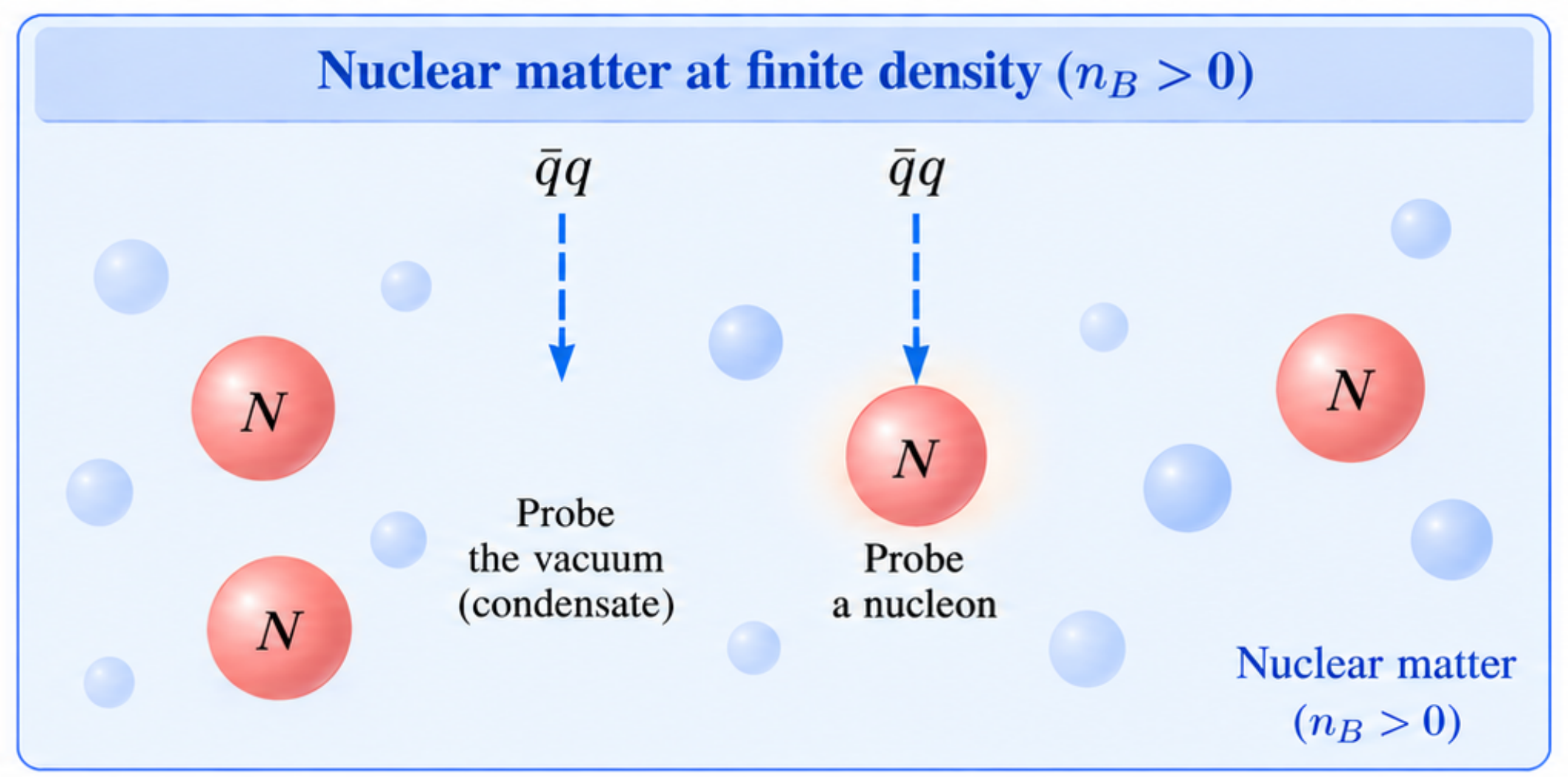}
\caption{A pictorial explanation of Eq.~(\ref{cond-rho}) using {non-covariantly
normalized nucleon states}.  The nucleon contribution 
$n_{B} \langle N|\bar uu+\bar dd|N\rangle 
=n_{B}\partial M_N/\partial\hat m=n_B \sigma_{\pi N}/\hat m$ is definitely positive.
It therefore reduces the magnitude of the negative vacuum chiral condensate
as the nucleon density increases.}
\label{Hellmann-Feynman-rho:fig}
\end{figure}

The physical origin of this decrease is the same as in the finite-temperature
case: the role played there by thermally excited pions (and other hadrons)
is now played by nucleons, as illustrated in
Fig.~\ref{Hellmann-Feynman-rho:fig}.  When the system is probed by
$\bar qq$ ($q=u,d$), the probe may act on the vacuum or on a nucleon, with
the latter contribution proportional to the nucleon density.
It is worth noting that the above argument based on the Hellmann--Feynman
theorem provides a more precise formulation of the intuitive picture
discussed in Ref.~(\cite{Brown:1987hj}), namely the
\textbf{Clearing-Out model}, in which the presence of nucleons in nuclear
matter ``clears out'' part of the vacuum contribution and thereby reduces
the magnitude of the chiral order parameter.

These results suggest that partial restoration of chiral symmetry in a
nuclear medium can be explored using heavy nuclei.  The chiral condensate
itself is, however, not directly observable, and experimental searches must
therefore rely on indirect probes such as in-medium hadron properties,
spectral functions, and symmetry relations among correlators.

\section{Phenomenological observables to probe chiral restoration}

We have seen in \S\ref{subsection4.2} that, through the
Hellmann--Feynman theorem, the presence of nucleons reduces the magnitude
of the chiral condensate by roughly $30$--$35\%$ at normal nuclear density
relative to the vacuum value. 
The chiral condensate itself is, however,
 not directly measurable in experiments. Searches for chiral restoration therefore rely on observables that are theoretically connected to the condensate or to symmetry relations among correlation functions.

Generally, a change of the QCD vacuum along with that of
 the environment  characterized by external variables such as 
 temperature $T$ and baryon density $\rho$ (or 
baryon chemical potential $\mu$)
 may cause some novel phenomena such as
modifications of spectral properties of hadrons or hadronic modes 
embedded in the system. 
 Important examples include deeply bound pionic atoms, in-medium vector spectral functions measured through dileptons, scalar-isoscalar correlations, possible $\eta'$-nucleus bound states, and parity doubling in the baryon sector.

%when some hadronic quantum mumbers are injected into
% QCD medium by imbedding a hadron in the system,

\subsection{Spectral functions}
\label{subsection:spectral function}
An appropriate way to describe the relevant system with 
some hadronic quantum numbers injected is 
to use the spectral function deduced from the 
response function (or the retarded Green's function),
because the hadron would acquire a finite, possibly large, 
width owing to the interactions with
the surrounding medium that it
 might even lose its identity.
% by forming a collective state with 
%together with other particles in the  medium, 
%even when the hadron has no or a tiny intrinsic width in the vacuum.
Then, we now proceed to define what the spectral function is.

Let $H_{\rm QCD}$ be QCD Hamiltonian  and  define
 $K$ by $K= H_{\rm QCD} -\mu  N$,
where $\mu$ is the chemical potential and $N$ the quark number operator.
 The statistical average $\lla  O\rra$ of an arbitrary operator $O$ 
is expressed as 
\beq
\lla O\rra = Z^{-1} \Tr [ e^{-\beta  K}  O],
\eeq
with $Z$ being the grand partition function, which is related to
 the thermodynamic potential $\Omega$ as
$Z=\Tr e^{-\beta  K}=e^{-\beta\Omega}$.
Then the retarded Green's function is  given by
\beq
\Pi_{\alpha \beta}(\omega,{\bf q})=-i\int \frac{d^4x}{(2\pi)^4}
e^{-iq\cdot x}\theta(t)\lla[ {O}_{K \alpha}(t,{\bf x}),
 {O}_{K \beta}(0,{\bf 0})]_{-}\rra
=
\int_{-\infty}^{\infty} {d\omega'}
\frac{\rho_{\alpha \beta}(\omega',\mathbf{q})}
{\omega - \omega' + i\epsilon}.
\eeq
where 
$\rho_{\alpha \beta}(\omega',\mathbf{q})$ denotes the spectral density or function, and
\[
{O}_{K \alpha}(t, {\bf x})=\bar{q}_{_K}(t,{\bf x})\Gamma _{\alpha}
q_{_K}(t,{\bf x})-\lla\bar{q}_{_K}(t,{\bf x})\Gamma _{\alpha}
q_{_K}(t,{\bf x})\rra,
\]
with $q_K(t, {\bf x})=\exp(-i {K}t)q(0,{\bf x})\exp (i {K}t)$ 
being the Heisenberg operator.
Here $\Gamma _{\alpha
}(\Gamma _{\beta})$ denotes  a product of  Dirac and flavor matrices 
specifying the quantum numbers of the hadronic modes;
for $\alpha=\sigma \ (\pi)$, 
$\Gamma_{\alpha}=\mathbf{1}_4 \ (i\gamma_5)$, for instance.
Conversely, the spectral density is expressed as
$\rho_{\alpha \beta}(q)=-\frac{1}{\pi}\,{\rm Im}\,\Pi_{\alpha \beta}^{R}(q)$.

Now, noting that there are no vacuum expectation values 
of any operators in the Wigner phase,
the equalities (\ref{eq:vac-sigma-pi-ccr}) and (\ref{eq:vac-vector-avector-ccr})
tell us that the spectral functions in the sigma(vector) and pion(axial vector)
 channels coincide with each other in  the vacuum  in the Wigner phase (normal phase):
\beq
\rho_{\sigma}(\omega, {\bf q})=\rho_{\pi}(\omega, {\bf q}),\quad
\rho_{V^a}(\omega, {\bf q})=\rho_{A^a}(\omega, {\bf q}).
\label{eq:spectral-fn-Wigner}
\eeq

%\begin{figure}[thb]
%\centering
%\includegraphics[width=.4\textwidth]{blankfig}
%\includegraphics[width=.4\textwidth]{rho-a1-pi-delta}
%\caption{Screening masses of light hadrons.
%Positive parity and negative parity states tend to come close to each other.}
%\label{Maezawa:fig}
%\end{figure}

The same is true even at finite temperature, as is easily found. In fact,
putting $[\hat{\pi}^b(x), \hat{\sigma}(0)]=:O^b(x)$,
\beq
Z\times \lla [O^b(x), Q_5^a(x_0)]\rra
&=&{\rm Tr}\left(\e^{-\beta K} [O^b(x), Q_5^a(x_0)]\right)
=\sum_n\la n\vert\e^{-\beta K} [O^b(x), Q_5^a(x_0)]\vert n\ra
\nonumber \\
&=&\sum_n\la n\vert \e^{-\beta K}O^b(x)Q_5^a(x_0)\vert n\ra-
\sum_n\la n\vert\e^{-\beta K}  Q_5^a(x_0)O^b(x)\vert n\ra.\nonumber
\eeq
However, since $H_{QCD}$ commutes with $Q_5^a(x_0)$ ($\forall a$) 
as well as the number operator $N$, $\forall \vert n\ra$ 
can be chosen to be a common eigenstate of all these operators.
Recalling that $Q_5^a$ is a Hermitian operator, we can write as
$Q_5^a\vert n\ra=q^a_n\vert n\ra$ and $\la n\vert Q_5^a=q^a_n\la n\vert$ with
$q_n^a$ being a real number. Then, we have
$Z\times\lla [O^b(x), Q_5^a(x_0)]\rra =\sum_n\, \e^{-\beta K_n}q_n^a
\la n\vert O^b(x)\vert n\ra-
\sum_n\, \e^{-\beta K_n}  q_n^a \la n\vert O^b(x)\vert n\ra=0$.
Thus, with the use of  the Jacobi identity as before,
$
0= \lla [[\hat{\pi}^b(x), \hat{\sigma}(0)], Q_5^a(x_0)]\rra
=- \lla [[\hat{\sigma}(0),Q_5^a(x_0)], \hat{\pi}^b(x)]\rra
-\lla [[Q_5^a(x_0)], \hat{\pi}^b(x)],\hat{\sigma}(0)]\rra,
$
which leads to
\beq
\lla 0\rvert[\hat{\sigma}(x), \hat{\sigma}(0)\rvert 0\rra=
\lla 0\rvert[\hat{\pi}^a(x), \hat{\pi}^a(0)]\rvert 0\rra,
\label{eq:s-p-corr-T}
\eeq
%\label{eq:vac-sigma-pi-ccr-T}
on account of  (\ref{eq:CCR-for-SSB}).
A quite similar argument gives the equality
\beq
\lla 0\rvert[V_{\mu}^a(x), V_{\nu}^a(0)\rvert 0\rra
=\lla 0\rvert[A_{\mu}^a(x), A_{\nu}^a(0)]\rvert 0\rra.
\label{eq:V-A-corr-T}
\eeq
%\label{eq:vac-vector-avector-ccr-T}
Thus, we confirm that the equalities of the spectral functions 
(\ref{eq:spectral-fn-Wigner}) {also hold} at finite temperature in the Wigner phase.

%=-\la 0\rvert[[ \hat{\sigma}(0), Q_5^a], \hat{\pi}^b(x)]\rvert 0\ra-
%\la 0\rvert[[Q_5^a, \hat{\pi}^b(x)], \hat{\sigma}(0)]\rvert 0\ra
%= i\la 0\rvert[\hat{\pi}^a(0), \hat{\pi}^b(x)]\rvert 0\ra+
%i\delta_{ab}\la 0\rvert[\hat{\sigma}(x), \hat{\sigma}(0)\rvert 0\ra,\nonumber
%\eeq

%for example, $\Gamma _{\alpha}=\Gamma _{\beta}
%=i\gamma _5\lambda _{4\pm i5}$ for kaons K$^{\pm}$.
%The poles of the response function (or the determinant of the response
%functions  for sigma
%and $\eta $ mesons) give the dispersion relations $\omega _{\alpha }
%=\omega _{\alpha }({\bf q})$ of the mode.

% with $\Gamma_{\alpha}=1,\, i\gamma_5\tau^a/2,\, \gamma_{\mu}\tau^a/2$,
%and $\gamma_{\mu}\gamma_5\tau^a$, for instance.

\subsection{Pionic atoms and the chiral condensate at finite density; experimental evidence}
\label{section:pionic atom}

It is a remarkable fact that precision spectroscopy of deeply bound 
pionic atoms with a heavy nucleus 
such as Sn with a large number of isotopes 
{provides experimental evidence consistent with} 
a reduction of the chiral condensate in nuclear matter.
Here, a deeply bound state means that the bound pion is in atomic 1s or 2p states, and thus
the pion is actually affected by the strong interaction with the nucleus as well 
as the Coulombic attraction: Since the heavy nuclei are neutron-rich, 
the $s$-wave $\pi^{-}$ optical potential  is repulsive due 
to the Tomozawa-Weinberg term, which reads 
$U_{TW}=\frac{1}{2m_{\pi}}\frac{1}{4\pi}\left[1+\frac{m_{\pi}}{m_N}\right]b_1\delta \rho$
with $\rho:=\rho_n+\rho_p$ and $\delta \rho:=\rho_n-\rho_p$,
 and hence the pion can avoid the strong absorption and becomes sufficiently long-lived that the deeply bound states become detectable with a good signal-to-noise ratio.
 
Historically, 
the deeply-bound pionic atoms were discovered in (\cite{Yamazaki:1996zb}) using the (d, He$^3$) reaction on Pb$^{208}$ target, which confirmed
the theoretical prediction (\cite{Hirenzaki:1991us}) with remarkable precision.
It was Weise who first suggested that the observed `missing repulsion' may be
attributed to the possible restoration of chiral symmetry 
in the nuclear medium as given 
by (\ref{cond-rho}) (\cite{Weise:2001sg})\footnote{These proceedings 
are also a brief but excellent account of 
the chiral symmetry and its restoration in hot and dense nuclear medium.}:
The pion decay constant $f_{\pi}^*$  in the Tomozawa-Weinberg term 
may become density dependent, and so does $b_1^{\ast}(\rho)$. 
%$\pi-N$ $\sigma$ term $\sigma_{\pi N}$.
The suggestion was subsequently given a foundation based on the
in-medium chiral perturbation theory at two-loop order in 
(\cite{Kolomeitsev:2002gc}).   
A more explicit formula that relates the in-medium enhancement of $b_1^{\ast}(\rho)$
was given by combining the wave-function renormalization of the pion 
(\cite{Jido:2008bk}) as
\beq
\la \bar{q}q\ra(\rho)/\la \bar{q}q\ra(0)=
(b_1/b_1^{\ast}(\rho))^{1/2}(1-\gamma \rho/\rho_0)
\eeq
with $\gamma\simeq 0.184$. 
The subsequent development of the experiments and 
their theoretical interpretation in terms of a reduction of the chiral condensate
%chiral restoration 
in the nuclear medium until 2012 are well summarized in 
(\cite{Yamazaki:2012zza}); see also (\cite{Hayano:2008vn}).
The latest precise measurement of the pionic atoms for extracting 
a precise amount of the rate of the reduction of the chiral condensate
%chiral  symmetry restoration
 in the nuclei
is reported in (\cite{piAF:2022gvw}), where they first showed that the pions are
effectively probing the density at $\rho_e= 0.58\rho_0$ using 
a precise elastic proton-Sn scattering data (\cite{Terashima:2008zza}), and concluded that 
chiral condensate is reduced as 
$\la \bar{q}q\ra(\rho_e)/\la \bar{q}q\ra(0)=77\pm 2\%$
at this density, which implies that the ratio can be
\beq
\la \bar{q}q\ra(\rho_0)/\la \bar{q}q\ra(0) \simeq 60 \%
\eeq
at the normal nuclear density according to some 
theoretical estimates 
(\cite{Kaiser:2007nv,Jido:2008bk,Lacour:2010ci,Goda:2013bka,Friedman:2019zhc,Hubsch:2021nih}).

A caviat is, however,  that the reduction of the chiral condensate itself does not
necessarily imply a partial restoration of chiral symmetry, 
since any (partial)  degeneracy 
of hadronic modes in a  chiral partner is not examined and accordingly not seen
 in the relevant systems.

\subsection{Symmetry properties in hadronic correlators and spectra}

%As was noted in the Introduction, a change of the vacuum, in particular, owing to a
%phase transition with an order parameter, elementary excitations on top of the vacuum 
%may change their properties. Thus, we may expect that the possible (partial) restoration
%of the chiral symmetry may lead to some changes in hadrons that are excited states 
%on top of the QCD vacuum.
In this subsection,  we shall pick up some hadrons or hadronic modes 
which may show a spectral change and hopefully degeneracy in association with 
(partial) chiral restoration in hot and/or dense matter.
%xspertinent to chiral transition.

\subsubsection{Vector and Axial vector mesons}
\label{subsection:V-A mesons}

As was stated at the beginning of this section, the vector and axial-vector modes
in the Wigner phase should be degenerate in the sense that the spectral functions
of them coincide with each other in the $\omega$-$q$ plane,
which is guarateed by the equality of the correlation functions of the vector and
axial-vector currents as given 
%by Eq.~(\ref{eq:vac-vector-avector-ccr}) for the vacuum
by Eq.~(\ref{eq:V-A-corr-T}) for the hot and/or dense medium\footnote{
As mentioned before,  there exists a mass gap in reality with some $0.5$ GeV between the $\rho$
and $a_1$ mesons in the vacuum, and the gap can be attributed to the spontaneous
breaking of chiral symmetry in the QCD vacuum as advocated by 
the Weinberg's sum rules that involve the pion decay constant $f_{\pi}$.}.
%Thus , 
It is, however, quite difficult to determine
the $a_1$ spectral function in the hot and dense hadronic 
medium in contrast to the $\rho$ meson,
though it is best if one could 
test the possible spectral degeneracy of the two
mesons by experiment.

Furthermore, as we have emphasized several times so far, the observable extracted 
from the experiment try to see the hadron properties in the medium 
is a spectral function as a function of the energy and momentum, 
which may have  or may not have a peak structure reminiscent of the 
hadron in the vacuum. Moreover, even when the spectral function
has a peak structure, it cannot be described solely by a 
mass $m^{\ast}$ (mass shift) at the peak position but 
it should be necessary also to use  a width $\Gamma^{\ast}$ in the medium to describe the 
peak structure, and $\Gamma^{\ast}$ can become much larger than that of the 
vacuum value (peak broadening). 
When the spectral function has no clear peak structure,
it is rather absurd to tell the whole spectral function in terms of 
a 'mass'.

With this warning in mind, we describe some historical developments in this
field. Indeed, it was conjectured by a heuristic scaling argument 
(\cite{Brown:1991kk}) that the mass of the non-strange vector mesons decreases 
along with the chiral restoration\footnote{See also \cite{Pisarski:1981mq} for an earlier argument for a decrease of the $\rho$-meson mass.}.
This conjecture was augmented by Georgi's vector manifestation
(\cite{Georgi:1989xy,Harada:2000kb}), in which the 
(longitudinal) $\rho$ meson is identified as the 
chiral partner of the pion, and they were considered to become  massless as
the chiral symmetry is restored. Moreover, the decrease of the $\rho$
meson mass as well as the a$_1$ was  shown in QCD sum rule 
approach (\cite{Hatsuda:1991ez,Hatsuda:1992bv}). 
For fairness, we must also add that 
a microscopic calculation with the use of a generalized NJL model
 (\cite{Bernard:1988db}) had shown that the $\rho$ meson hardly 
changes its mass against temperature increase, whereas
the a$_1$ does show a decrease in its mass as temperature is increased.
These results motivated studies aimed at deducing the spectral function 
in the {\em vector channel} in hot and dense matter, 
as if forgetting about the axial-vector channel for the moment.

\paragraph{\textbf{Lepton-pair production in heavy-ion collisions 
and vector spectral function in hot QCD medium}}

Precise measurements of lepton-pair production from 
the ultra-relativistic heavy-ion collisions could 
provide us with a good measure for 
 deducing the spectral functions in the vector channel
in hot and dense matter.; see nice reviews and article 
(\cite{Rapp:2009yu,Hayano:2008vn,CERES:2005uih})
 for an instructive  account of the physics background of  the 
intimate relevance of the lepton-pair production 
to the spectral properties in the vector channel 
in QCD medium.  

A S-Au collision experiment (NA45) (\cite{CERES:1995vll,Agakichiev:1998kip})
 at CERN with an innovative detector, CERES showed
a substantial enhancement of the di-electron pairs per charged particles 
in the mass region $0.2$-$1.5$ GeV with a factor $5$ 
in comparison with those in $p$-Be and $p$-Au collisions, 
which are reproduced within errors by Dalitz and direct decays 
of neutral mesons $\pi^0$, $\eta$, and $\eta'$ as known from $p$-$p$
collisions; see also (\cite{CERES:2005uih}) for a summary with new data 
on the Pb-Au collisions at 158A GeV and their detailed analyses.
Subsequent measurements of the lepton pairs 
from the relativistic heavy-ion collisions  
include 
(\cite{HELIOS3:1998xeb,NA60:2006ymb,NA60:2007cjy,NA60:2007lzy,CERES:2006wcq,NA60:2006ymb,STAR:2013pwb,PHENIX:2015vek}).
For example, the NA60 collaboration (\cite{NA60:2006ymb})
succeeded in extracting the $\rho$ spectral function from a 
precision measurement of low-mass muon pairs
in 158A GeV In-In collisions at the CERN SPS.

It seems that the prevailing view of these experiments
is that the $\rho$ spectral function in the hot and/or dense medium
is best characterized by a strong broadening, which shows no
 clear mass shift of the meson. 
Moreover, the spectral shape extracted from the experiments
are well reproduced by hadron many-body 
calculations (\cite{Rapp:1999us,Rapp:2000pe,Rapp:2009yu}) without
an explicit recourse to chiral restoration.

Nevertheless, the complicated many-body calculations might implicitly
simulate the dynamics in the quark many-body system owing to 
the quark-hadron duality or continuity (\cite{Rapp:1999us}) provided 
that the quark-hadron duality would get realized at a much lower energy 
region at finite temperature.
Then it is found interesting that QCD sum rules at finite temperature
(\cite{Hatsuda:1992bv}) shows that the threshold energy to 
the quark continuum rapidly decreases at finite temperature
\footnote{See also a recent review (\cite{Shuryak:2026pqt}) 
on the hadron-parton bridge}.

\begin{figure}[thb]
\centering
\includegraphics[width=.65\textwidth]{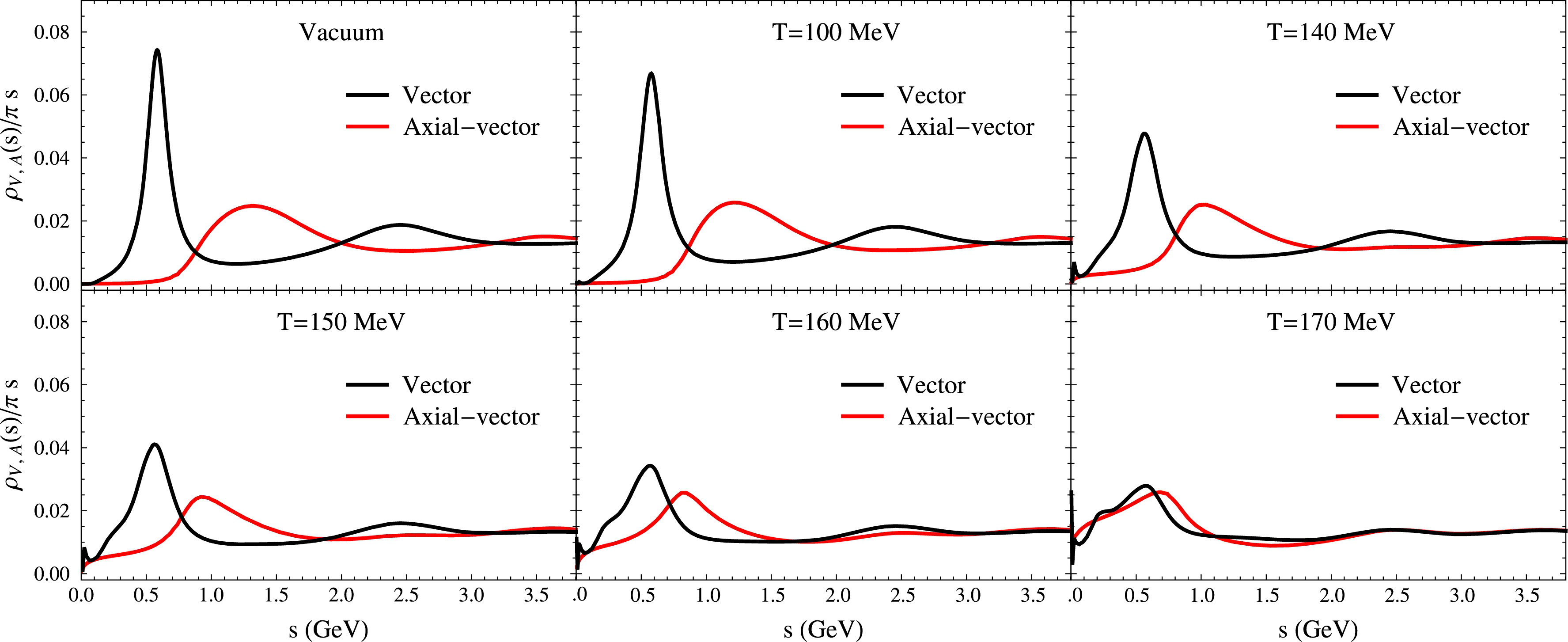}
\caption{Temperature dependence of the $\rho$ (black curve) and $a_1$
meson spectral functions: Taken from Fig.~3 of (\cite{Hohler:2013eba}).
The position of the $\rho$ meson peak hardly changes, but with increasing width
when the temperature is raised toward the pseudo-critical temperature $T_c=170$ MeV 
of the chiral transition.
On the other hand, the spectral function in  the axial-vector channel 
shows a significant decrease in the peak position as well as the width broadening, and 
one sees a merging of the $\rho$-$a_1$ spectral function at  $T_c$.}
\label{A1-rho-Gohler-Rapp:fig}
\end{figure}

We emphasize again that the analysis of the spectral
functions solely in the vector channel is not perfectly 
conclusive to tell anything about the possible chiral restoration 
in the hot and/or dense QCD medium, and it would be highly desirable to get some experimental hints on the spectral function 
in the {\em axial-vector channel} in hot and/or dense medium.
A rather heuristic theoretical argument
 (\cite{Hohler:2013eba}) based on the various sum rules, such as 
the ones given in (\cite{Kapusta:1993hq,Hatsuda:1992bv} together  with
the  lattice QCD data and the hadron gas model
provided a phenomenological reconstruction on the spectral
function in the axial-vector channel that the a$_1$ meson would 
show a mass reduction toward that of the $\rho$ accompanied by
a large spreading of the width; see Fig.~\ref{A1-rho-Gohler-Rapp:fig}. 
A similar result is 
also obtained as a 
possible scenario of the chiral symmetry restoration on the
basis of the generalized hidden local symmetry approach (\cite{Harada:2008hj}). 
It is also remarkable that the scenario that 
no mass shift of the $\rho$ but the mass reduction of a$_1$ 
had been suggested in the Nambu-Jona-Lasinio model 
(\cite{Bernard:1988db}), as mentioned before.
Then one may be inclined to speculate that the chiral quark picture
(\cite{Manohar:1983md})
as realized by GNJL model (\ref{eq:GNJL}) with the vector interaction
might reflect some reality of QCD medium 
in view of the suggestion of the quark-hadron 
duality made in (\cite{Rapp:1999us}).

\paragraph{\textbf{Vector spectral functions in cold matter 
as probed by elementary reactions on nuclei}}
 
The E325 collaboration successfully measured the invariant-mass spectra of the 
 e$^+$e$^{-}$ pairs produced by 12 GeV proton-C and p-Cu reactions, and 
concluded that the both $\rho$ and $\omega$ mesons show some 9 \% reduction
in their mass but with no clear broadening of the width (\cite{Naruki:2005kd}).
A similar result was also reported on the $\phi$ meson by E325 experiment
(\cite{KEK-PS-E325:2005wbm}).  However, a photon-induced reaction
on the C and Cu targets to produce the $\rho$ showed only a small mass shift 
consistent with zero, and the broadening of the width is in accord
with standard nuclear many-body effects such as 
collisional broadening and Fermi motions.

A summary is given in Table.1 of 
(\cite{Metag:2010zza}) on 
the experimental results on medium modifications of vector mesons;
one sees (\cite{Hayano:2008vn,Metag:2010zza}) that 
almost all experiments find a softening of the vector meson spectral
functions, 
and the earlier claims of 
mass shifts (\cite{Naruki:2005kd,KEK-PS-E325:2005wbm}), unfortunately, 
 have not been confirmed.

%%%%%17:15%%%%%%%
\subsubsection{Possible softening of the spectral function of the 
$\sigma$ meson in the hot and/or dense medium}
\label{subsection:sigma mesons}

As was discussed in \S\ref{subsection:Wigner-NG phases},
the equality of the commutation relations (\ref{eq:vac-sigma-pi-ccr}) 
imply that the spectral functions with quantum numbers of the $\sigma$ and
the pion coincides in the Wigner phase where the chiral symmetry is restored,
in the chiral limit; in reality, the degeneracy would be
slightly lifted with a small current quark mass causing an explicit breaking of chiral symmetry.

Even if the chiral restoration is a crossover transition close to a second order 
in hot and/or dense QCD matter (\cite{Aoki:2006we, Ding:2015ona}), 
there is a window in the $(T, \mu$) plane where the 
fluctuation of the order parameter of the chiral transition, 
i.e., $\lla (\bar{q}q-\lla\bar{q}q\rra)^2\rra$ is getting large, implying
the lowering of the mass of the corresponding meson excitation, which has 
the quantum number of $J^{PC}~0^{++}$, the scalar meson
that one is inclined to call the $\sigma$ meson.
Actually, as stated in \S\ref{section:NJL-2f}, the nature of the
low-lying scalar meson is intricate, and its relation to the quantum fluctuations
of the order parameter of the chiral transition is obscure;
hence, it is quite a challenge to clarify how the crossover transition 
of the restoration of the chiral symmetry manifests itself in the spectral functions
in the scalar channel with the same quantum number of the $\sigma$.

In 1996, using the $\pi^{+}A\,\to\, \pi^{+}\pi^{-}X$ reaction, the CHAOS collaboration 
(\cite{CHAOS:1996nql,CHAOS:2000fjr}) 
reported a downward shift of the spectral functions
in the $\sigma$ channel. Similar shapes of the spectral function with a peak near
the $\pi\pi$ threshold were obtained in several chiral models with a drop of the 
$\sigma$ meson mass incorporated (\cite{Hatsuda:1998vb,Aouissat:1999ss,Davesne:1999qj}).
However, there were caveats with the CHAOS experiment\footnote{
see \S IV of (\cite{Hayano:2008vn}) for
a detailed account of this subject, both in the 
theoretical and experimental aspects of this issue.}: Firstly, 
the incident particle is the pion that suffers from large absorption by 
the nuclear medium, and hence it should only probe the surface region
of the nuclei. Secondly, the acceptance of the CHAOS is as small as
about 10\,\% of the whole solid angle.

 Indeed, a similar experiment by the Crystal Ball collaboration
(\cite{CrystalBall:2000tgu})  using the pion beam
 But a much larger acceptance showed no sharp enhancement
 near the threshold as shown in CHAOS experiment.
 A photon incident beam that is less absorptive and can probe more inside
of nuclei was made by the TAPS collaboration (\cite{Messchendorp:2002au,Bloch:2007ka});
the $\pi^0\pi^0$ as well as  $\pi^0\pi^{\pm}$ spectra were measured for
various nuclear targets: Although  only the former shows a light softening, 
they can be described by the hadron dynamics described by a chiral Lagrangian without 
recourse to the restoration of the chiral symmetry or the 
Boltzmann-Uehling-Uhkenbeck transport dynamics (\cite{Buss:2006vh}) 
which include the charge-exchange $\pi$-N scattering in nuclei.
 
Since the low-energy scalar resonance $f_0(500)$ 
may be well described solely by the multiple-pion dynamics given by
the nonlinear realization of chiral symmetry  
(\cite{Pelaez:2015qba}),
 some downward shift of the spectral function near the 2 $\pi$ threshold 
 seen in the 
scalar channel can be just a result of 
hadron dynamics involving the pion and the nuclear medium, without
a recourse to a possible change of the QCD vacuum.
Nevertheless, the intricate nature of the spectral properties of the 
hadronic excitations in the $\sigma$ channel may 
manifest itself in a specific pattern in the change of the spectral 
function when the chiral symmetry were to be largely restored (\cite{Hyodo:2010jp}).

% At finite temperature and density, the instanton
%density is known to gets smaller (\cite{Gross:1980br,Schafer:1996wv}),
% and hence physical manifestation of the U(1)$_A$
%breaking may tends {\em effectively restored}
%(\cite{Pisarski:1983ms,Aoki:2021qws}).

\subsubsection{%$\eta'$ in hot and dense medium and related issue:
Restoration of chiral symmetry versus effective restoration of U(1)$_A$ symmetry}
\label{subsection:etaprime-nuclei}

The fact that the large mass of $\eta'(958)$ and 
the mixing angles are a reflection of the axial anomaly and the QCD
vacuum is the $\theta$-vacuum due to instantons implies that
those quantities  may change in
hot and dense medium (\cite{Pisarski:1983ms,Kunihiro:1989my,Hatsuda:1994pi}) 
where the instanton density is known to become smaller
(\cite{Gross:1980br,Schafer:1996wv,Aoki:2021qws}).

At high temperature and/or density, observable manifestations of U(1)$_A$
breaking may be suppressed (\cite{Gross:1980br,Schafer:1996wv,Aoki:2021qws}),
although the anomalous operator identity itself remains valid.  This is
commonly called {\em effective restoration} of U(1)$_A$ symmetry.  Such a
statement concerns U(1)$_A$-sensitive correlation functions and should not
be inferred from an $\eta'$ mass shift alone.  Indeed, the model result
Eq.~(\ref{eq:eta-0-mass-chiral-limit}) shows that the singlet mass can
decrease simply because the chiral order parameter decreases, even if the
anomaly coupling is kept fixed.

 Although restoration of $SU(N_f)_L\times SU(N_f)_R$ and effective restoration of $U(1)_A$ 
are logically distinct, 
one of the most essential issues is the role of {\em instantons} in the breaking 
not only of the U(1)$_A$ symmetry  but also of
the  $SU(N_f)_L\times SU(N_f)_R$ chiral symmetry
 (\cite{Callan:1977gz,Gross:1980br,Diakonov:1984vw,Aoki:2021qws,Pisarski:2024esv}) 
 in the hot and/or dense medium as
well as in the vacuum.
%they may be related dynamically near the chiral transition,
%although .
 For instance, it is shown (\cite{Diakonov:1984vw}) 
that the vacuum chiral condensate is given in terms of the number
  density $N/V$ of instantons, the average size $\bar{R}$ of the instantons and
the average distance of instantons as $\la\bar{q}q\ra\propto (N/V)^{3/4}\bar{R}/\bar{\rho}$.
{Both symmetries may be related dynamically} 
around the crossover region of chiral restoration through 
{instantons} (\cite{Aoki:2006we, Ding:2015ona}).
%although .

\subsubsection*{$\eta'$-mesic nuclei and in-medium $\eta'$ spectroscopy}

The possibility of an in-medium modification of the $\eta'$ has motivated
searches for $\eta'$--nucleus bound states and direct measurements of its
spectral properties.  A proposal was made to produce $\eta'$-mesic nuclei
with the $(p,d)$ reaction using missing-mass spectroscopy
(\cite{Nagahiro:2012aq}).  Recent $(p,d)$ data have been analyzed in terms
of an attractive real optical potential of order $60$ MeV
(\cite{e-PRiME:2025fer}); see also Ref.~(\cite{Friedman:2025avs}) for a
theoretical interpretation.

More recently, the LEPS2/BGOegg Collaboration reported a direct measurement
of the $\eta'(958)$ invariant-mass spectrum in photoproduction on carbon
through the $\eta'\to\gamma\gamma$ decay channel
(\cite{LEPS2BGOegg:2026xxj}).  For low-momentum events
($P_{\gamma\gamma}<1$ GeV/$c$), an enhancement was observed in the lower
tail of the $\eta'$ peak with a statistical significance above $3.7\sigma$.
A spectral-function fit indicated an in-medium mass reduction of
$57.5^{+5.7}_{-27.8}$ MeV/$c^2$.  
%This is the first evidence for an
%in-medium spectral modification of the $\eta'$ obtained from a direct
%invariant-mass measurement. 
 As emphasized above, however, such a mass
shift by itself does not establish either chiral restoration or effective
restoration of U(1)$_A$ symmetry.

\subsubsection{Possible observables involving baryons for restoration of 
chiral symmetry}
\label{section:baryon-neutron-star}

When the baryon density $\rho$ {and/or} temperature $T$ increase, 
and thereby 
%the order parameter $\sigma_0$ of 
the chiral symmetry is being restored,
{what would be expected to occur in baryonic matter and in baryons embedded in it?}
A clue for this may be provided by  the parity-doublet model of baryons discussed 
in \S\ref{section:parity-doublet baryons}, in which 
%a restoration of the chiral symmetry gives rise to a decrease both in the masses of 
$N$ and $N^*(1535)$ {would tend to become degenerate} 
{with both masses decreasing, but with a more pronounced decrease for the $N^*(1535)$ than for the nucleon.}
%and eventually the two tend to degenerate asymptotically.
{Matter in which substantial chiral restoration may occur could be realized in}
 the core of neutron stars, be created 
in the intermediate stage of relativistic-heavy-ion collisions,
and might be realized in (cold) heavy nuclei.

\subsubsection*{Parity-doubling in baryon sector  and chiral restoration in hot and/or dense QCD matter}

As is nicely summarized in (\cite{Minamikawa:2023eky}), 
the past decade has witnessed dramatic progress in the measurements of 
the mass-radius ($M_{\star}$-$R_{\star}$) relations of neutron stars (NS) 
(\cite{Miller:2019cac,Riley:2021pdl}) as well as the 
observation of massive NS heavier than 2$M_{\odot}$ 
(\cite{NANOGrav:2019jur}), which have provided us with a constraint on 
 QCD equation of state (EoS), i.e.,
 the energy density as a function of the density and temperature $E(\rho, T)$).

Except for the pioneering work  (\cite{Hatsuda:1988mv}), 
only relatively recently has the construction of QCD-motivated EoS become an active subject
 of the isospin-symmetric as well as 
asymmetric nuclear/quark matter: 
For instance, 
in (\cite{Mukherjee:2017jzi,Kong:2023nue,Eser:2023oii,Minamikawa:2023eky,Eser:2024xil}),
 the parity doubling 
in the baryon and quark sectors as well 
as other ingredients, such as the axial anomaly
 are taken into account for constructing the EOS.

It is noteworthy in particular that the EoS in the 
Quark-Hadron Chiral Parity Doublet Model (\cite{Mukherjee:2017jzi})
successfully reproduces  the temperature dependence
of $M_N^{\ast}$ as well as $M_N$
  obtained in a lattice simulation (\cite{PhysRevD.92.014503}).
As for the relevance to neutron-star phenomenology, the EoS given in (\cite{Kong:2023nue})
not only gives reasonable predictions for isospin-symmetric as well 
 as neutron matter, but also 
satisfies important criteria accommodating a maximum mass larger 
than 2.19$M_{\odot}$ and the radius $R=12.9$\, km of a neutron star with the mass
 $1.4M_{\odot}$ , as noted in (\cite{Brodie:2025nww}).   

The parity-doublet model is also applied to other important phenomena
of neutron stars,
% It is one of the important subjects to clarify 
such as the {\em cooling mechanism} of
neutron stars, which are created by a supernova and hence born as hot matter.
In the original parity-doublet model discussed in 
\S\ref{section:parity-doublet baryons},
 both the axial charges and the couplings to the pion decrease
as chiral symmetry is restored, which may affect the cooling rate of neutron stars
(\cite{Detar:1988kn}). 
Moreover, the chiral restoration as given by the EoS that is based on the parity-doublet models
(\cite{Mukherjee:2017jzi,Kong:2023nue,Eser:2023oii,Minamikawa:2023eky,Eser:2024xil})
can allow the appearance of the Fermi seas of negative-parity nucleons so as to
make a new type of direct Urca process possible, and thereby  cause a rapid cooling of
neutron stars (\cite{Brodie:2025nww,Negreiros:2026ode}). Thus, detailed analyses of 
the cooling curves of neutron stars 
may provide constraints on parity-doubling scenarios associated with chiral restoration in dense QCD matter.
%decrease of 
%$M_{N^{\ast}}$ associated to the chiral symmetry restoration would 
%realize matter at some baryon density region in chemical equilibrium
% $\mu_{n_{+}}=\mu_{n_{-}}$ and $\mu_{p_{+}}=\mu_{p_{-}}$,
%involving the positive- and negative-parity neutrons and protons $(n_+, n_{-})$ and $(p_{;}, p_{-})$, 
%then a novel direct Urca processes become possible, and hence a rapid cooling
%of neutron stars. 

 In relativistic heavy-ion collisions,
an enhancement of  the spectral function along with a peak shift of $N^{\ast}(1535)$
might be observed (\cite{Zhang:2015fda}), provided that it is 
experimentally possible to identify the negative-parity baryons coming from 
the hot and dense matter created by relativistic heavy-ion collisions at all. 
If N$^{\ast}(1535)$ is produced in a cold heavy nucleus, the observed life 
might be elongated due to the decrease in the phase-space volume 
owing to the mass shift of the N$^{\ast}(1535)$ (\cite{Suenaga:2017wbb}); see
also (\cite{Olbrich:2017fsd}), which includes a discussion on the glueball search in 
{the future antiProton ANnihilations at DArmstadt (PANDA) experiment}
at FAIR (\cite{PANDA:2009yku,PANDA:2021ozp}).

\section{Brief summary and concluding remarks}

%The key qunatities characterising the symmetry properties of the QCD vacuum
% is the quark chiral condensate,
%which is directly related to the pion decay constant $f_{\pi}$.
One of the most important messages of this article is 
that the pion is a pseudoscalar particle with a rather small mass, a fact that plays quite a significant role in nuclear physics, and originates from the 
intricate nature of the QCD vacuum as well as the symmetry properties of QCD:
The (approximate) chiral symmetry possessed by QCD is spontaneously broken in the QCD vacuum.

Precision spectroscopy of deeply bound pionic atoms provides quantitative information on the in-medium pion decay constant and, through chiral low-energy relations, evidence for a reduction of the chiral order parameter in nuclei.
This result may be regarded as one of the clearest experimental indications consistent with a reduction of the chiral order parameter in nuclear matter.
%it connects directly to the in-medium pion decay constant and condensate.

{Ongoing experiments on $\eta'$--nucleus systems} 
might also provide information on the in-medium modification of anomaly-related and chiral-symmetry-breaking effects.
However, one must be cautious about the interpretation of the 
extracted value of the chiral condensate, because
the Hellmann--Feynman theorem tells us that it may merely represent 
an averaged value of the nucleon matrix element multiplied by the nuclear density and 
the unchanged vacuum condensate in QCD.
 
The most direct signature of chiral restoration would be the degeneracy of spectral functions in channels related by chiral transformations.
However, no compelling experimental evidence for such spectral degeneracy is presently available:
Although tremendous efforts have been made to extract the spectral function in the vector channel,
in particular, the $\rho$ meson channel, it remains an experimental challenge to 
extract the in-medium spectral functions in the axial-vector channel. 
%Such modes togive a hint of the possible retoration of
%the chiral symmetry are , chiral partners of hadrons that are interconnected by the chiral
%transformation. Historically, the spectral change of the vector mesons have been
%focused, and it is now almost established that the $\rho$ meson spectral function 
%only shows a brodening with a slight, if any, 
% change of th peak position. However, the compelling evidence 
%of the restoration of the chiral symmetry in this case should be 
%given by the spectral degeneracy of the $\rho$-$a_1$ mesons.
%It is, unfortunately, still  quite a challenge to measure the in-medium 
%spectral change in the $a_1$ meson channel, although intersting and 
%suggesttive theoretical works have been done  on this subject.

Such a difficulty is also present in the scalar channel: 
First of all, the physical low-energy scalar resonance $f_0(500)$, identified through sophisticated analyses of $\pi$-$\pi$ scattering data, cannot in general be identified with a simple $q\bar q$ chiral partner of the pion.
The scalar excitation associated with fluctuations of the chiral order parameter need not coincide with the physical $f_0(500)$ and may mix strongly with other scalar configurations in QCD, including tetraquark components, meson--meson correlations, and glueball degrees of freedom.   

As for the baryon sector,  the recent progress in the measurements of 
the mass-radius ($M_{\star}$-$R_{\star}$) relations of neutron stars (NS)  
 as well as the 
observation of NS's heavier than 2$M_{\odot}$ may provide us with
 some direct or indirect information on the properties of the dense QCD matter 
that may undergo the phase transition of chiral restoration, and also the 
way of the chiral restoration in the baryon sector. 

Thus one sees that chiral symmetry provides a bridge between the microscopic structure 
of the QCD vacuum, the macroscopic properties of nuclear and neutron-star matter, 
and experimental searches for in-medium modifications of hadrons.

%\end{Large}
\begin{ack}[Acknowledgments]

The author thanks Kenji Fukushima for inviting him to contribute to this
project and for his suggestions on making the manuscript more accessible
to undergraduate students.
He is also grateful to Yuya Tanizaki for his invaluable comments on
the 't Hooft anomaly matching condition and related topics, including
the fundamental significance of the $\ln\det$ terms in chiral effective
theories incorporating the axial anomaly of QCD, as well as for carefully
reading the first draft.
He is also indebted to Yoshimasa Hidaka for his helpful comments and
for pointing out typos in the original manuscript.
He thanks Masayasu Harada for his valuable comments on parity-doublet
models of baryons.
Figures\,1, 4, and 5 were created with the help of ChatGPT.
This work was supported in part by JSPS KAKENHI (No. 24K07049).
\end{ack}

\seealso{%article title article title}
Effective Restoration of the Axial U(1) Symmetry; 
Generalized {Nambu--Goldstone} Theorem; 
Lattice QCD at Finite Temperature and Density;
Low-energy QCD Effective Models;
Magnetic Catalysis and Inverse Magnetic Catalysis in QCD;
Quark-Gluon Plasma (QGP);
Relativistic Heavy-ion Collision and QCD Phase Diagram}

\bibliographystyle{Harvard}
%\bibliographystyle{plain}
%\addbibresource{reference.bib}
\bibliography{reference}

@article{Kunihiro:1990ts,
    author = "Kunihiro, T. and Hatsuda, T.",
    title = "{{OZI} Violation in Nucleon, $\Delta$ and Hyperons}",
    reportNumber = "SUNY-NTG-90-5",
    doi = "10.1016/0370-2693(90)90435-9",
    journal = "Phys. Lett. B",
    volume = "240",
    pages = "209--214",
    year = "1990"
}

@article{Hatsuda:1990uw,
    author = "Hatsuda, T. and Kunihiro, T.",
    title = "{Strange quark, heavy quarks, and gluon contents of light hadrons}",
    reportNumber = "CERN-TH-5838-90, TPR-90-46",
    doi = "10.1016/0550-3213(92)90212-T",
    journal = "Nucl. Phys. B",
    volume = "387",
    pages = "715--742",
    year = "1992"
}

@article{Gao:2026scv,
    author = "Gao, Bikai",
    title = "{Chiral symmetry restoration and hyperon suppression in neutron stars}",
    eprint = "2602.12503",
    archivePrefix = "arXiv",
    primaryClass = "nucl-th",
    doi = "10.1103/vb5r-vdm7",
    journal = "Phys. Rev. D",
    volume = "113",
    number = "8",
    pages = "083012",
    year = "2026"
}

@book{Preston:1975,
  title={Structure of the Nucleus},
  author={M. A. Preston and R. K. Bhaduri},
  year={1975},
  publisher={Addison-Wesley Publishing Co.}
}

@article{Bethe:1971xm,
    author = "Bethe, H. A.",
    title = "{Theory of nuclear matter}",
    doi = "10.1146/annurev.ns.21.120171.000521",
    journal = "Ann. Rev. Nucl. Part. Sci.",
    volume = "21",
    pages = "93--244",
    year = "1971"
}

@article{Callan:1977gz,
    author = "Callan, Jr., Curtis G. and Dashen, Roger F. and Gross, David J.",
    editor = "Shifman, Mikhail A.",
    title = "{Toward a Theory of the Strong Interactions}",
    reportNumber = "COO-2220-115",
    doi = "10.1103/PhysRevD.17.2717",
    journal = "Phys. Rev. D",
    volume = "17",
    pages = "2717",
    year = "1978"
}

@article{Gross:1980br,
    author = "Gross, David J. and Pisarski, Robert D. and Yaffe, Laurence G.",
    title = "{QCD and Instantons at Finite Temperature}",
    reportNumber = "PRINT-80-0538 (PRINCETON)",
    doi = "10.1103/RevModPhys.53.43",
    journal = "Rev. Mod. Phys.",
    volume = "53",
    pages = "43",
    year = "1981"
}

@article{Schechter:1971qa,
    author = "Schechter, J. and Ueda, Y.",
    title = "{Symmetry breaking and spin-zero mass spectrum}",
    doi = "10.1103/PhysRevD.3.168",
    journal = "Phys. Rev. D",
    volume = "3",
    pages = "168--176",
    year = "1971"
}

@article{Pisarski:1983ms,
    author = "Pisarski, Robert D. and Wilczek, Frank",
    title = "{Remarks on the Chiral Phase Transition in Chromodynamics}",
    reportNumber = "NSF-ITP-83-152",
    doi = "10.1103/PhysRevD.29.338",
    journal = "Phys. Rev. D",
    volume = "29",
    pages = "338--341",
    year = "1984"
}

@article{Diakonov:1984vw,
    author = "Diakonov, Dmitri and Petrov, V. Yu.",
    title = "{CHIRAL CONDENSATE IN THE INSTANTON VACUUM}",
    doi = "10.1016/0370-2693(84)90132-1",
    journal = "Phys. Lett. B",
    volume = "147",
    pages = "351--356",
    year = "1984"
}

@book{povh2008particles,
  title={Particles and nuclei: an introduction to the physical concepts},
  author={Povh, Bogdan and Rith, Klaus and Scholz, Christoph and Zetsche, Frank},
  year={2008},
  publisher={Springer}
}

@book{Peskin:1995ev,
    author = "Peskin, Michael E. and Schroeder, Daniel V.",
    title = "{An Introduction to quantum field theory}",
    isbn = "978-0-201-50397-5",
    publisher = "Addison-Wesley",
    address = "Reading, USA",
    year = "1995"
}

@book{Sakurai:1967,
    author = "Sakurai, J.J.",
    title = "{Advanced Quantum Mechanics}",
     isbn = "978-8177589160",
    publisher = "Addison-Wesley",
    address = "Reading, USA",
    year = "1967"
}

@book{cheng1994gauge,
  title={Gauge Theory of Elementary Particle Physics},
  author={Cheng, Ta-Pei and Li, Ling-Fong},
  year={1994},
  publisher={Oxford University Press}
}

@book{georgi2009weak,
  title={Weak interactions and modern particle theory, revised and updated},
  author={Georgi, Howard},
   publisher={Mineola, NY: Dover},
  year={2009}
}

@article{ParticleDataGroup:2024cfk,
    author = "{Navas et al.}",
    collaboration = "Particle Data Group",
    title = "{Review of particle physics}",
    doi = "10.1103/PhysRevD.110.030001",
    journal = "Phys. Rev. D",
    volume = "110",
    number = "3",
    pages = "030001",
    year = "2024"
}

@article{Nambu:1961tp,
    author = "Nambu, Yoichiro and Jona-Lasinio, G.",
    editor = "Eguchi, T.",
    title = "{Dynamical Model of Elementary Particles Based on an Analogy with Superconductivity. 1.}",
    doi = "10.1103/PhysRev.122.345",
    journal = "Phys. Rev.",
    volume = "122",
    pages = "345--358",
    year = "1961"
}

@article{Goldstone:1961eq,
    author = "Goldstone, J.",
    title = "{Field Theories with Superconductor Solutions}",
    doi = "10.1007/BF02812722",
    journal = "Nuovo Cim.",
    volume = "19",
    pages = "154--164",
    year = "1961"
}

@article{Goldstone:1962es,
    author = "Goldstone, Jeffrey and Salam, Abdus and Weinberg, Steven",
    title = "{Broken Symmetries}",
    doi = "10.1103/PhysRev.127.965",
    journal = "Phys. Rev.",
    volume = "127",
    pages = "965--970",
    year = "1962"
}

@book{tinkham2004introduction,
  title={Introduction to superconductivity},
  author={Tinkham, Michael},
  year={2004},
  publisher={Courier Corporation}
}

@book{schrieffer2018theory,
  title={Theory of superconductivity},
  author={Schrieffer, J Robert},
  year={2018},
  publisher={CRC press}
}

@book{sakurai1969currents,
  title={Currents and mesons},
  author={Sakurai, Jun John},
  year={1969},
  publisher={University of Chicago press}
}

@article{Klevansky:1992qe,
    author = "Klevansky, S. P.",
    title = "{The Nambu-Jona-Lasinio model of quantum chromodynamics}",
    doi = "10.1103/RevModPhys.64.649",
    journal = "Rev. Mod. Phys.",
    volume = "64",
    pages = "649--708",
    year = "1992"
}

@article{Kunihiro:1983ej,
    author = "Kunihiro, Teiji and Hatsuda, Tetsuo",
    title = "{A Selfconsistent Mean Field Approach to the Dynamical Symmetry Breaking: The Effective Potential of the {Nambu-Jona-Lasinio} Model}",
    reportNumber = "KUNS-705",
    doi = "10.1143/PTP.71.1332",
    journal = "Prog. Theor. Phys.",
    volume = "71",
    pages = "1332",
    year = "1984"
}

@article{Hatsuda:1985ey,
    author = "Hatsuda, T. and Kunihiro, T.",
    title = "{Critical Phenomena Associated with Chiral Symmetry Breaking and Restoration in QCD}",
    reportNumber = "KUNS-778",
    doi = "10.1143/PTP.74.765",
    journal = "Prog. Theor. Phys.",
    volume = "74",
    pages = "765",
    year = "1985"
}

@article{Kunihiro:1989my,
    author = "Kunihiro, Teiji",
    title = "{Effects of the UA (1) anomaly on the quark condensates and meson properties at finite temperature}",
    doi = "10.1016/0370-2693(89)90405-X",
    journal = "Phys. Lett. B",
    volume = "219",
    pages = "363--368",
    year = "1989",
    note = "[Erratum: Phys.Lett.B 245, 687 (1990)]"
}

@book{Fujikawa:2004cx,
    author = "Fujikawa, K. and Suzuki, H.",
    title = "{Path integrals and quantum anomalies}",
    doi = "10.1093/acprof:oso/9780198529132.001.0001",
    year = "2004"
}

@article{Weinberg:1967kj,
    author = "Weinberg, Steven",
    title = "{Precise relations between the spectra of vector and axial vector mesons}",
    doi = "10.1103/PhysRevLett.18.507",
    journal = "Phys. Rev. Lett.",
    volume = "18",
    pages = "507--509",
    year = "1967"
}

@article{Kawarabayashi:1966kd,
    author = "Kawarabayashi, Ken and Suzuki, Mahiko",
    title = "{Partially conserved axial vector current and the decays of vector mesons}",
    doi = "10.1103/PhysRevLett.16.255",
    journal = "Phys. Rev. Lett.",
    volume = "16",
    pages = "255",
    year = "1966"
}

@article{riazuddin1966algebra,
  title={Algebra of current components and decay widths of $\rho$ and K* mesons},
  author={Riazuddin, Fayyazuddin},
  journal={Physical Review},
  volume={147},
  number={4},
  pages={1071--1073},
  year={1966},
  publisher={American Physical Society (APS)}
}

@article{Das:1967zze,
    author = "Das, T. and Mathur, V. S. and Okubo, S.",
    title = "{SYMMETRY, SUPERCONVERGENCE AND SUM RULES FOR SPECTRAL FUNCTIONS}",
    journal = "Phys. Rev. Lett.",
    volume = "18",
    pages = "761--763",
    year = "1967"
}

@article{Das:1967ek,
    author = "Das, T. and Mathur, V. S. and Okubo, S.",
    title = "{Low-energy theorem in the radiative decays of charged pions}",
    doi = "10.1103/PhysRevLett.19.859",
    journal = "Phys. Rev. Lett.",
    volume = "19",
    pages = "859--861",
    year = "1967"
}

@article{Kapusta:1993hq,
    author = "Kapusta, Joseph I. and Shuryak, Edward V.",
    title = "{Weinberg type sum rules at zero and finite temperature}",
    eprint = "hep-ph/9312245",
    archivePrefix = "arXiv",
    reportNumber = "NSF-ITP-93-127, NUC-MINN-93-25-T, SUNY-NTG-93-45",
    doi = "10.1103/PhysRevD.49.4694",
    journal = "Phys. Rev. D",
    volume = "49",
    pages = "4694--4704",
    year = "1994"
}

@article{Hayano:2008vn,
    author = "Hayano, Ryugo S. and Hatsuda, Tetsuo",
    title = "{Hadron properties in the nuclear medium}",
    eprint = "0812.1702",
    archivePrefix = "arXiv",
    primaryClass = "nucl-ex",
    doi = "10.1103/RevModPhys.82.2949",
    journal = "Rev. Mod. Phys.",
    volume = "82",
    pages = "2949",
    year = "2010"
}

@article{Oller:1998hw,
    author = "Oller, J. A. and Oset, E. and Pelaez, J. R.",
    title = "{Meson meson interaction in a nonperturbative chiral approach}",
    eprint = "hep-ph/9804209",
    archivePrefix = "arXiv",
    reportNumber = "SLAC-PUB-7787",
    doi = "10.1103/PhysRevD.59.074001",
    journal = "Phys. Rev. D",
    volume = "59",
    pages = "074001",
    year = "1999",
    note = "[Erratum: Phys.Rev.D 60, 099906 (1999), Erratum: Phys.Rev.D 75, 099903 (2007)]"
}

@article{Oller:1998zr,
    author = "Oller, J. A. and Oset, E.",
    title = "{N/D description of two meson amplitudes and chiral symmetry}",
    eprint = "hep-ph/9809337",
    archivePrefix = "arXiv",
    doi = "10.1103/PhysRevD.60.074023",
    journal = "Phys. Rev. D",
    volume = "60",
    pages = "074023",
    year = "1999"
}

@article{Truong:1988zp,
    author = "Truong, Tran N.",
    title = "{Chiral Perturbation Theory and Final State Theorem}",
    reportNumber = "PRINT-88-0530 (ECOLE POLY)",
    doi = "10.1103/PhysRevLett.61.2526",
    journal = "Phys. Rev. Lett.",
    volume = "61",
    pages = "2526",
    year = "1988"
}

@article{Truong:1991gv,
    author = "Truong, Tran N.",
    title = "{Remarks on the unitarization methods}",
    reportNumber = "CPTH-A074-0891",
    doi = "10.1103/PhysRevLett.67.2260",
    journal = "Phys. Rev. Lett.",
    volume = "67",
    pages = "2260--2263",
    year = "1991"
}

@article{Hatsuda:1998vb,
    author = "Hatsuda, T. and Kunihiro, T. and Shimizu, H.",
    title = "{Precursor of chiral symmetry restoration in the nuclear medium}",
    eprint = "nucl-th/9810022",
    archivePrefix = "arXiv",
    reportNumber = "RYUTHP-98-2",
    doi = "10.1103/PhysRevLett.82.2840",
    journal = "Phys. Rev. Lett.",
    volume = "82",
    pages = "2840--2843",
    year = "1999"
}

@article{Aouissat:1999ss,
    author = "Aouissat, Z. and Chanfray, G. and Schuck, P. and Wambach, J.",
    title = "{Reduced alpha meson mass and in-medium S wave pi pi correlations}",
    doi = "10.1103/PhysRevC.61.012202",
    journal = "Phys. Rev. C",
    volume = "61",
    pages = "012202",
    year = "2000"
}

@article{Davesne:1999qj,
    author = "Davesne, D. and Zhang, Y. J. and Chanfray, G.",
    title = "{Medium modification of the pion pion interaction at finite density}",
    eprint = "nucl-th/9909032",
    archivePrefix = "arXiv",
    reportNumber = "LYCEN-99-63",
    doi = "10.1103/PhysRevC.62.024604",
    journal = "Phys. Rev. C",
    volume = "62",
    pages = "024604",
    year = "2000"
}

@article{CHAOS:1996nql,
    author = "{Bonutti et al.}",
    collaboration = "CHAOS",
    title = "{A dependence of the (pi+,pi+ pi+-) reaction near the 2m(pi) threshold}",
    doi = "10.1103/PhysRevLett.77.603",
    journal = "Phys. Rev. Lett.",
    volume = "77",
    pages = "603--606",
    year = "1996"
}

@article{CHAOS:2000fjr,
    author = "{Bonutti et al.}",
    collaboration = "CHAOS",
    title = "{The pi pi interaction in nuclear matter from a study of the pi+ A ---{\ensuremath{>}} pi+ pi+- A-prime reactions}",
    eprint = "nucl-ex/0007017",
    archivePrefix = "arXiv",
    doi = "10.1016/S0375-9474(00)00286-4",
    journal = "Nucl. Phys. A",
    volume = "677",
    pages = "213--240",
    year = "2000"
}

@article{CrystalBall:2000tgu,
    author = "{Starostin et al.}",
    collaboration = "Crystal Ball",
    title = "{Measurement of pi0 pi0 production in the nuclear medium by pi- interactions at 0.408-GeV/c}",
    doi = "10.1103/PhysRevLett.85.5539",
    journal = "Phys. Rev. Lett.",
    volume = "85",
    pages = "5539--5542",
    year = "2000"
}

@article{Messchendorp:2002au,
    author = "{Messchendorp et al.}",
    title = "{In-medium modifications of the pi pi interaction in photon induced reactions}",
    eprint = "nucl-ex/0205009",
    archivePrefix = "arXiv",
    doi = "10.1103/PhysRevLett.89.222302",
    journal = "Phys. Rev. Lett.",
    volume = "89",
    pages = "222302",
    year = "2002"
}

@article{Bloch:2007ka,
    author = "{Bloch et al.}",
    title = "{Double pion photoproduction off Ca-40}",
    eprint = "nucl-ex/0703037",
    archivePrefix = "arXiv",
    doi = "10.1140/epja/i2006-10383-2",
    journal = "Eur. Phys. J. A",
    volume = "32",
    pages = "219",
    year = "2007"
}

@article{Buss:2006vh,
    author = "Buss, Oliver and Alvarez-Ruso, Luis and Muhlich, Pascal and Mosel, Ulrich",
    title = "{Low-energy pions in nuclear matter and pi pi photoproduction within a BUU transport model}",
    eprint = "nucl-th/0603003",
    archivePrefix = "arXiv",
    doi = "10.1140/epja/i2006-10075-y",
    journal = "Eur. Phys. J. A",
    volume = "29",
    pages = "189--207",
    year = "2006"
}

@article{Detar:1987kae,
    author = "Detar, Carleton E. and Kogut, John B.",
    title = "{The Hadronic Spectrum of the Quark Plasma}",
    reportNumber = "ILL-TH-87-27, UU-HEP-87-3",
    doi = "10.1103/PhysRevLett.59.399",
    journal = "Phys. Rev. Lett.",
    volume = "59",
    pages = "399",
    year = "1987"
}

@article{Detar:1987hib,
    author = "Detar, Carleton E. and Kogut, John B.",
    title = "{Measuring the Hadronic Spectrum of the Quark Plasma}",
    reportNumber = "ILL-TH-87-35, UU-HEP-87-5",
    doi = "10.1103/PhysRevD.36.2828",
    journal = "Phys. Rev. D",
    volume = "36",
    pages = "2828",
    year = "1987"
}

@article{Gottlieb:1987gz,
    author = "Gottlieb, Steven A. and Liu, W. and Toussaint, D. and Renken, R. L. and Sugar, R. L.",
    title = "{Hadronic Screening Lengths in the High Temperature Plasma}",
    reportNumber = "UCSD-PTH-87-16",
    doi = "10.1103/PhysRevLett.59.1881",
    journal = "Phys. Rev. Lett.",
    volume = "59",
    pages = "1881",
    year = "1987"
}

@article{Detar:1988kn,
    author = "Detar, Carleton E. and Kunihiro, Teiji",
    title = "{Linear $\sigma$ Model With Parity Doubling}",
    reportNumber = "RIFP-748",
    doi = "10.1103/PhysRevD.39.2805",
    journal = "Phys. Rev. D",
    volume = "39",
    pages = "2805",
    year = "1989"
}

@article{tHooft:1979rat,
    author = "'t Hooft, Gerard",
    editor = "'t Hooft, Gerard and Itzykson, C. and Jaffe, A. and Lehmann, H. and Mitter, P. K. and Singer, I. M. and Stora, R.",
    title = "{Naturalness, chiral symmetry, and spontaneous chiral symmetry breaking}",
    reportNumber = "PRINT-80-0083 (UTRECHT)",
    doi = "10.1007/978-1-4684-7571-5_9",
    journal = "NATO Sci. Ser. B",
    volume = "59",
    pages = "135--157",
    year = "1980"
}

@article{Frishman:1980dq,
    author = "Frishman, Y. and Schwimmer, A. and Banks, Tom and Yankielowicz, S.",
    title = "{The Axial Anomaly and the Bound State Spectrum in Confining Theories}",
    reportNumber = "WIS-80/27-Ph",
    doi = "10.1016/0550-3213(81)90268-6",
    journal = "Nucl. Phys. B",
    volume = "177",
    pages = "157--171",
    year = "1981"
}

@article{Coleman:1982yg,
    author = "Coleman, Sidney R. and Grossman, Bernard",
    title = "{'t Hooft's Consistency Condition as a Consequence of Analyticity and Unitarity}",
    reportNumber = "HUTP-82/A009",
    doi = "10.1016/0550-3213(82)90028-1",
    journal = "Nucl. Phys. B",
    volume = "203",
    pages = "205--220",
    year = "1982"
}

@article{Itoyama:1982up,
    author = "Itoyama, Hiroshi and Mueller, Alfred H.",
    title = "{The Axial Anomaly at Finite Temperature}",
    reportNumber = "CU-TP-256",
    doi = "10.1016/0550-3213(83)90370-X",
    journal = "Nucl. Phys. B",
    volume = "218",
    pages = "349--365",
    year = "1983"
}

@article{Pisarski:1997bq,
    author = "Pisarski, Robert D. and Trueman, T. L. and Tytgat, Michel H. G.",
    title = "{How pi0 ---{\ensuremath{>}} gamma gamma changes with temperature}",
    eprint = "hep-ph/9702362",
    archivePrefix = "arXiv",
    doi = "10.1103/PhysRevD.56.7077",
    journal = "Phys. Rev. D",
    volume = "56",
    pages = "7077--7088",
    year = "1997"
}

@article{Hsu:2000by,
    author = "Hsu, Stephen D. H. and Sannino, Francesco and Schwetz, Myck",
    title = "{Anomaly matching in gauge theories at finite matter density}",
    eprint = "hep-ph/0006059",
    archivePrefix = "arXiv",
    reportNumber = "YCTP-P5-00, BUHEP-00-11",
    doi = "10.1142/S0217732301005217",
    journal = "Mod. Phys. Lett. A",
    volume = "16",
    pages = "1871--1880",
    year = "2001"
}

@article{Kummer:2025kch,
    author = "Kummer, Christian and Leupold, Stefan and von Smekal, Lorenz",
    title = "{Kinetic mixing and axial charges in the parity-doublet model}",
    eprint = "2512.03894",
    archivePrefix = "arXiv",
    primaryClass = "hep-ph",
    doi = "10.1103/qc6c-wg6h",
    journal = "Phys. Rev. D",
    volume = "113",
    number = "11",
    pages = "116010",
    year = "2026"
}

@article{Brodie:2025nww,
    author = "Brodie, Liam and Pisarski, Robert D.",
    title = "{Parity-Doubled Nucleons Can Rapidly Cool Neutron Stars}",
    eprint = "2501.02055",
    archivePrefix = "arXiv",
    primaryClass = "nucl-th",
    doi = "10.1103/bf66-1hr5",
    journal = "Phys. Rev. Lett.",
    volume = "135",
    number = "15",
    pages = "152702",
    year = "2025"
}

@article{Negreiros:2026ode,
    author = "Negreiros, Rodrigo and Brodie, Liam and Steinheimer, Jan and Dexheimer, Veronica and Pisarski, Robert D.",
    title = "{Enhanced Neutrino Cooling from Parity-Doubled Nucleons in Neutron Star Cooling Simulations}",
    eprint = "2603.06789[astro-ph.HE]",
    archivePrefix = "arXiv",
    primaryClass = "astro-ph.HE",
    month = "3",
    year = "2026"
}

@article{PhysRevD.92.014503,
  title = {Nucleons and parity doubling across the deconfinement transition},
  author = {Aarts, Gert and Allton, Chris and Hands, Simon and J\"ager, Benjamin and Praki, Chrisanthi and Skullerud, Jon-Ivar},
  journal = {Phys. Rev. D},
  volume = {92},
  issue = {1},
  pages = {014503},
  numpages = {6},
  year = {2015},
  month = {Jul},
  publisher = {American Physical Society},
  doi = {10.1103/PhysRevD.92.014503},
  url = {https://link.aps.org/doi/10.1103/PhysRevD.92.014503}
}

@article{Gao:2025eax,
    author = "Gao, Bikai and Hosaka, Atsushi",
    title = "{Linear realization of an SU(3) parity doublet model for octet baryons with a bad diquark}",
    eprint = "2512.01192",
    archivePrefix = "arXiv",
    primaryClass = "hep-ph",
    doi = "10.1103/s39n-mlw2",
    journal = "Phys. Rev. D",
    volume = "113",
    number = "3",
    pages = "036008",
    year = "2026"
}

@article{PhysRevLett.84.3252,
  title = {Chiral-Symmetry Realization for Even- and Odd-Parity Baryon Resonances},
  author = {Jido, D. and Hatsuda, T. and Kunihiro, T.},
  journal = {Phys. Rev. Lett.},
  volume = {84},
  issue = {15},
  pages = {3252--3255},
  numpages = {0},
  year = {2000},
  month = {Apr},
  publisher = {American Physical Society},
  doi = {10.1103/PhysRevLett.84.3252},
  url = {https://link.aps.org/doi/10.1103/PhysRevLett.84.3252}
}

@article{Olbrich:2015gln,
    author = "Olbrich, Lisa and Z{\'e}t{\'e}nyi, Mikl{\'o}s and Giacosa, Francesco and Rischke, Dirk H.",
    title = "{Three-flavor chiral effective model with four baryonic multiplets within the mirror assignment}",
    eprint = "1511.05035",
    archivePrefix = "arXiv",
    primaryClass = "hep-ph",
    doi = "10.1103/PhysRevD.93.034021",
    journal = "Phys. Rev. D",
    volume = "93",
    number = "3",
    pages = "034021",
    year = "2016"
}

@article{Olbrich:2017fsd,
    author = "Olbrich, Lisa and Z{\'e}t{\'e}nyi, Mikl{\'o}s and Giacosa, Francesco and Rischke, Dirk H.",
    title = "{Influence of the axial anomaly on the decay $N(1535) \rightarrow N\eta $}",
    eprint = "1708.01061",
    archivePrefix = "arXiv",
    primaryClass = "hep-ph",
    doi = "10.1103/PhysRevD.97.014007",
    journal = "Phys. Rev. D",
    volume = "97",
    number = "1",
    pages = "014007",
    year = "2018"
}

@article{PANDA:2009yku,
    author = "{Lutz et al.}",
    collaboration = "PANDA",
    title = "{Physics Performance Report for PANDA: Strong Interaction Studies with Antiprotons}",
    eprint = "0903.3905[hep-ex]",
    archivePrefix = "arXiv",
    primaryClass = "hep-ex",
    month = "3",
    year = "2009"
}

@article{PANDA:2021ozp,
    author = "{Barucca et al.}",
    collaboration = "PANDA",
    title = "{PANDA Phase One}",
    eprint = "2101.11877",
    archivePrefix = "arXiv",
    primaryClass = "hep-ex",
    doi = "10.1140/epja/s10050-021-00475-y",
    journal = "Eur. Phys. J. A",
    volume = "57",
    number = "6",
    pages = "184",
    year = "2021"
}

@article{PhysRevD.92.054022,
  title = {Extended Goldberger-Treiman relation in a three-flavor parity doublet model},
  author = {Nishihara, Hiroki and Harada, Masayasu},
  journal = {Phys. Rev. D},
  volume = {92},
  issue = {5},
  pages = {054022},
  numpages = {10},
  year = {2015},
  month = {Sep},
  publisher = {American Physical Society},
  doi = {10.1103/PhysRevD.92.054022},
  url = {https://link.aps.org/doi/10.1103/PhysRevD.92.054022}
}

@article{Mukherjee:2017jzi,
    author = "Mukherjee, A. and Schramm, S. and Steinheimer, J. and Dexheimer, V.",
    title = "{The application of the Quark-Hadron Chiral Parity-Doublet Model to neutron star matter}",
    eprint = "1706.09191",
    archivePrefix = "arXiv",
    primaryClass = "nucl-th",
    doi = "10.1051/0004-6361/201731505",
    journal = "Astron. Astrophys.",
    volume = "608",
    pages = "A110",
    year = "2017"
}

@article{Gao:2022klm,
    author = "Gao, Bikai and Minamikawa, Takuya and Kojo, Toru and Harada, Masayasu",
    title = "{Impacts of the U(1)A anomaly on nuclear and neutron star equation~of state based on a parity doublet model}",
    eprint = "2207.05970",
    archivePrefix = "arXiv",
    primaryClass = "nucl-th",
    doi = "10.1103/PhysRevC.106.065205",
    journal = "Phys. Rev. C",
    volume = "106",
    number = "6",
    pages = "065205",
    year = "2022"
}

@article{Kong:2023nue,
    author = "Kong, Yuk Kei and Minamikawa, Takuya and Harada, Masayasu",
    title = "{Neutron star matter based on a parity doublet model including the a0(980) meson}",
    eprint = "2306.08140",
    archivePrefix = "arXiv",
    primaryClass = "nucl-th",
    doi = "10.1103/PhysRevC.108.055206",
    journal = "Phys. Rev. C",
    volume = "108",
    number = "5",
    pages = "055206",
    year = "2023"
}

@article{Minamikawa:2023eky,
    author = "Minamikawa, Takuya and Gao, Bikai and Kojo, Toru and Harada, Masayasu",
    title = "{Chiral Restoration of Nucleons in Neutron Star Matter: Studies Based on a Parity Doublet Model}",
    eprint = "2302.00825",
    archivePrefix = "arXiv",
    primaryClass = "nucl-th",
    doi = "10.3390/sym15030745",
    journal = "Symmetry",
    volume = "15",
    number = "3",
    pages = "745",
    year = "2023"
}

@article{NANOGrav:2019jur,
    author = "{Cromartie et al.}",
    collaboration = "NANOGrav",
    title = "{Relativistic Shapiro delay measurements of an extremely massive millisecond pulsar}",
    eprint = "1904.06759",
    archivePrefix = "arXiv",
    primaryClass = "astro-ph.HE",
    doi = "10.1038/s41550-019-0880-2",
    journal = "Nature Astron.",
    volume = "4",
    number = "1",
    pages = "72--76",
    year = "2019"
}

@article{Miller:2019cac,
    author = "{Miller et al.}",
    title = "{PSR J0030+0451 Mass and Radius from $NICER$ Data and Implications for the Properties of Neutron Star Matter}",
    eprint = "1912.05705",
    archivePrefix = "arXiv",
    primaryClass = "astro-ph.HE",
    doi = "10.3847/2041-8213/ab50c5",
    journal = "Astrophys. J. Lett.",
    volume = "887",
    number = "1",
    pages = "L24",
    year = "2019"
}

@article{Riley:2021pdl,
    author = "{Riley et al.}",
    title = "{A NICER View of the Massive Pulsar PSR J0740+6620 Informed by Radio Timing and XMM-Newton Spectroscopy}",
    eprint = "2105.06980",
    archivePrefix = "arXiv",
    primaryClass = "astro-ph.HE",
    doi = "10.3847/2041-8213/ac0a81",
    journal = "Astrophys. J. Lett.",
    volume = "918",
    number = "2",
    pages = "L27",
    year = "2021"
}

@article{Eser:2023oii,
    author = {Eser, J{\"u}rgen and Blaizot, Jean-Paul},
    title = "{Thermodynamics of the parity-doublet model: Symmetric nuclear matter and the chiral transition}",
    eprint = "2309.06566",
    archivePrefix = "arXiv",
    primaryClass = "nucl-th",
    doi = "10.1103/PhysRevC.109.045201",
    journal = "Phys. Rev. C",
    volume = "109",
    number = "4",
    pages = "045201",
    year = "2024"
}

@article{Eser:2024xil,
    author = {Eser, J{\"u}rgen and Blaizot, Jean-Paul},
    title = "{Thermodynamics of the parity-doublet model. II. Asymmetric and neutron matter}",
    eprint = "2408.01302",
    archivePrefix = "arXiv",
    primaryClass = "nucl-th",
    doi = "10.1103/PhysRevC.110.065205",
    journal = "Phys. Rev. C",
    volume = "110",
    number = "6",
    pages = "065205",
    year = "2024"
}

@article{Kunihiro:1987bb,
    author = "Kunihiro, Teiji and Hatsuda, Tetsuo",
    title = "{Effects of Flavor Mixing Induced by Axial Anomaly on the Quark Condensates and Meson Spectra}",
    reportNumber = "KEK-TH-190, KEK-Preprint-87-135",
    doi = "10.1016/0370-2693(88)91596-1",
    journal = "Phys. Lett. B",
    volume = "206",
    pages = "385--390",
    year = "1988",
    note = "[Erratum: Phys.Lett.B 210, 278--278 (1988)]"
}

@article{Bernard:1987sg,
    author = "Bernard, Veronique and Jaffe, R. L. and Meissner, Ulf G.",
    title = "{Strangeness Mixing and Quenching in the Nambu-Jona-Lasinio Model}",
    reportNumber = "MIT-CTP-1547",
    doi = "10.1016/0550-3213(88)90127-7",
    journal = "Nucl. Phys. B",
    volume = "308",
    pages = "753--790",
    year = "1988"
}

@article{Hatsuda:1994pi,
    author = "Hatsuda, Tetsuo and Kunihiro, Teiji",
    title = "{QCD phenomenology based on a chiral effective Lagrangian}",
    eprint = "hep-ph/9401310",
    archivePrefix = "arXiv",
    reportNumber = "UTHEP-270, RYUTHP-94-1",
    doi = "10.1016/0370-1573(94)90022-1",
    journal = "Phys. Rept.",
    volume = "247",
    pages = "221--367",
    year = "1994"
}

@article{Takizawa:1989sv,
    author = "Takizawa, M. and Tsushima, K. and Kohyama, Y. and Kubodera, K.",
    title = "{Study of Meson Flavor Mixing in the {Nambu-Jona-Lasinio} Model Including Instanton Effects}",
    doi = "10.1143/PTP.82.481",
    journal = "Prog. Theor. Phys.",
    volume = "82",
    pages = "481--486",
    year = "1989"
}

@article{Ebert:1994mf,
    author = "Ebert, D. and Reinhardt, H. and Volkov, M. K.",
    title = "{Effective hadron theory of QCD}",
    doi = "10.1016/0146-6410(94)90043-4",
    journal = "Prog. Part. Nucl. Phys.",
    volume = "33",
    pages = "1--120",
    year = "1994"
}

@article{Vogl:1991qt,
    author = "Vogl, U. and Weise, W.",
    title = "{The Nambu and Jona Lasinio model: Its implications for hadrons and nuclei}",
    reportNumber = "TPR-91-6",
    doi = "10.1016/0146-6410(91)90005-9",
    journal = "Prog. Part. Nucl. Phys.",
    volume = "27",
    pages = "195--272",
    year = "1991"
}

@article{Jido:1998av,
    author = "Jido, D. and Nemoto, Y. and Oka, M. and Hosaka, A.",
    title = "{Chiral symmetry for positive and negative parity nucleons}",
    eprint = "hep-ph/9805306",
    archivePrefix = "arXiv",
    reportNumber = "TIT-HEP-393-NP",
    doi = "10.1016/S0375-9474(99)00844-1",
    journal = "Nucl. Phys. A",
    volume = "671",
    pages = "471--480",
    year = "2000"
}

@article{Jido:2001nt,
    author = "Jido, Daisuke and Oka, Makoto and Hosaka, Atsushi",
    title = "{Chiral symmetry of baryons}",
    eprint = "hep-ph/0110005",
    archivePrefix = "arXiv",
    doi = "10.1143/PTP.106.873",
    journal = "Prog. Theor. Phys.",
    volume = "106",
    pages = "873--908",
    year = "2001"
}

@article{Takahashi:2008fy,
    author = "Takahashi, Toru T. and Kunihiro, Teiji",
    title = "{Axial charges of N(1535) and N(1650) in lattice QCD with two flavors of dynamical quarks}",
    eprint = "0801.4707",
    archivePrefix = "arXiv",
    primaryClass = "hep-lat",
    doi = "10.1103/PhysRevD.78.011503",
    journal = "Phys. Rev. D",
    volume = "78",
    pages = "011503",
    year = "2008"
}

@article{Gerber:1988tt,
    author = "Gerber, P. and Leutwyler, H.",
    title = "{Hadrons Below the Chiral Phase Transition}",
    reportNumber = "BUTP-88/30-BERN",
    doi = "10.1016/0550-3213(89)90349-0",
    journal = "Nucl. Phys. B",
    volume = "321",
    pages = "387--429",
    year = "1989"
}

@article{Bijnens:1995ww,
    author = "Bijnens, Johan",
    title = "{Chiral Lagrangians and Nambu-Jona-Lasinio - like models}",
    eprint = "hep-ph/9502335",
    archivePrefix = "arXiv",
    reportNumber = "NORDITA-95-10-N-P",
    doi = "10.1016/0370-1573(95)00051-8",
    journal = "Phys. Rept.",
    volume = "265",
    pages = "369--446",
    year = "1996"
}

@article{Shizuya:1979bv,
    author = "Shizuya, Ken-ichi",
    title = "{1/N Expansion and the Theory of Composite Particles}",
    reportNumber = "FERMILAB-PUB-79-024-T",
    doi = "10.1103/PhysRevD.21.2327",
    journal = "Phys. Rev. D",
    volume = "21",
    pages = "2327",
    year = "1980"
}

@article{RuizArriola:1991gc,
    author = "Ruiz Arriola, E.",
    title = "{The Low-energy expansion of the generalized SU(3) NJL model}",
    doi = "10.1016/0370-2693(91)91746-I",
    journal = "Phys. Lett. B",
    volume = "253",
    pages = "430--435",
    year = "1991"
}

@article{Gasser:1983yg,
    author = "Gasser, J. and Leutwyler, H.",
    title = "{Chiral Perturbation Theory to One Loop}",
    reportNumber = "CERN-TH-3689",
    doi = "10.1016/0003-4916(84)90242-2",
    journal = "Annals Phys.",
    volume = "158",
    pages = "142",
    year = "1984"
}

@article{Gasser:1984gg,
    author = "Gasser, J. and Leutwyler, H.",
    title = "{Chiral Perturbation Theory: Expansions in the Mass of the Strange Quark}",
    reportNumber = "CERN-TH-3798",
    doi = "10.1016/0550-3213(85)90492-4",
    journal = "Nucl. Phys. B",
    volume = "250",
    pages = "465--516",
    year = "1985"
}

@article{Pelaez:2015qba,
    author = "Pelaez, J. R.",
    title = "{From controversy to precision on the sigma meson: a review on the status of the non-ordinary $f_0(500)$ resonance}",
    eprint = "1510.00653",
    archivePrefix = "arXiv",
    primaryClass = "hep-ph",
    doi = "10.1016/j.physrep.2016.09.001",
    journal = "Phys. Rept.",
    volume = "658",
    pages = "1",
    year = "2016"
}

@article{Hyodo:2010jp,
    author = "Hyodo, Tetsuo and Jido, Daisuke and Kunihiro, Teiji",
    title = "{Nature of the $\sigma$ meson as revealed by its softening process}",
    eprint = "1007.1718",
    archivePrefix = "arXiv",
    primaryClass = "hep-ph",
    doi = "10.1016/j.nuclphysa.2010.09.016",
    journal = "Nucl. Phys. A",
    volume = "848",
    pages = "341--365",
    year = "2010"
}

@article{Jaffe:1976ig,
    author = "Jaffe, Robert L.",
    title = "{Multi-Quark Hadrons. 1. The Phenomenology of (2 Quark 2 anti-Quark) Mesons}",
    reportNumber = "SLAC-PUB-1772",
    doi = "10.1103/PhysRevD.15.267",
    journal = "Phys. Rev. D",
    volume = "15",
    pages = "267",
    year = "1977"
}

@article{Maiani:2004uc,
    author = "Maiani, L. and Piccinini, F. and Polosa, A. D. and Riquer, V.",
    title = "{A New look at scalar mesons}",
    eprint = "hep-ph/0407017",
    archivePrefix = "arXiv",
    doi = "10.1103/PhysRevLett.93.212002",
    journal = "Phys. Rev. Lett.",
    volume = "93",
    pages = "212002",
    year = "2004"
}

@article{Kunihiro:2003yj,
    author = "Kunihiro, Teiji and Muroya, Shin and Nakamura, Atsushi and Nonaka, Chiho and Sekiguchi, Motoo and Wada, Hiroaki",
    collaboration = "SCALAR",
    title = "{Scalar mesons in lattice QCD}",
    eprint = "hep-ph/0310312",
    archivePrefix = "arXiv",
    doi = "10.1103/PhysRevD.70.034504",
    journal = "Phys. Rev. D",
    volume = "70",
    pages = "034504",
    year = "2004"
}

@article{Yamazaki:1996zb,
    author = "{Yamazaki et al.}",
    title = "{Discovery of deeply bound pi- states in the Pb-208 (d, He-3) reaction}",
    reportNumber = "INS-1144, GSI-PREPRINT-96-32",
    doi = "10.1007/s002180050101",
    journal = "Z. Phys. A",
    volume = "355",
    pages = "219--221",
    year = "1996"
}

@article{Hirenzaki:1991us,
    author = "Hirenzaki, S. and Toki, H. and Yamazaki, T.",
    title = "{(d, He-3) reactions for the formation of deeply bound pionic atoms}",
    doi = "10.1103/PhysRevC.44.2472",
    journal = "Phys. Rev. C",
    volume = "44",
    pages = "2472--2479",
    year = "1991"
}

@article{Yamazaki:2012zza,
    author = "Yamazaki, Toshimitsu and Hirenzaki, Satoru and Hayano, Ryugo S. and Toki, Hiroshi",
    title = "{Deeply bound pionic states in heavy nuclei}",
    doi = "10.1016/j.physrep.2012.01.003",
    journal = "Phys. Rept.",
    volume = "514",
    pages = "1--87",
    year = "2012"
}

@article{piAF:2022gvw,
    author = "{Nishi et al.}",
    collaboration = "piAF",
    title = "{Chiral symmetry restoration at high matter density observed in pionic atoms}",
    eprint = "2204.05568",
    archivePrefix = "arXiv",
    primaryClass = "nucl-ex",
    doi = "10.1038/s41567-023-02001-x",
    journal = "Nature Phys.",
    volume = "19",
    number = "6",
    pages = "788--793",
    year = "2023"
}

@article{Kolomeitsev:2002gc,
    author = "Kolomeitsev, Evgeni E. and Kaiser, N. and Weise, W.",
    title = "{Chiral dynamics of deeply bound pionic atoms}",
    eprint = "nucl-th/0207090",
    archivePrefix = "arXiv",
    reportNumber = "ECT-02-16",
    doi = "10.1103/PhysRevLett.90.092501",
    journal = "Phys. Rev. Lett.",
    volume = "90",
    pages = "092501",
    year = "2003"
}

@article{Weise:2001sg,
    author = "Weise, W.",
    editor = "Braun-Munzinger, P. and Zilges, A.",
    title = "{Hadronic excitations and chiral symmetry in nuclear systems}",
    doi = "10.1016/S0375-9474(01)00934-4",
    journal = "Nucl. Phys. A",
    volume = "690",
    pages = "98--109",
    year = "2001"
}

@article{Jido:2008bk,
    author = "Jido, D. and Hatsuda, T. and Kunihiro, T.",
    title = "{In-medium Pion and Partial Restoration of Chiral Symmetry}",
    eprint = "0805.4453",
    archivePrefix = "arXiv",
    primaryClass = "nucl-th",
    reportNumber = "YITP-07-23",
    doi = "10.1016/j.physletb.2008.10.034",
    journal = "Phys. Lett. B",
    volume = "670",
    pages = "109--113",
    year = "2008"
}

@article{Terashima:2008zza,
    author = "{Terashima et al.}",
    title = "{Proton elastic scattering from tin isotopes at 295-MeV and systematic change of neutron density distributions}",
    doi = "10.1103/PhysRevC.77.024317",
    journal = "Phys. Rev. C",
    volume = "77",
    pages = "024317",
    year = "2008"
}

@article{Hubsch:2021nih,
    author = {H{\"u}bsch, Stephan and Jido, Daisuke},
    title = "{Density dependence of the quark condensate in isospin-asymmetric nuclear matter}",
    eprint = "2103.08823",
    archivePrefix = "arXiv",
    primaryClass = "nucl-th",
    doi = "10.1103/PhysRevC.104.015202",
    journal = "Phys. Rev. C",
    volume = "104",
    number = "1",
    pages = "015202",
    year = "2021"
}

@article{Friedman:2019zhc,
    author = "Friedman, E. and Gal, A.",
    title = "{The pion-nucleon {\ensuremath{\sigma}} term from pionic atoms}",
    eprint = "1901.03130",
    archivePrefix = "arXiv",
    primaryClass = "nucl-th",
    doi = "10.1016/j.physletb.2019.03.036",
    journal = "Phys. Lett. B",
    volume = "792",
    pages = "340--344",
    year = "2019"
}

@article{Kaiser:2007nv,
    author = "Kaiser, N. and de Homont, P. and Weise, W.",
    title = "{In-medium chiral condensate beyond linear density approximation}",
    eprint = "0711.3154",
    archivePrefix = "arXiv",
    primaryClass = "nucl-th",
    doi = "10.1103/PhysRevC.77.025204",
    journal = "Phys. Rev. C",
    volume = "77",
    pages = "025204",
    year = "2008"
}

@article{Goda:2013bka,
    author = "Goda, Soichiro and Jido, D.",
    title = "{Chiral condensate at finite density using the chiral Ward identity}",
    eprint = "1308.2660",
    archivePrefix = "arXiv",
    primaryClass = "nucl-th",
    doi = "10.1103/PhysRevC.88.065204",
    journal = "Phys. Rev. C",
    volume = "88",
    number = "6",
    pages = "065204",
    year = "2013"
}

@article{Lacour:2010ci,
    author = "Lacour, A. and Oller, J. A. and Meissner, U. -G.",
    title = "{The Chiral quark condensate and pion decay constant in nuclear matter at next-to-leading order}",
    eprint = "1007.2574",
    archivePrefix = "arXiv",
    primaryClass = "nucl-th",
    doi = "10.1088/0954-3899/37/12/125002",
    journal = "J. Phys. G",
    volume = "37",
    pages = "125002",
    year = "2010"
}

@article{Bernard:1988db,
    author = "Bernard, Veronique and Meissner, Ulf G.",
    title = "{Properties of Vector and Axial Vector Mesons from a Generalized Nambu-Jona-Lasinio Model}",
    reportNumber = "MIT-CTP-1581",
    doi = "10.1016/0375-9474(88)90114-5",
    journal = "Nucl. Phys. A",
    volume = "489",
    pages = "647--670",
    year = "1988"
}

@article{Brown:1987hj,
    author = "Brown, G. E. and Kubodera, K. and Rho, Mannque",
    title = "{Strangeness Condensation and 'Clearing' of the Vacuum}",
    doi = "10.1016/0370-2693(87)90104-3",
    journal = "Phys. Lett. B",
    volume = "192",
    pages = "273--278",
    year = "1987"
}

@article{Brown:1991kk,
    author = "Brown, G. E. and Rho, Mannque",
    title = "{Scaling effective Lagrangians in a dense medium}",
    doi = "10.1103/PhysRevLett.66.2720",
    journal = "Phys. Rev. Lett.",
    volume = "66",
    pages = "2720--2723",
    year = "1991"
}

@article{Pisarski:1981mq,
    author = "Pisarski, Robert D.",
    title = "{Phenomenology of the Chiral Phase Transition}",
    reportNumber = "NSF-ITP-81-92",
    doi = "10.1016/0370-2693(82)91025-5",
    journal = "Phys. Lett. B",
    volume = "110",
    pages = "155--158",
    year = "1982"
}

@article{Georgi:1989xy,
    author = "Georgi, Howard",
    title = "{Vector Realization of Chiral Symmetry}",
    reportNumber = "HUTP-89/A042",
    doi = "10.1016/0550-3213(90)90210-5",
    journal = "Nucl. Phys. B",
    volume = "331",
    pages = "311--330",
    year = "1990"
}

@article{Harada:2000kb,
    author = "Harada, Masayasu and Yamawaki, Koichi",
    title = "{Vector manifestation of the chiral symmetry}",
    eprint = "hep-ph/0010207",
    archivePrefix = "arXiv",
    reportNumber = "DPNU-00-38",
    doi = "10.1103/PhysRevLett.86.757",
    journal = "Phys. Rev. Lett.",
    volume = "86",
    pages = "757--760",
    year = "2001"
}

@article{Hatsuda:1991ez,
    author = "Hatsuda, Tetsuo and Lee, Su Houng",
    title = "{QCD sum rules for vector mesons in the nuclear medium}",
    reportNumber = "DOE-ER-40561-05, INT91-00-02, YSTP-91-10",
    doi = "10.1103/PhysRevC.46.R34",
    journal = "Phys. Rev. C",
    volume = "46",
    number = "1",
    pages = "R34",
    year = "1992"
}

@article{Hatsuda:1992bv,
    author = "Hatsuda, Tetsuo and Koike, Yuji and Lee, Su-Houng",
    title = "{Finite temperature QCD sum rules reexamined: rho, omega and A1 mesons}",
    reportNumber = "UMD-PP-92-203, DOE-ER-40322-156, YSTP-92-011",
    doi = "10.1016/0550-3213(93)90107-Z",
    journal = "Nucl. Phys. B",
    volume = "394",
    pages = "221--266",
    year = "1993"
}

@article{Rapp:1999us,
    author = "Rapp, Ralf and Wambach, Jochen",
    title = "{Low mass dileptons at the CERN SPS: Evidence for chiral restoration?}",
    eprint = "hep-ph/9907502",
    archivePrefix = "arXiv",
    reportNumber = "SUNY-NTG-99-25",
    doi = "10.1007/s100500050364",
    journal = "Eur. Phys. J. A",
    volume = "6",
    pages = "415--420",
    year = "1999"
}

@article{Rapp:2000pe,
    author = "Rapp, R.",
    title = "{Signatures of thermal dilepton radiation at RHIC}",
    eprint = "hep-ph/0010101",
    archivePrefix = "arXiv",
    reportNumber = "SUNY-NTG-00-41",
    doi = "10.1103/PhysRevC.63.054907",
    journal = "Phys. Rev. C",
    volume = "63",
    pages = "054907",
    year = "2001"
}

@article{Rapp:2009yu,
    author = "Rapp, R. and Wambach, J. and van Hees, H.",
    editor = "Stock, R.",
    title = "{The Chiral Restoration Transition of QCD and Low Mass Dileptons}",
    eprint = "0901.3289",
    archivePrefix = "arXiv",
    primaryClass = "hep-ph",
    doi = "10.1007/978-3-642-01539-7_6",
    journal = "Landolt-Bornstein",
    volume = "23",
    pages = "134",
    year = "2010"
}

@article{Manohar:1983md,
    author = "Manohar, Aneesh and Georgi, Howard",
    title = "{Chiral Quarks and the Nonrelativistic Quark Model}",
    reportNumber = "HUTP-83/A042a",
    doi = "10.1016/0550-3213(84)90231-1",
    journal = "Nucl. Phys. B",
    volume = "234",
    pages = "189--212",
    year = "1984"
}

@article{Carruthers:1971ldw,
    author = "Carruthers, P. and Haymaker, R. W.",
    title = "{Radius of convergence for perturbation expansions in the su(3) sigma model}",
    doi = "10.1103/PhysRevD.4.1808",
    journal = "Phys. Rev. D",
    volume = "4",
    pages = "1808--1815",
    year = "1971"
}

@article{Kobayashi:1970ji,
    author = "Kobayashi, M. and Maskawa, T.",
    title = "{Chiral symmetry and eta-x mixing}",
    doi = "10.1143/PTP.44.1422",
    journal = "Prog. Theor. Phys.",
    volume = "44",
    pages = "1422--1424",
    year = "1970"
}

@article{Kobayashi:1971qz,
    author = "Kobayashi, M. and Kondo, H. and Maskawa, T.",
    title = "{Symmetry breaking of the chiral u(3) x u(3) and the quark model}",
    doi = "10.1143/PTP.45.1955",
    journal = "Prog. Theor. Phys.",
    volume = "45",
    pages = "1955--1959",
    year = "1971"
}

@article{tHooft:1976snw,
    author = "'t Hooft, Gerard",
    editor = "Shifman, Mikhail A.",
    title = "{Computation of the Quantum Effects Due to a Four-Dimensional Pseudoparticle}",
    reportNumber = "PRINT-76-0551 (HARVARD)",
    doi = "10.1103/PhysRevD.14.3432",
    journal = "Phys. Rev. D",
    volume = "14",
    pages = "3432--3450",
    year = "1976",
    note = "[Erratum: Phys.Rev.D 18, 2199 (1978)]"
}

@article{Mirelli:1976ww,
    author = "Mirelli, V. and Schechter, J.",
    title = "{An Effective Strong Interaction Lagrangian}",
    reportNumber = "COO-3533-81, COO-4210-81",
    doi = "10.1103/PhysRevD.15.1361",
    journal = "Phys. Rev. D",
    volume = "15",
    pages = "1361",
    year = "1977"
}

@article{Adler:1969gk,
    author = "Adler, Stephen L.",
    title = "{Axial vector vertex in spinor electrodynamics}",
    doi = "10.1103/PhysRev.177.2426",
    journal = "Phys. Rev.",
    volume = "177",
    pages = "2426--2438",
    year = "1969"
}

@article{Bell:1969ts,
    author = "Bell, J. S. and Jackiw, R.",
    title = "{A PCAC puzzle: $\pi^0 \to \gamma \gamma$ in the $\sigma$ model}",
    doi = "10.1007/BF02823296",
    journal = "Nuovo Cim. A",
    volume = "60",
    pages = "47--61",
    year = "1969"
}

@article{Vafa:1983tf,
    author = "Vafa, C. and Witten, Edward",
    title = "{Restrictions on Symmetry Breaking in Vector-Like Gauge Theories}",
    reportNumber = "PRINT-83-0910 (PRINCETON)",
    doi = "10.1016/0550-3213(84)90230-X",
    journal = "Nucl. Phys. B",
    volume = "234",
    pages = "173--188",
    year = "1984"
}

@article{Schafer:1996wv,
    author = {Sch{\"a}fer, Thomas and Shuryak, Edward V.},
    title = "{Instantons in QCD}",
    eprint = "hep-ph/9610451",
    archivePrefix = "arXiv",
    reportNumber = "DOE-ER-40561-293, INT-96-00-150",
    doi = "10.1103/RevModPhys.70.323",
    journal = "Rev. Mod. Phys.",
    volume = "70",
    pages = "323--426",
    year = "1998"
}

@article{Fujikawa:1979ay,
    author = "Fujikawa, Kazuo",
    title = "{Path Integral Measure for Gauge Invariant Fermion Theories}",
    reportNumber = "INS-328",
    doi = "10.1103/PhysRevLett.42.1195",
    journal = "Phys. Rev. Lett.",
    volume = "42",
    pages = "1195--1198",
    year = "1979"
}

@article{Fujikawa:1980eg,
    author = "Fujikawa, Kazuo",
    title = "{Path Integral for Gauge Theories with Fermions}",
    reportNumber = "INS-370",
    doi = "10.1103/PhysRevD.21.2848",
    journal = "Phys. Rev. D",
    volume = "21",
    pages = "2848",
    year = "1980",
    note = "[Erratum: Phys.Rev.D 22, 1499 (1980)]"
}

@article{Aoki:2006we,
    author = "Aoki, Y. and Endrodi, G. and Fodor, Z. and Katz, S. D. and Szabo, K. K.",
    title = "{The Order of the quantum chromodynamics transition predicted by the standard model of particle physics}",
    eprint = "hep-lat/0611014",
    archivePrefix = "arXiv",
    doi = "10.1038/nature05120",
    journal = "Nature",
    volume = "443",
    pages = "675--678",
    year = "2006"
}

@article{Aoki:2020noz,
    author = "Aoki, S. and Aoki, Y. and Cossu, G. and Fukaya, H. and Hashimoto, S. and Kaneko, T. and Rohrhofer, C. and Suzuki, K.",
    collaboration = "JLQCD",
    title = "{Study of the axial $U(1)$ anomaly at high temperature with lattice chiral fermions}",
    eprint = "2011.01499",
    archivePrefix = "arXiv",
    primaryClass = "hep-lat",
    reportNumber = "YITP-20-126, OU-HET-905, KEK-CP-0378",
    doi = "10.1103/PhysRevD.103.074506",
    journal = "Phys. Rev. D",
    volume = "103",
    number = "7",
    pages = "074506",
    year = "2021"
}

@article{Ding:2015ona,
    author = "Ding, Heng-Tong and Karsch, Frithjof and Mukherjee, Swagato",
    title = "{Thermodynamics of strong-interaction matter from Lattice QCD}",
    eprint = "1504.05274",
    archivePrefix = "arXiv",
    primaryClass = "hep-lat",
    doi = "10.1142/S0218301315300076",
    journal = "Int. J. Mod. Phys. E",
    volume = "24",
    number = "10",
    pages = "1530007",
    year = "2015"
}

@article{Tomiya:2016jwr,
    author = "Tomiya, A. and Cossu, G. and Aoki, S. and Fukaya, H. and Hashimoto, S. and Kaneko, T. and Noaki, J.",
    title = "{Evidence of effective axial U(1) symmetry restoration at high temperature QCD}",
    eprint = "1612.01908",
    archivePrefix = "arXiv",
    primaryClass = "hep-lat",
    doi = "10.1103/PhysRevD.96.034509",
    journal = "Phys. Rev. D",
    volume = "96",
    number = "3",
    pages = "034509",
    year = "2017",
    note = "[Addendum: Phys.Rev.D 96, 079902 (2017)]"
}

@article{Aoki:2021qws,
    author = "Aoki, S. and Aoki, Y. and Fukaya, H. and Hashimoto, S. and Rohrhofer, C. and Suzuki, K.",
    collaboration = "JLQCD",
    title = "{Role of the axial U(1) anomaly in the chiral susceptibility of QCD at high temperature}",
    eprint = "2103.05954",
    archivePrefix = "arXiv",
    primaryClass = "hep-lat",
    reportNumber = "YITP-21-18, OU-HET-1089, KEK-CP-0380",
    doi = "10.1093/ptep/ptac001",
    journal = "PTEP",
    volume = "2022",
    number = "2",
    pages = "023B05",
    year = "2022"
}

@article{Pisarski:2019upw,
    author = "Pisarski, Robert D. and Rennecke, Fabian",
    title = "{Multi-instanton contributions to anomalous quark interactions}",
    eprint = "1910.14052",
    archivePrefix = "arXiv",
    primaryClass = "hep-ph",
    doi = "10.1103/PhysRevD.101.114019",
    journal = "Phys. Rev. D",
    volume = "101",
    number = "11",
    pages = "114019",
    year = "2020"
}

@article{Pisarski:2024esv,
    author = "Pisarski, Robert D. and Rennecke, Fabian",
    title = "{Conjectures about the Chiral Phase Transition in QCD from Anomalous Multi-Instanton Interactions}",
    eprint = "2401.06130",
    archivePrefix = "arXiv",
    primaryClass = "hep-ph",
    doi = "10.1103/PhysRevLett.132.251903",
    journal = "Phys. Rev. Lett.",
    volume = "132",
    number = "25",
    pages = "251903",
    year = "2024"
}

@article{CERES:1995vll,
    author = "{Agakichiev et al.}",
    collaboration = "CERES",
    title = "{Enhanced production of low mass electron pairs in 200-GeV/u S - Au collisions at the CERN SPS}",
    reportNumber = "CERN-PPE-95-026, CERN-PPE-95-26",
    doi = "10.1103/PhysRevLett.75.1272",
    journal = "Phys. Rev. Lett.",
    volume = "75",
    pages = "1272--1275",
    year = "1995"
}

@article{Agakichiev:1998kip,
    author = "{Agakichiev et al.}",
    title = "{Systematic study of low mass electron pair production in p Be and p Au collisions at 450-GeV/c}",
    doi = "10.1007/PL00021659",
    journal = "Eur. Phys. J. C",
    volume = "4",
    pages = "231--247",
    year = "1998"
}

@article{CERES:2005uih,
    author = "{Agakichiev et al.}",
    collaboration = "CERES",
    title = "{e+ e- pair production in Pb - Au collisions at 158-GeV per nucleon}",
    eprint = "nucl-ex/0506002",
    archivePrefix = "arXiv",
    doi = "10.1140/epjc/s2005-02272-3",
    journal = "Eur. Phys. J. C",
    volume = "41",
    pages = "475--513",
    year = "2005"
}

@article{HELIOS3:1998xeb,
    author = "{Angelis et al.}",
    collaboration = "HELIOS/3",
    title = "{Excess of continuum dimuon production at masses between threshold and the J / Psi in S - W interactions at 200-GeV/c/nucleon}",
    reportNumber = "CERN-EP-98-082, CERN-EP-98-82",
    doi = "10.1007/s100520050707",
    journal = "Eur. Phys. J. C",
    volume = "13",
    pages = "433--452",
    year = "2000"
}

@article{NA60:2006ymb,
    author = "{Arnaldi et al.}",
    collaboration = "NA60",
    title = "{First measurement of the rho spectral function in high-energy nuclear collisions}",
    eprint = "nucl-ex/0605007",
    archivePrefix = "arXiv",
    doi = "10.1103/PhysRevLett.96.162302",
    journal = "Phys. Rev. Lett.",
    volume = "96",
    pages = "162302",
    year = "2006"
}

@article{NA60:2007cjy,
    author = "{Damjanovic et al.}",
    editor = "Jacobs, P. and Wang, Xin-Nian",
    collaboration = "NA60",
    title = "{NA60 results on the rho spectral function in In-In collisions}",
    eprint = "nucl-ex/0701015",
    archivePrefix = "arXiv",
    doi = "10.1016/j.nuclphysa.2006.11.015",
    journal = "Nucl. Phys. A",
    volume = "783",
    pages = "327--334",
    year = "2007"
}

@article{NA60:2007lzy,
    author = "{Arnaldi et al.}",
    collaboration = "NA60",
    title = "{Evidence for radial flow of thermal dileptons in high-energy nuclear collisions}",
    eprint = "0711.1816",
    archivePrefix = "arXiv",
    primaryClass = "nucl-ex",
    doi = "10.1103/PhysRevLett.100.022302",
    journal = "Phys. Rev. Lett.",
    volume = "100",
    pages = "022302",
    year = "2008"
}

@article{CERES:2006wcq,
    author = "{Adamova et al.}",
    collaboration = "CERES",
    title = "{Modification of the rho-meson detected by low-mass electron-positron pairs in central Pb-Au collisions at 158-A-GeV/c}",
    eprint = "nucl-ex/0611022",
    archivePrefix = "arXiv",
    doi = "10.1016/j.physletb.2008.07.104",
    journal = "Phys. Lett. B",
    volume = "666",
    pages = "425--429",
    year = "2008"
}

@article{ALEPH:1998rgl,
    author = "{Barate et al.}",
    collaboration = "ALEPH",
    title = "{Measurement of the spectral functions of axial - vector hadronic tau decays and determination of alpha(S)(M**2(tau))}",
    reportNumber = "CERN-EP-98-012, CERN-EP-98-12",
    doi = "10.1007/s100520050217",
    journal = "Eur. Phys. J. C",
    volume = "4",
    pages = "409--431",
    year = "1998"
}

@article{OPAL:1998rrm,
    author = "{Ackerstaff et al.}",
    collaboration = "OPAL",
    title = "{Measurement of the strong coupling constant alpha(s) and the vector and axial vector spectral functions in hadronic tau decays}",
    eprint = "hep-ex/9808019",
    archivePrefix = "arXiv",
    reportNumber = "CERN-EP-98-102",
    doi = "10.1007/s100529901061",
    journal = "Eur. Phys. J. C",
    volume = "7",
    pages = "571--593",
    year = "1999"
}

@article{STAR:2013pwb,
    author = "{Adamczyk et al.}",
    collaboration = "STAR",
    title = "{Dielectron Mass Spectra from Au+Au Collisions at $\sqrt{s_{\rm NN}}$ = 200 GeV}",
    eprint = "1312.7397",
    archivePrefix = "arXiv",
    primaryClass = "hep-ex",
    doi = "10.1103/PhysRevLett.113.022301",
    journal = "Phys. Rev. Lett.",
    volume = "113",
    number = "2",
    pages = "022301",
    year = "2014",
    note = "[Addendum: Phys.Rev.Lett. 113, 049903 (2014)]"
}

@article{PHENIX:2015vek,
    author = "{Adare et al.}",
    collaboration = "PHENIX",
    title = "{Dielectron production in Au$+$Au collisions at $\sqrt{s_{NN}}$=200 GeV}",
    eprint = "1509.04667",
    archivePrefix = "arXiv",
    primaryClass = "nucl-ex",
    doi = "10.1103/PhysRevC.93.014904",
    journal = "Phys. Rev. C",
    volume = "93",
    number = "1",
    pages = "014904",
    year = "2016"
}

@article{Hohler:2013eba,
    author = "Hohler, Paul M. and Rapp, Ralf",
    title = "{Is $\rho$-Meson Melting Compatible with Chiral Restoration?}",
    eprint = "1311.2921",
    archivePrefix = "arXiv",
    primaryClass = "hep-ph",
    doi = "10.1016/j.physletb.2014.02.021",
    journal = "Phys. Lett. B",
    volume = "731",
    pages = "103--109",
    year = "2014"
}

@article{Harada:2008hj,
    author = "Harada, Masayasu and Sasaki, Chihiro and Weise, Wolfram",
    title = "{Vector-axialvector mixing from a chiral effective field theory at finite temperature}",
    eprint = "0807.1417",
    archivePrefix = "arXiv",
    primaryClass = "hep-ph",
    doi = "10.1103/PhysRevD.78.114003",
    journal = "Phys. Rev. D",
    volume = "78",

    pages = "114003",
    year = "2008"
}

@article{Naruki:2005kd,
    author = "{Naruki et al.}",
    title = "{Experimental signature of the medium modification for rho and omega mesons in 12-GeV p + A reactions}",
    eprint = "nucl-ex/0504016",
    archivePrefix = "arXiv",
    doi = "10.1103/PhysRevLett.96.092301",
    journal = "Phys. Rev. Lett.",
    volume = "96",
    pages = "092301",
    year = "2006"
}

@article{KEK-PS-E325:2005wbm,
    author = "{Muto et al.}",
    collaboration = "KEK-PS-E325",
    title = "{Evidence for in-medium modification of the phi meson at normal nuclear density}",
    eprint = "nucl-ex/0511019",
    archivePrefix = "arXiv",
    doi = "10.1103/PhysRevLett.98.042501",
    journal = "Phys. Rev. Lett.",
    volume = "98",
    pages = "042501",
    year = "2007"
}

@article{Metag:2010zza,
    author = "Metag, Volker",
    editor = "Choudhury, R. K. and Mohanty, A. K. and Saxena, Alok and Mahata, K. and Santra, S.",
    title = "{Vector mesons in strongly interacting matter}",
    doi = "10.1007/s12043-010-0108-6",
    journal = "Pramana",
    volume = "75",
    pages = "195--206",
    year = "2010"
}

@article{Shuryak:2026pqt,
    author = "Shuryak, Edward and Zahed, Ismail",
    title = "{The Hadron-Parton Bridge, From the QCD Vacuum to Partons}",
    eprint = "2601.15085",
    archivePrefix = "arXiv",
    primaryClass = "hep-ph",
    month = "1",
    year = "2026"
}

@article{Friedman:2025avs,
    author = "Friedman, E. and Gal, A.",
    title = "{Revisiting {\ensuremath{\eta}}'(958) nuclear states}",
    eprint = "2507.05832",
    archivePrefix = "arXiv",
    primaryClass = "nucl-th",
    doi = "10.1016/j.physletb.2025.139916",
    journal = "Phys. Lett. B",
    volume = "870",
    pages = "139916",
    year = "2025"
}

@article{Nagahiro:2012aq,
    author = "Nagahiro, Hideko and Jido, Daisuke and Fujioka, Hiroyuki and Itahashi, Kenta and Hirenzaki, Satoru",
    title = "{Formation of $\eta'(958)$-mesic nuclei by ($p,d$) reaction}",
    eprint = "1211.2506",
    archivePrefix = "arXiv",
    primaryClass = "nucl-th",
    reportNumber = "YITP-12-81, J-PARC-TH-0019",
    doi = "10.1103/PhysRevC.87.045201",
    journal = "Phys. Rev. C",
    volume = "87",
    number = "4",
    pages = "045201",
    year = "2013"
}

@article{e-PRiME:2025fer,
    author = "{Sekiya et al.}",
    collaboration = "{\ensuremath{\eta}}-PRiME, Super-FRS",
    title = "{Excitation Spectra of the C12(p,d) Reaction near the {\ensuremath{\eta}}'-Meson Emission Threshold Measured in Coincidence with High-Momentum Protons}",
    eprint = "2509.07824",
    archivePrefix = "arXiv",
    primaryClass = "nucl-ex",
    doi = "10.1103/6vsl-ng7x",
    journal = "Phys. Rev. Lett.",
    volume = "136",
    number = "14",
    pages = "142501",
    year = "2026"
}

@article{Kohno:2015hha,
    author = "Kohno, M.",
    title = "{Nuclear saturation in lowest-order Brueckner theory with two- and three-nucleon forces in view of chiral effective field theory}",
    eprint = "1510.07469",
    archivePrefix = "arXiv",
    primaryClass = "nucl-th",
    doi = "10.1093/ptep/ptv166",
    journal = "PTEP",
    volume = "2015",
    number = "12",
    pages = "123D02",
    year = "2015"
}

@article{Hatsuda:1988mv,
    author = "Hatsuda, T. and Prakash, M.",
    title = "{Parity Doubling of the Nucleon and First Order Chiral Transition in Dense Matter}",
    reportNumber = "NTG-88-62",
    doi = "10.1016/0370-2693(89)91040-X",
    journal = "Phys. Lett. B",
    volume = "224",
    pages = "11--15",
    year = "1989"
}

@article{Nemoto:1998um,
    author = "Nemoto, Y. and Jido, D. and Oka, M. and Hosaka, A.",
    title = "{Decays of 1/2- baryons in chiral effective theory}",
    eprint = "hep-ph/9710445",
    archivePrefix = "arXiv",
    reportNumber = "TIT-HEP-379-NP",
    doi = "10.1103/PhysRevD.57.4124",
    journal = "Phys. Rev. D",
    volume = "57",
    pages = "4124--4135",
    year = "1998"
}

@article{Steinheimer:2011ea,
    author = "Steinheimer, J. and Schramm, S. and Stocker, H.",
    title = "{The hadronic SU(3) Parity Doublet Model for Dense Matter, its extension to quarks and the strange equation of state}",
    eprint = "1108.2596",
    archivePrefix = "arXiv",
    primaryClass = "hep-ph",
    doi = "10.1103/PhysRevC.84.045208",
    journal = "Phys. Rev. C",
    volume = "84",
    pages = "045208",
    year = "2011"
}

@article{LEPS2BGOegg:2026xxj,
    author = "Matsumura, Y. and others",
    collaboration = "LEPS2/BGOegg",
    title = "{Direct Measurement of the In-Medium \(\eta'\) Mass Spectrum through the \(\gamma\gamma\) Decay Channel}",
    eprint = "2608.25074",
    archivePrefix = "arXiv",
    primaryClass = "nucl-ex",
    doi = "10.1103/x414-n6g9",
    journal = "Phys. Rev. Lett.",
    volume = "137",
    number = "5",
    pages = "051901",
    year = "2026"
}

@book{Hellmann1937,
  author    = {Hans Hellmann},
  title     = {Einf{\"u}hrung in die Quantenchemie},
  publisher = {Franz Deuticke},
  address   = {Leipzig and Vienna},
  year      = {1937}
}

@article{Feynman1939,
  author  = {Richard P. Feynman},
  title   = {Forces in Molecules},
  journal = {Physical Review},
  volume  = {56},
  number  = {4},
  pages   = {340--343},
  year    = {1939},
  doi     = {10.1103/PhysRev.56.340}
}

@article{Suenaga:2017wbb,
    author = "Suenaga, Daiki",
    title = "{Examination of $N^*(1535)$ as a probe to observe the partial restoration of chiral symmetry in nuclear matter}",
    eprint = "1704.03630",
    archivePrefix = "arXiv",
    primaryClass = "nucl-th",
    doi = "10.1103/PhysRevC.97.045203",
    journal = "Phys. Rev. C",
    volume = "97",
    number = "4",
    pages = "045203",
    year = "2018"
}

@article{Zhang:2015fda,
    author = "Zhang, Zhao and Kunihiro, Teiji",
    title = "{Combined chiral and diquark fluctuations along QCD critical line and enhanced baryon production with parity doubling}",
    eprint = "1510.04417",
    archivePrefix = "arXiv",
    primaryClass = "hep-ph",
    doi = "10.1140/epja/i2016-16230-y",
    journal = "Eur. Phys. J. A",
    volume = "52",
    number = "8",
    pages = "230",
    year = "2016"
}

@article{Witten:1979vv,
    author = "Witten, Edward",
    title = "{Current Algebra Theorems for the U(1) ``Goldstone Boson''}",
    doi = "10.1016/0550-3213(79)90031-2",
    journal = "Nucl. Phys. B",
    volume = "156",
    pages = "269--283",
    year = "1979"
}

@article{Veneziano:1979ec,
    author = "Veneziano, Gabriele",
    title = "{U(1) Without Instantons}",
    reportNumber = "CERN-TH-2651",
    doi = "10.1016/0550-3213(79)90332-8",
    journal = "Nucl. Phys. B",
    volume = "159",
    pages = "213--224",
    year = "1979"
}

@misc{Bonanno:2026topology,
    author = "Bonanno, Claudio and Bonati, Claudio and D'Elia, Massimo",
    title = "{Topological Susceptibility and QCD at Finite Theta Angle}",
    year = "2026",
    eprint = "2604.28035",
    archivePrefix = "arXiv",
    primaryClass = "hep-lat",
    note = "Companion chapter for the Elsevier Encyclopedia of Nuclear Physics"
}

@article{Kawarabayashi:1980dp,
    author = "Kawarabayashi, Ken and Ohta, Norisuke",
    title = "{The Problem of Eta in the Large N Limit: Effective Lagrangian Approach}",
    journal = "Nucl. Phys. B",
    volume = "175",
    pages = "477--492",
    year = "1980",
    doi = "10.1016/0550-3213(80)90024-3"
}

@article{DiVecchia:1981a,
    author = "Di Vecchia, P. and Nicodemi, G. and Pettorino, R. and Veneziano, G.",
    title = "{Large N, Chiral Approach to Pseudoscalar Masses, Mixings and Decays}",
    journal = "Nucl. Phys. B",
    volume = "181",
    pages = "318--334",
    year = "1981",
    doi = "10.1016/0550-3213(81)90356-4"
}

@article{Christos:1984tu,
    author = "Christos, G. A.",
    title = "{Chiral Symmetry and the U(1) Problem}",
    journal = "Phys. Rept.",
    volume = "116",
    pages = "251--336",
    year = "1984",
    doi = "10.1016/0370-1573(84)90025-5"
}

@article{Smit:1987un,
    author = "Smit, J. and Vink, J. C.",
    title = "{Neutral Pseudoscalar Masses in Lattice QCD}",
    journal = "Nucl. Phys. B",
    volume = "284",
    pages = "234--252",
    year = "1987",
    doi = "10.1016/0550-3213(87)90034-4"
}

@article{Kunihiro:1988sqc,
    author = "Kunihiro, Teiji",
    title = "{Strange-Quark Condensate in Nucleon Based on an Effective Lagrangian Incorporating U$_A$(1) Anomaly}",
    journal = "Prog. Theor. Phys.",
    volume = "80",
    pages = "34--38",
    year = "1988",
    doi = "10.1143/PTP.80.34"
}

@article{Nelson:1989sgc,
    author = "Nelson, Ann E.",
    title = "{Strange Goings on in the Chiral Quark Model}",
    journal = "Phys. Lett. B",
    volume = "221",
    pages = "60--66",
    year = "1989",
    doi = "10.1016/0370-2693(89)90192-5"
}

@article{Hatsuda:1991fml,
    author = "Hatsuda, Tetsuo and Kunihiro, Teiji",
    title = "{Flavor Mixing in the Low-Energy Hadron Dynamics: Interplay of the SU(3)$_f$ Breaking and the U(1)$_A$ Anomaly}",
    journal = "Z. Phys. C",
    volume = "51",
    pages = "49--68",
    year = "1991",
    doi = "10.1007/BF01579559"
}

@article{Wess:1971yu,
  author  = {Wess, J. and Zumino, B.},
  title   = {Consequences of anomalous Ward identities},
  journal = {Phys. Lett. B},
  volume  = {37},
  pages   = {95--97},
  year    = {1971},
  doi     = {10.1016/0370-2693(71)90582-X}
}

@article{Witten:1983tw,
  author  = {Witten, Edward},
  title   = {Global Aspects of Current Algebra},
  journal = {Nucl. Phys. B},
  volume  = {223},
  pages   = {422--432},
  year    = {1983},
  doi     = {10.1016/0550-3213(83)90063-9}
}

@article{Yamazaki:2018stk,
  author        = {Yamazaki, Takahiro and Harada, Masayasu},
  title         = {Chiral partner structure of light nucleons in an extended parity doublet model},
  journal       = {Phys. Rev. D},
  volume        = {99},
  number        = {3},
  pages         = {034012},
  year          = {2019},
  doi           = {10.1103/PhysRevD.99.034012},
  eprint        = {1809.02359},
  archivePrefix = {arXiv},
  primaryClass  = {hep-ph}
}

\end{document}